\documentclass[%
 amsmath,amssymb,
 a4paper,
published,accepted=2024-07-01 
]{quantumarticle}

\pdfoutput=1

\usepackage{graphicx}

\usepackage{microtype}
\usepackage{hyperref}
\usepackage{amsthm}

\usepackage{dsfont}
\usepackage{mathrsfs}
\usepackage{mathtools}
\usepackage{array}
\usepackage{braket}
\usepackage{qcircuit}
\usepackage[caption=false]{subfig}
\usepackage[export]{adjustbox}
\usepackage[dvipsnames]{xcolor}
\usepackage{makecell}
\usepackage{array}
\newcolumntype{?}{!{\vrule width 1pt}}

\usepackage[square, comma, sort&compress, numbers]{natbib}

\def\tcm{T.C.M. Group, Cavendish Laboratory, University of Cambridge, J.J. Thomson Avenue, Cambridge, CB3 0HE, UK}
\def\DAMTP{DAMTP, University of Cambridge, Wilberforce Road, Cambridge, CB3 0WA, UK}

\begin{document}

\title{A new twist on the Majorana surface code: Bosonic and fermionic defects for fault-tolerant quantum computation}

\author{Campbell McLauchlan}
\affiliation{\DAMTP}
\author{Benjamin B\'eri}
\affiliation{\DAMTP}
\affiliation{\tcm}

\date{May 2023}
             
\begin{abstract}
Majorana
zero modes (MZMs) are promising candidates for topologically-protected quantum computing hardware, %
however their large-scale use will likely require quantum error correction.
Majorana surface codes (MSCs) have been proposed to achieve this.
However, many MSC properties remain unexplored.
We present a unified framework for MSC ``twist defects" -- anyon-like objects encoding quantum information.
We show that twist defects in MSCs can encode twice the amount of topologically protected information as in qubit-based codes or other MSC encoding schemes.
This is due to twists encoding both logical qubits and ``logical MZMs," with the latter enhancing the protection microscopic MZMs can offer.
We explain how to perform universal computation with logical qubits and logical MZMs while potentially using far fewer resources than in other MSC schemes.
All Clifford gates can be implemented on logical qubits by braiding twist defects.
We introduce lattice-surgery-based techniques for computing with logical MZMs and logical qubits, achieving the effect of Clifford gates with zero time overhead.
We also show that logical MZMs may result in improved spatial overheads for sufficiently low rates of quasi-particle poisoning.
Finally, we introduce a novel MSC analogue of transversal gates that
achieves encoded Clifford gates in small codes by braiding microscopic MZMs.
MSC twist defects thus open new paths towards fault-tolerant quantum computation.
\end{abstract}

\maketitle

\section{Introduction}

Quantum
computers promise considerable advantages over their classical counterparts, such as the ability to efficiently simulate large quantum systems~\cite{nielsen_chuang}.
In order to achieve these advantages, one requires quantum systems that are well-shielded from environmental noise, and on which one can perform operations fault-tolerantly.

Majorana zero modes (MZMs) in topological superconductors~\cite{Volovik_Vortex, Kitaev_chain2001, MjFerms_Surface_TIs,von_Oppen_Majorana_Nanowires2010, Das_Sarma_Majorana_Nanowire_2010, Alicea_Review_2012} have been the focus of intense experimental efforts recently~\cite{Nature_Maj_Experiment2012, Science_Maj_Experiment_2012, Albrecht-et-al-exponential-2016, Maj_Experiment_2_2017, Non-Locality_Experiment_2018, Maj_Experiment_3, vaitiekenas2021zero, Lutchyn_Review_2018, Nanowire_Review_2021}.
MZMs promise many advantages for quantum computing, including topological protection of quantum information~\cite{Kitaev_chain2001,MZM_TQC_Review2015} and fault-tolerant generation of Clifford gates~\cite{Gottesman_thesis} by braiding operations~\cite{MZMs_in_pwave_superconds2001,Kitaev_Anyons2006,Coulomb_assisted_braiding2012,MZM_TQC_Review2015,Karzig_Majorana2017,Optimizing_Clifford_Gates_2020}. 
While braiding is not sufficient for universal quantum computation, techniques exist for implementing high-fidelity non-Clifford gates in MZM qubits~\cite{Karzig_geom_magic_2016, Karzig_geom_magic_meas2019}.
Despite these advantages, the ultimately limited coherence times~\cite{MZM_Coherence_Times2018}, the large timescales for braids~\cite{Diabatic_errors2016, Time_scales_for_braiding2016} and the effects of quasi-particle poisoning~\cite{Karzig_QPP_Maj_Qubits_2021} likely mean that MZMs must be combined with quantum error correction to achieve fault-tolerant quantum computation.

Surface codes~\cite{Kitaev_toric_code, Bravyi_Kitaev_Codes_with_Bdry,Top_quant_memory, Surface_Codes_Fowler_2012,Bombin_Color_Codes} are quantum error-correcting codes with many promising features.
They are naturally suited to physical implementation due to the locality of their stabilizers (operators to be measured for error correction), and their high error thresholds (the maximum tolerable single-qubit error rate)~\cite{High_Thresh_Raussendorf_2007,Terhal_QEC_Review}.
They also support various schemes for fault-tolerant state preparation~\cite{Top_quant_memory,Surface_Code_Lattice_Surg} and logic gate implementation~\cite{High_Thresh_Raussendorf_2007,Transversal_gates_folded_surface_codes_2016,Holes_Twists_SurfaceCode,Lattice_Surg_with_Twist,Brown_Non-Clifford_Gate_2020,Twist-Free_Lattice_Surgery_2022,Twist-Based_Lattice_Surgery_2022}.
A leading paradigm stores encoded qubits in surface code patches with holes~\cite{Punctures_Raussendorf_2006, High_Thresh_Raussendorf_2007, Bombin_Code_Deformation} and/or extrinsic defects called twists~\cite{Bombin_Twists, Bombin_twists_code_deformation,Twists_genons,Dua_Twists_Majorana_stats}.
These twists and holes can be braided to fault-tolerantly enact gates on encoded qubits~\cite{Bombin_twists_code_deformation,Holes_Twists_SurfaceCode}, and can reduce error-correction overheads~\cite{Reduced_ST_Costs,Yoder_Surface_code_twist,BBrown_Twists_CC}.
Recent experimental advances include proof-of-principle demonstrations of error correction~\cite{krinner2022realizing,Google_SC} and twist braiding~\cite{Google_braid} in surface codes.

Surface codes also have a fermionic counterpart, built from MZMs rather than qubits~\cite{Maj_Ferm_Codes, MjFerm_Surface_Code, Maj_Triangle_Code2018, Realistic_MSC,Roadmap_to_MSCs, QC_with_MFCs}. 
These Majorana surface codes (MSCs) utilize the topological protection of MZMs and also enjoy partial protection of quantum information by fermion parity conservation~\cite{Maj_Ferm_Codes, Maj_Triangle_Code2018}.
In certain physical setups, stabilizer measurements can also be performed in a single shot~\cite{MjFerm_Surface_Code, FPBC}; this is in contrast to qubit-based surface codes where this requires multiple gates and ancillas.
Setups with large charging energies require fewer stabilizer measurements~\cite{Karzig_Majorana2017,QC_with_MFCs}.
Additionally, MSCs have favourable error thresholds~\cite{Maj_Triangle_Code2018,Ferm_Error_Corr_2019},
and can be used for fault-tolerant fermionic computation~\cite{Maj_Triangle_Code2018} that allows considerable resource savings on fermionic simulation tasks~\cite{Bravyi_Kitaev_Fermionic_QC2002, Maj_Based_FQC2018}.

The full range of MSC features are yet to be explored. 
Analogously to the bosonic (i.e., qubit-based) surface code, storing logical qubits in holes can achieve a logical CNOT gate by braiding the holes~\cite{Surface_Codes_Fowler_2012,MjFerm_Surface_Code,Roadmap_to_MSCs}.
But other Clifford gates require measuring high-weight operators, or distilling ancilla states and implementing gadget circuits~\cite{MjFerm_Surface_Code,Roadmap_to_MSCs}. 
While braiding holes alone cannot achieve all Clifford gates, the study of twist defects may expand the gate set: 
In the bosonic surface code, braiding twists and holes can achieve all Clifford gates~\cite{Holes_Twists_SurfaceCode}, and in the bosonic Color Code~\cite{Bombin_Color_Codes} (closely related to the MSC~\cite{Maj_Ferm_Codes}), braiding twists alone can achieve the same~\cite{Bombin_twists_code_deformation}.
This suggests that twists in MSCs are features worthy of exploration.

Motivated by this, in this work we establish the fundamental features and classification of twist defects in the MSC, and show how to use them for quantum computation.
Besides highlighting features analogous to bosonic codes, we show that MSC
twists can include ``logical MZMs"~\cite{Maj_Ferm_Codes,Maj_Triangle_Code2018}:
fermion-parity-odd objects that can enhance the protection from ``microscopic" MZMs and enable fault-tolerant fermionic quantum computation.
(Objects related to logical MZMs appeared also in other contexts~\cite{Akhmerov2010,Goldstein2012,Behrends2020,FPBC}.)
The presence of both bosonic and fermionic twist features opens up two modes of fault-tolerant quantum computation that could be profitably combined. 
Our study of MSC twists also complements results on the classification and use of topological defects in the context of topological order~\cite{Defects_Abelian_states,Twist_Liquid,Twists_genons,Barkeshli_defects_gauging2019}.

The rest of this work is structured as follows. 
We start, in Section~\ref{sec:Background_MSC}, with the main ingredients of MSCs, including their anyons, and fault-tolerant quantum computation.
We also establish here the types of boundaries supported by MSCs.
In Section~\ref{sec:Domain_Walls_Twists}, we classify MSC twists and detail their properties, including their logical Hilbert space dimensions (i.e., their ``quantum dimensions").
As in bosonic codes, twists are classified via the symmetries of the MSC's anyons. 
But logical MZMs are inherently fermionic and hence have no counterparts in bosonic codes.
We find that a phase transition along the domain wall connecting two twists can result in logical MZMs emerging at the twists.
This expands on how one-dimensional physics can enrich the classification of twist defects~\cite{Defects_Abelian_states}, by showing how this may lead to non-local logical MZMs.
While mentioned in the MSC literature~\cite{Maj_Ferm_Codes,Maj_Triangle_Code2018}, the range of scenarios in which one expects to find logical MZMs and their link to twists have not yet been explored. 

In Section~\ref{sec:FT_Comp_with_Twists}, we discuss how universal fault-tolerant quantum computation can be performed with twists in the MSC, simultaneously utilizing both logical qubits and logical MZMs. 
This allows for roughly twice the information to be stored and manipulated per twist as would be possible in bosonic  approaches. 
The twists we discuss provide independent ways to protect information from fermion-parity-violating errors (so-called quasi-particle poisoning or QPP events~\cite{MZM_TQC_Review2015}) and fermion-parity-conserving (FPC) 
errors.

In Section~\ref{sec:Low_Over_Ferm_QC} we explore the spatial overheads of computing with logical MZMs, and show that these may be considerably reduced if the QPP probability is suitably small.
We focus on a fault-tolerant implementation of ``Pauli-based computation" (PBC)~\cite{Bravyi_PBC2016, Mithuna_Magic_State_PBC2019,FPBC} with the MSC;
PBC is a measurement-based model of quantum computation that optimizes the use of quantum resources, at the cost of additional, but efficient, classical computing. 

Finally, in Section~\ref{sec:Braids_Small_Codes}, we present a new method to implement fault-tolerant gates in MSCs. %
This is a fermionic analogue of transversal gates in bosonic codes.
It implements encoded Clifford gates by braiding the MSC's constituent MZMs.
This technique is particularly suitable for small codes, and hence could be utilized in early-generation Majorana devices.
In Section~\ref{sec:Conclusion} we conclude and discuss avenues for future work.

\section{Basic MSC Ingredients}\label{sec:Background_MSC}

In this section, we briefly review the basics of Majorana fermion codes~\cite{Maj_Ferm_Codes}, fault-tolerant quantum computing~\cite{nielsen_chuang}, and the construction of MSCs and their anyons~\cite{MjFerm_Surface_Code,Roadmap_to_MSCs,QC_with_MFCs}. 
We also establish the types of boundaries MSCs can support and describe their lattice realization. 

Majorana fermion codes, and as such the MSC, are based on Majorana operators $\gamma_i$ satisfying 
\begin{align}\label{eqn:Maj_comm_single}
\gamma_j^\dagger = \gamma_j,\quad \lbrace \gamma_i, \gamma_j \rbrace = 2\delta_{i,j},
\end{align}
where $\lbrace \, , \rbrace$ is the anti-commutator. 
MZMs in topological superconductors realize this algebra. 
In addition, they commute with and are absent from the Hamiltonian, %
hence the term zero mode. 
In the MSC, the latter property will not hold for constituent MZMs, thus we will sometimes call them simply Majoranas. 
Logical MZMs however, as we will see, enjoy both the Majorana [Equation~\eqref{eqn:Maj_comm_single}] and the zero mode properties.

\subsection{Majorana Fermion Codes}\label{subsec:MFCs}
Let Maj$(2n)$ denote the group of Majorana strings generated by Majorana operators $\gamma_1,\ldots,\gamma_{2n}$ and the phase factor $i$. 
For index subsets $A$ and $B$, take $\Gamma_A$ and $\Gamma_B$ to be $\Gamma_{A\, (B)} = \prod_{j\in A\, (B)}\gamma_j$. 
Equation~\eqref{eqn:Maj_comm_single} implies
\begin{align}\label{eqn:MajModeCommutation}
\Gamma_A \Gamma_B = (-1)^{|A|\cdot |B| + |A\cap B|} \Gamma_B \Gamma_A,
\end{align}
where $|A|$ is the weight of $\Gamma_A$, i.e., the cardinality of $A$.  

\renewcommand{\arraystretch}{1.5}
\begin{table*}
    \centering
    \begin{tabular}{c|c}
        \textbf{Majorana fermion code object} & \textbf{Definition}\\
        \hline
        Total fermion parity $\Gamma$ & $\Gamma = i^n \prod_{j=1}^{2n}\gamma_j$\\
        \hline
        Stabilizer group $\mathcal{S}$ & Abelian subgroup $\mathcal{S}\subset \text{Maj}(2n)$ s.t.  $-I\notin \mathcal{S}$ and $[s,\Gamma]=0\; \forall s\in\mathcal{S}$ \\
        \hline
        Logical operator $L$
                & 
        \begin{tabular}{c}
        Equivalence class $[L] = \lbrace Ls|s\in\mathcal{S}\rbrace$ of operators,\\
        s.t. $[L,s] = 0\; \forall s\in\mathcal{S}$, $L\notin \mathcal{S}$,
        and $[L,\Gamma]=0$
        \end{tabular}\\
        \hline
        Logical MZM $\Upsilon$ & 
        \begin{tabular}{c}
        Equivalence class $[\Upsilon] = \lbrace \Upsilon s | s\in\mathcal{S}\rbrace$ of operators, \\
        s.t. $[\Upsilon, s] = 0 \; \forall s\in\mathcal{S}$ and $\lbrace \Upsilon, \Gamma\rbrace = 0$
        \end{tabular}\\
        \hline
        Code distance $d$ & $d = \min_{L,s}|Ls|$, where $|Ls|$ is the weight of the logical $Ls$\\
        \hline
        Code diameter $l$ & 
        \begin{tabular}{c}
        $l = \min_{L,s}\text{diam}(Ls)$, \\
        where diam$(Ls)$ is the geometric diameter of the support of $Ls$
        \end{tabular}\\
        \hline
    \end{tabular}
    \caption{Summary of important objects and quantities of Majorana fermion codes.}\label{tab:Summary}
\end{table*}

From the $2n$ Majorana operators we can construct $n$ fermionic creation operators $c_{j}^\dagger = (\gamma_{2j-1} - i \gamma_{2j})/2$. 
The $2n$ Majorana operators thus act on a $2^n$-dimensional space, with basis states labeled by $c_j^\dagger c_j$.
Let $\Gamma = (-1)^{\sum_{j}c_j^\dagger c_j} = i^{n}\prod_{j=1}^{2n} \gamma_j$ denote the total fermion parity operator.
Half of the states will have even fermion parity, $\Gamma \ket{\psi} = \ket{\psi}$, and the other half will have odd fermion parity.

A Majorana fermion code is specified by its stabilizer group $\mathcal{S}$, an Abelian subgroup of $\text{Maj}(2n)$~\cite{Gottesman_thesis, Maj_Ferm_Codes}.
The stabilizer group satisfies $-I\notin \mathcal{S}$ and all $s\in\mathcal{S}$ have even weight, hence $[\Gamma, s]=0$. 
(We assume for now that $\Gamma\in\mathcal{S}$, although this assumption will be dropped below.)
Logical (or code) states $\ket{\psi}$ span the code space. They satisfy $s\ket{\psi} = \ket{\psi} \; \forall s\in\mathcal{S}$. 
Since $\Gamma\in\mathcal{S}$, code states have definite (namely even) parity, hence they can be superposed~\cite{Bravyi_Kitaev_Fermionic_QC2002,Maj_Ferm_Codes}.  
Suppose that $\mathcal{S}$ has $m$ independent generators. 
This leaves a $2^{n-m}$-dimensional code space. 
Hence $k = n-m$ is the number of logical qubits.

The code's logical operators are even-weight elements of Maj$(2n)$ that commute with all $s\in\mathcal{S}$ but are not themselves in $\mathcal{S}$. 
We define logical operators $\bar{X}_j$ and $\bar{Z}_j$ such that they act as Pauli operators on logical qubit $j$.
Note that a logical operator $L$ and $Ls$ act equivalently on code states for any $s\in \mathcal{S}$. 
Hence we define the equivalence class of logical operator $L$ to be $[L] = \lbrace Ls | s\in \mathcal{S}\rbrace$.
Some Majorana fermion codes also have parity-non-preserving (odd-weight) operators that commute with all members of $\mathcal{S}$ apart from $\Gamma$ (in such cases $\Gamma$ is removed from $\mathcal{S}$, and we separately require that logical operators commute with $\Gamma$). 
We refer to these operators as logical MZMs, for reasons mentioned above (see also Section~\ref{sec:Domain_Walls_Twists}).

Suppose the code is defined on a $D$-dimensional lattice with one Majorana operator per site. 
Suppose also that $\mathcal{S}$ has generators $\mathcal{O}_p$ whose support (i.e., sites where $\mathcal{O}_p$ differs from identity) is local. 
While logical operators in $[L]$ act the same way on code states, they can have different weights.
The distance of the code is defined as the smallest among the weights of logical operators, $\min_{L,s}|Ls|$.
To further characterize codes, we define the diameter $\text{diam}(\Gamma_A)$ of a logical operator $\Gamma_A$ to be the geometrical diameter of its support. 
Similarly to the distance, the code diameter is defined as  $\min_{L,s}\text{diam}(Ls)$. 
As noted in Ref.~\cite{Maj_Ferm_Codes}, and as we explain in Section~\ref{subsec:LogMZM_fermion_twist}, a code with logical MZMs can have large diameter but small distance. 
See Table~\ref{tab:Summary} for a summary of the objects and quantities introduced above.

Error correction is performed in the same way in Majorana fermion codes and bosonic stabilizer codes~\cite{Gottesman_thesis,Maj_Ferm_Codes}. 
First, one measures the generators of $\mathcal{S}$; the returned outcomes form the ``syndrome". 
If no error has occurred, then all of these outcomes are $+1$; if some of them are $-1$, then an error has occurred (assuming for now perfect measurements).
Then one can use classical algorithms, called decoders, to diagnose the syndrome, and determine the correction operation to map the post-measurement state back to the original code space. 
Methods also exist for performing stabilizer measurements in a way that is resilient to measurement errors, and hence error correction can be performed fault-tolerantly~\cite{Top_quant_memory,nielsen_chuang,Terhal_QEC_Review}.
Error correction results in a logical error rate approaching zero with increasing code distance when the error probability (both for storage and measurements) is below the error threshold~\cite{Top_quant_memory,Terhal_QEC_Review}.

\subsection{Universal Fault-Tolerant Quantum Computation}\label{subsec:Universal_FT_QC}

Beyond just storing quantum information we also require methods for performing gates on encoded qubits in ways resilient to faults or pre-existing errors in the system.
A universal set of logic gates allows one to enact an arbitrary quantum circuit~\cite{nielsen_chuang}. 
One such universal gate set is given by the Clifford gates combined with the $T$ gate. 
Suppose we have $k$ qubits. 
Let $\mathcal{P}_k$ be the $k$-qubit Pauli group and $\mathcal{C}_k$ be the $k$-qubit Clifford group.
The group $\mathcal{C}_k$ consists of Clifford gates: those unitary operations that preserve $\mathcal{P}_k$ under conjugation.
That is, for all $P\in\mathcal{P}_k$ and $U\in \mathcal{C}_k$, $U P U^\dagger \in \mathcal{P}_k$.
$\mathcal{C}_k$ is generated by the single-qubit Hadamard and phase gates and the two-qubit CNOT gate.  
The Hadamard gate $H_i = H_i^\dagger$ acting on qubit $i$ maps the Pauli operator $X_i$ (also acting on qubit $i$) to $H_i X_i H_i^\dagger = Z_i$.
The phase gate $S_i$ obeys $S_iX_iS_i^\dagger = Y_i$ and $S_iY_iS_i^\dagger = -X_i$.
The CNOT gate, controlled on qubit $i$ and targeted on qubit $j$ sends $X_i \mapsto X_i X_j$ and $Z_j \mapsto Z_iZ_j$ (while preserving $Z_i$ and $X_j$) under conjugation.
The single-qubit $T$ gate is a non-Clifford gate which acts as diag$(1,e^{i\pi/4})$ in the computational basis.

We now describe a universal set of gates for fermionic quantum computation with Majorana operators, in which $n$ qubits are ``densely" encoded~\cite{MZM_TQC_Review2015} using $2n+2$ Majorana operators into a fixed total fermion parity sector (either even or odd). 
This model of computation is ultimately equivalent to qubit-based computation, but it conveys advantages for tasks such as simulating other fermionic systems, as it avoids the overheads associated with non-local fermion-qubit mappings (e.g., Jordan-Wigner transformations)~\cite{Bravyi_Kitaev_Fermionic_QC2002,Maj_Based_FQC2018,OptimalFQmap}.

Braid operations generate some of the gates required for quantum computation~\cite{Bravyi_Kitaev_Fermionic_QC2002,MZM_TQC_Review2015}. 
These are unitary transformations of the form $\exp{(\frac{\pi}{4}\gamma_a\gamma_b)}$ for $a,b\in\lbrace 1,\ldots, 2n+2\rbrace$, sending $\gamma_a\mapsto -\gamma_b$ and $\gamma_b\mapsto \gamma_a$ under conjugation.
They can be performed in a (e.g., two-dimensional) physical system by braiding (physically or via measurements) two MZMs in a clockwise direction. 
Braids are naturally fault-tolerant, since the unitary transformation depends only on the topology of the path along which the state is taken, rather than the details of this path.
However, braids alone do not generate a universal gate set.

A universal gate set with $2n+2$ Majorana operators ($n\geq 1$) is given by two types of operators, denoted $B_4$ and $T_2$~\cite{Bravyi_Kitaev_Fermionic_QC2002}.
$B_4$ operators are fermionic variants of Clifford gates~\cite{FPBC},
and they have the form $B_{4,abcd} = \exp{(i\frac{\pi}{4}\gamma_a\gamma_b\gamma_c\gamma_d)}$
for distinct indices $a,b,c,d\in \lbrace 1,\ldots, 2n+2\rbrace$. 
We call these operators ``logical braids" since they have the same form as a braid operator, but between Majorana $\gamma_a$ and the logical Majorana operator $\Upsilon_{bcd}=i\gamma_b\gamma_c\gamma_d$.
(This logical Majorana is not necessarily a logical MZM for it may not commute with all stabilizer operators; when all the $\gamma_a$ are used for computation, the stabilizer group is $\lbrace \mathds{1},\Gamma\rbrace$).
If multiple logical Majorana operators $\Upsilon_j$ (with multi-index $j$) are defined on non-overlapping sites, they obey the same relations as Majorana operators [cf. Equation~\eqref{eqn:Maj_comm_single}]: $\Upsilon_j^\dagger = \Upsilon_j$ and $\lbrace \Upsilon_j,\Upsilon_k\rbrace = 2\delta_{j,k}$ for all $j,k$.
$T_2$ operators are non-Clifford gates, with the form $T_{2,ab} = \exp{(\frac{\pi}{8}\gamma_a\gamma_b)}$ for indices $a,b\in\lbrace 1,\ldots, 2n+2\rbrace$.
$B_4$ gates can be implemented fault-tolerantly via a combination of ancilla preparation, braids and fermion parity measurement~\cite{Bravyi_Kitaev_Fermionic_QC2002}, whereas the fault-tolerant implementation of $T_2$ gates requires magic state distillation (see below).

There are many proposed methods for performing logic gates fault-tolerantly on information stored in quantum codes.
For MSCs, previous proposals have suggested replacing Clifford gates with measurements, which can be performed fault-tolerantly via ``lattice surgery"~\cite{QC_with_MFCs,Surface_Code_Lattice_Surg,Maj_Triangle_Code2018}.
These involve fault-tolerantly measuring extra stabilizer operators, which results in the fusion of two or more patches of code. 
In the bosonic surface code, Clifford gates can be performed via the introduction of holes or twist defects (see Section~\ref{sec:Domain_Walls_Twists}) in the lattice.
Using ``code deformation", one can braid these lattice features, thereby achieving gates on the logical qubits~\cite{Bombin_Code_Deformation,Bombin_twists_code_deformation,Holes_Twists_SurfaceCode}.
Holes can also be introduced to the MSC and braided to achieve CNOT gates~\cite{MjFerm_Surface_Code}.
The Hadamard gate can also be applied to these qubits, but this involves measuring a high-weight Majorana operator, which may be challenging in physical setups~\cite{MjFerm_Surface_Code,Roadmap_to_MSCs}.
The $S$ gate can be implemented by preparing an ancilla logical qubit in a high-fidelity Pauli-$Y$ eigenstate, and then performing $H$ and CNOT gates to the ancilla and target qubits~\cite{MjFerm_Surface_Code,Roadmap_to_MSCs}.
Thus, all Clifford generators can be enacted in MSCs with holes.

$T$ gates may be enacted by preparing an ancilla logical qubit in a ``magic state" such as  $\frac{1}{\sqrt{2}}(\ket{0} + e^{i\pi/4}\ket{1})$.
By performing Clifford gates and measurements on the ancilla and target qubits, one can generate a $T$ gate acting on the target~\cite{nielsen_chuang}. 
To prepare these magic states fault-tolerantly, we can prepare many (noisy) copies in our quantum code, and distill a single purified copy, again via Clifford gates and measurements (this procedure is called magic state distillation)~\cite{Bravyi5p2_2006,Magic_State_Dist,MSD_Low_Overhead2012,MSD_Low_Overhead2013,MSD_Low_Overhead2017,MSD_Not_Costly2019}.
Magic states can also be used to enact $T_2$ gates (or other non-Clifford gates) on Majorana-based qubits~\cite{Maj_Based_FQC2018,FPBC}.

In addition to logic gates, we require the ability to fault-tolerantly prepare logical qubits in some state (e.g., the state $\ket{0}^{\otimes N}$ for $N$ logical qubits) and measure the logical qubits in the computational basis. 
In bosonic codes, these processes involve performing single-qubit Pauli measurements and stabilizer measurements~\cite{Terhal_QEC_Review}.
We will discuss how they are performed in the MSC below.

\subsection{Majorana Surface Code}

\begin{figure}
\centering
\includegraphics[width=\linewidth]{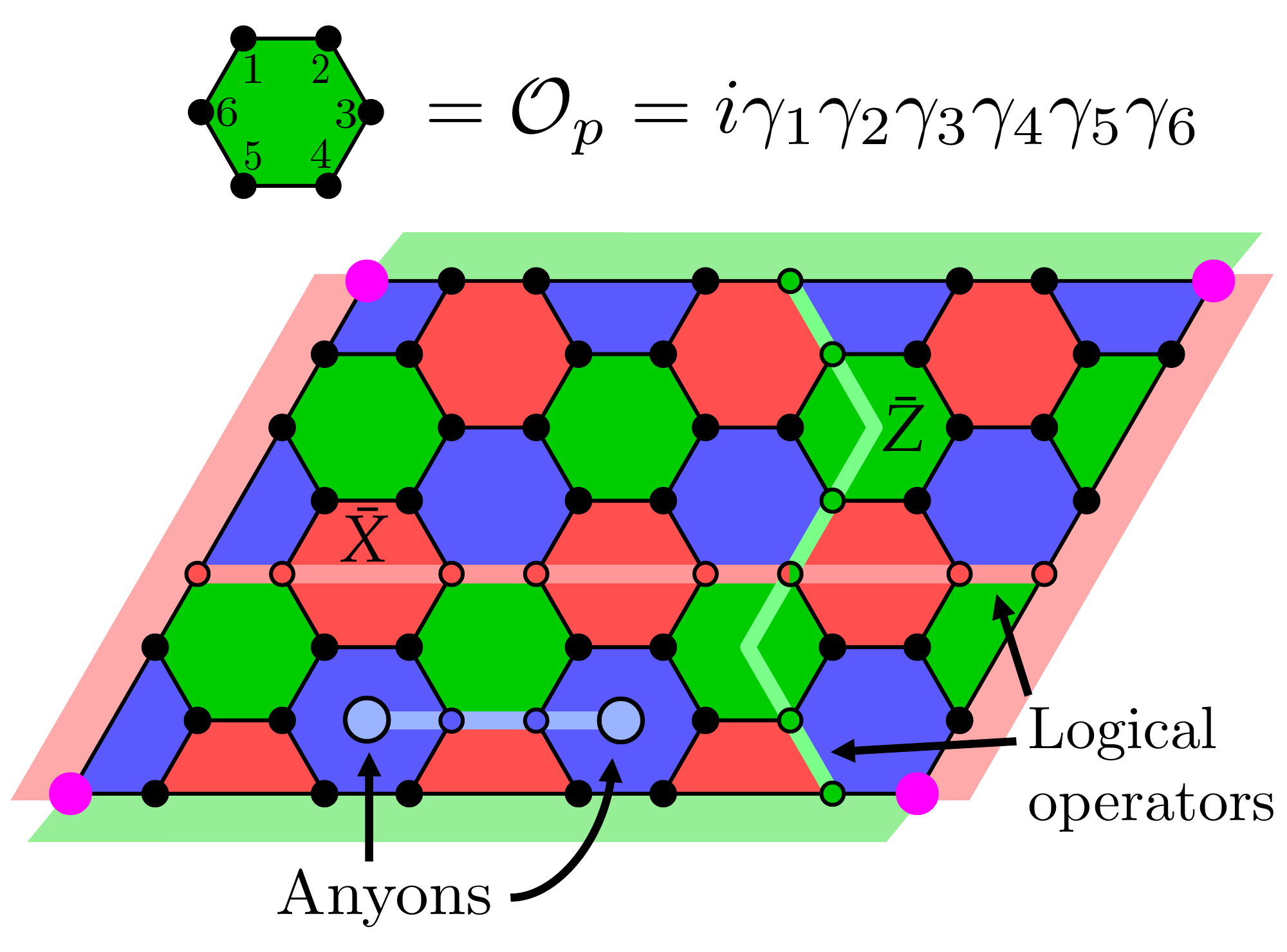}
\caption{MSC patch encoding a single logical qubit. Plaquettes are colored red, green or blue. Plaquette operators $\mathcal{O}_p$ are defined as shown for the example of a green plaquette. An example of a blue bond operator is highlighted. This operator creates two anyons (blue dots) on the blue plaquettes at its endpoints. The boundaries of the patch are red or green as shown; they are assigned the colors not featured in boundary plaquettes. Logical operators $\bar{X}$ and $\bar{Z}$ correspond to red and green string operators respectively, which terminate on same-colored, opposite boundaries. %
Corners between boundaries of different types are indicated by pink dots. 
}
\label{fig:Patch}
\end{figure} 

We define the MSC on a trivalent, three-colorable lattice~\cite{MjFerm_Surface_Code} (i.e., a lattice whose plaquettes can each be assigned one of three colors, such that no two adjacent plaquettes have the same color). 
In general, the plane has several trivalent, three-colorable tilings~\cite{Bombin_Color_Codes}. 
Among these, we focus on the honeycomb lattice, a patch of which is shown in Figure~\ref{fig:Patch}. 
Majorana operators $\gamma_i$  are assigned to sites of the lattice. Plaquettes  are assigned one of three colors, red, green or blue, as shown in Figure~\ref{fig:Patch}. 
Each plaquette $p$ has an associated plaquette operator $\mathcal{O}_p = i^{|\partial p|/2} \prod_{j\in \partial p}\gamma_j$, where $\partial p$ is $p$'s boundary and where we choose an arbitrary (but definite) ordering of the product (see Figure~\ref{fig:Patch}).
The stabilizer group $\mathcal{S}$ is then generated by the complete set of $\mathcal{O}_p$ in the system. 
In Figure~\ref{fig:Patch}, there are $2n=60$ Majoranas and $m=29$ independent stabilizer generators. Hence there is $k=1$ logical qubit encoded in the patch.

Bonds are also assigned a color, namely that of the plaquettes at their ends. 
A bond operator associated with bond $e$, connecting Majoranas $\gamma_j$ and $\gamma_k$, is defined as $\mathcal{O}_e = i\gamma_j\gamma_k$.
A string $S$ is a set of bonds that all have the same color. String operators are then given by the product of all associated bond operators: $\mathcal{O}_S = \prod_{e\in S}\mathcal{O}_e$.
Logical operators of the MSC are string operators that commute with all plaquette operators. 
For the patch in Figure~\ref{fig:Patch}, these correspond to strings  running between opposite boundaries. 
In general, boundary conditions determine the number of logical qubits in the code. The patch in Figure~\ref{fig:Patch} contains one logical qubit, which has logical operators given by the string operators $\bar{X}$ and $\bar{Z}$ shown.
To see that $\bar{X}$ and $\bar{Z}$ anticommute note that their strings intersect an odd number of times; this  renders $|A\cap B|$ odd in Equation~\eqref{eqn:MajModeCommutation}, while $|A|$ and $|B|$ are both even.
As noted earlier, logical operators can be deformed by stabilizers ($L\mapsto Ls$ for $s\in \mathcal{S}$) while still acting in the same manner on logical qubits. 
In the MSC, these deformations map between different string operators running between the same boundaries. 
Thus, in Figure~\ref{fig:Patch}, the equivalence class $[\bar{X}]$ contains all string operators terminating on horizontally opposite boundaries and similarly, $[\bar{Z}]$ contains all string operators terminating on vertically opposite boundaries.

\subsection{Anyons in the MSC}\label{subsec:Anyons_MSC}

Anyons are quasi-particle excitations of two-dimensional, topologically ordered systems~\cite{Anyon_Stats_Wilczek1984,Wen_FQH_GS_Degen,Bravyi_Kitaev_Codes_with_Bdry,Kitaev_Anyons2006,Non-Abelian_TQC_Review}. 
In topological quantum codes, we can define a stabilizer Hamiltonian according to which errors in the code correspond to anyons~\cite{Kitaev_toric_code, Kitaev_Anyons2006}. 
The MSC, owing to its fermionic nature, has an anyon content that differs significantly from bosonic codes~\cite{MjFerm_Surface_Code}.
We review the MSC's anyon model here.

One can assign a stabilizer Hamiltonian to the MSC by $H= -\frac{u}{2}\sum_{p} \mathcal{O}_p$, where $u>0$ and the sum runs over all plaquettes in the lattice.
Thus the ground states of the Hamiltonian are code states of the MSC. 
Excited states are created by applying string operators to one of the ground states $\ket{\psi}$. 
If string operator $\mathcal{O}_S$ anti-commutes with plaquette operators $\mathcal{O}_{p_1},\ldots, \mathcal{O}_{p_k}$, then $\mathcal{O}_S \ket{\psi}$ has energy $ku$ above the ground state energy.
We interpret plaquettes $p_1,\ldots, p_k$ as hosting anyons and $\mathcal{O}_S$ as creating these anyons (or annihilating anyons already located at those sites, if $\mathcal{O}_S$ acts on an excited state). An anyon is labeled with the same color as the plaquette that hosts it: $\mathsf{R}$, $\mathsf{G}$ or $\mathsf{B}$ for those hosted on red, green or blue plaquettes respectively.
The simplest string operator is a bond operator; this creates a pair of anyons at its endpoints, such as those indicated by blue dots in Figure~\ref{fig:Patch}.
If we consider two blue bonds $e_1$ and $e_2$ that have endpoint plaquettes $p,q$ and $q,r$ respectively, applying $\mathcal{O}_{e_1}$ to $\ket{\psi}$ is interpreted as creating anyons on $p$ and $q$. Then subsequently applying $\mathcal{O}_{e_2}$ is interpreted as annihilating the anyon on $q$ and creating one on $r$. 
Alternatively, it can be thought of as moving the anyon from $q$ to $r$. 
Thus applying string operators to an excited state can move the anyons to any other plaquette of the same color as the anyon. 

There also exist parity-non-conserving processes that create anyons in triplets, rather than pairs.
For example, applying a Majorana operator $\gamma_j$ to a ground state creates three anyons located on the plaquettes surrounding $\gamma_j$.
Due to the 3-colorability of the lattice, these anyons will all be colored differently. 

\begin{figure}
    \centering
    \includegraphics[width=0.9\linewidth]{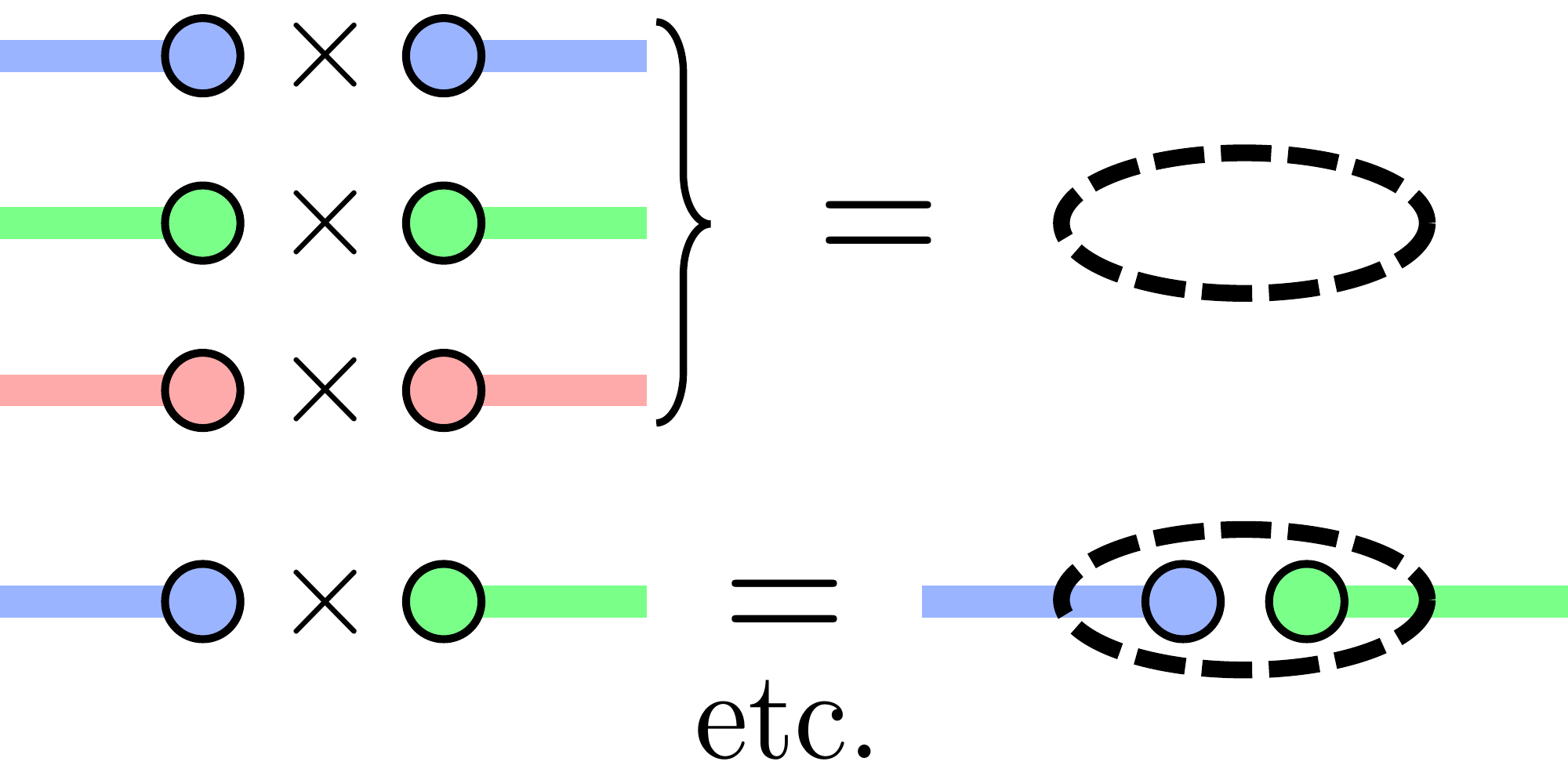}
    \caption{Non-trivial anyon fusion processes in the MSC. Any two anyons of the same color may be fused to the vacuum (top). Two anyons of different colors $\mathsf{A}$ and $\mathsf{B}$ may be fused into the composite $\mathsf{AB}$ (the example of $\mathsf{B}$ and $\mathsf{G}$ fusion is shown at the bottom).}
    \label{fig:anyon_fusion}
\end{figure}

Along with the three anyons mentioned, we also interpret the vacuum (no anyons present) as a quasi-particle, labeled $\mathbf{1}$. 
This will allow us to formalize fusion: the process of bringing two anyons close together and viewing the result as a separate anyon. 
For example, the fusion of $\mathsf{R}$ and $\mathsf{B}$ can be viewed as the anyon $\mathsf{BR}$.
A triplet of anyons created by a single Majorana operator is also considered as a distinct anyon, $\mathsf{RGB}$.
Note that creating a single $\mathsf{RGB}$ from $\mathbf{1}$ is not a fermion parity conserving process. 
To highlight this, we break the anyon content of the model into two sets, $\lbrace \mathbf{1}, \mathsf{R}, \mathsf{G}, \mathsf{RG}\rbrace$ and $\lbrace \mathsf{RGB}, \mathsf{GB}, \mathsf{BR}, \mathsf{B}\rbrace $~\cite{MjFerm_Surface_Code}, 
where the second set is obtained from the first by fusion with $\mathsf{RGB}$. 
We have the following (commutative and associative) fusion rules, dictating how anyons can be combined or split:
\begin{align}\label{eqn:Fusion_Rules}
    &\mathbf{1}\times \mathsf{A} = \mathsf{A},\; \mathsf{A}\times \mathsf{A} = \mathbf{1}\quad (\text{for all anyons } \mathsf{A})\nonumber\\
    &\mathsf{A} \times \mathsf{C} = \mathsf{AC} \quad (\text{for all anyons } \mathsf{A}\neq \mathsf{C}).
\end{align}
The non-trivial rules are summarized in Figure~\ref{fig:anyon_fusion}.
The fusion rules imply that anyons can be created (via splitting from an initial set) only in color-neutral combinations. All anyons in the system must be able to annihilate back to the initial set of anyons. 
There are two initial sets, $\mathbf{1}$ and $\mathsf{RGB}$, which are inequivalent because $\mathsf{RGB}$ cannot transform to $\mathbf{1}$ under any fermion parity conserving process. 
These anyons and fusion rules thus describe a system with a $Z_2$ fermion parity grading~\cite{MjFerm_Surface_Code}. 
We discuss the braiding statistics of these anyons  in Appendix~\ref{app:Anyon_Statistics}.

\begin{figure}
    \centering
    \subfloat[\label{subfig:String_Fusion_1}]{\includegraphics[height=130pt]{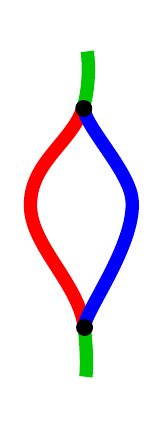}}\hspace{20pt}
    \subfloat[\label{subfig:String_Fusion_2}]{\includegraphics[height=130pt]{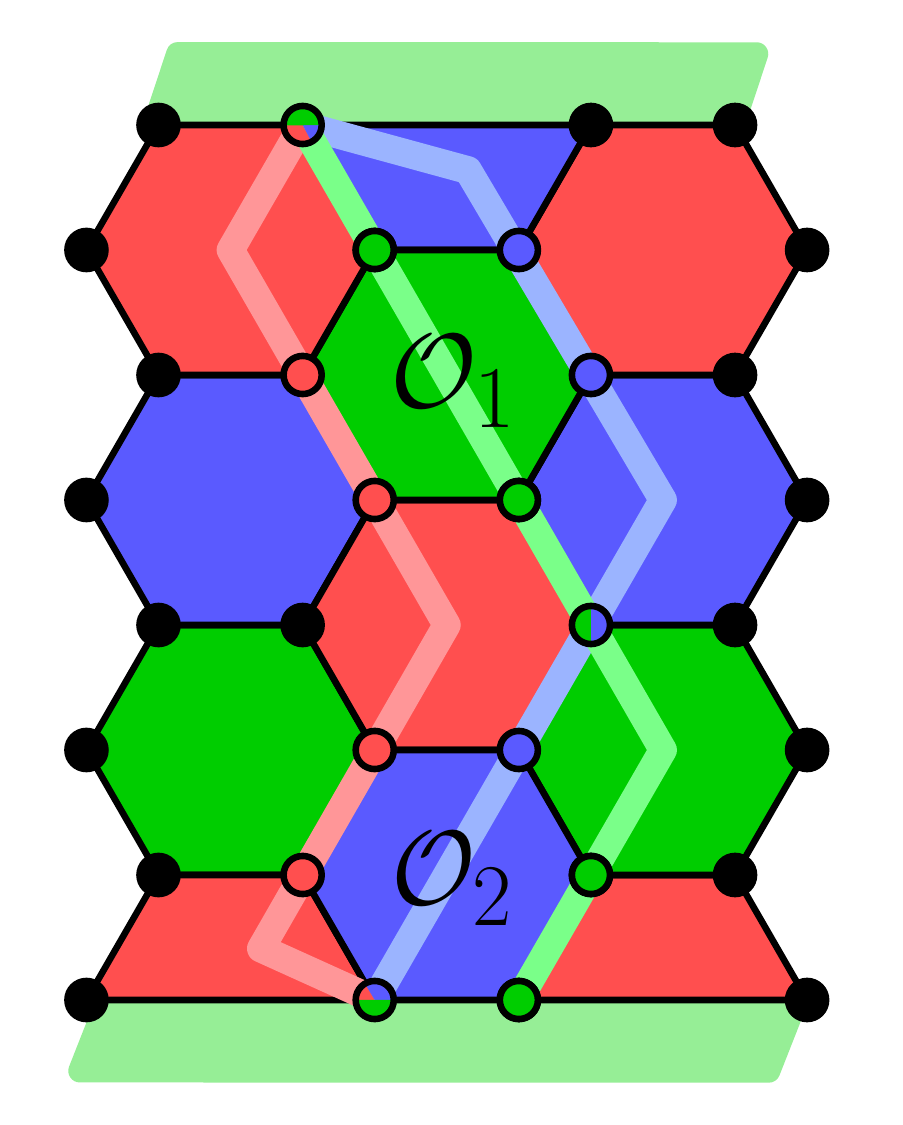}}
    \caption{An example of string operator fusion. (a) Schematic illustration of a green string splitting into blue and red strings away from its endpoints. Black dots indicate single Majorana operators, at which points strings of three colors fuse. (b) A lattice-based illustration of the same phenomenon. The blue and red strings arise from the green string via multiplication by stabilizers $\mathcal{O}_1$ and $\mathcal{O}_2$.
    }
    \label{fig:String_Fusion}
\end{figure}

Similarly to anyons, string operators also obey fusion rules.  
Viewing $\mathsf{C}$ as the endpoint of string operator of color $\mathsf{C}\in \lbrace \mathsf{R}, \mathsf{G}, \mathsf{B}\rbrace$ and $\mathbf{1}$ as the identity operator, we can apply Equations~\eqref{eqn:Fusion_Rules} to string operators even away from their endpoints.
(We equate $\mathsf{RGB}$ with $\mathbf{1}$ in this context because an $\mathsf{RGB}$ pair can be created by a Majorana pair without string operators.)
Fusion and splitting of string operators is implemented by multiplication with stabilizers.
For example, the green string operator $\bar{Z}$ in Figure~\ref{fig:Patch} can be split, by multiplication with green stabilizers, into blue and red strings (i.e., products of blue and red bond operators) along its length.
This is shown in Figure~\ref{fig:String_Fusion}.
Hence the non-trivial rules from Equations~\eqref{eqn:Fusion_Rules} for string operators become $\mathsf{R}= \mathsf{G}\times \mathsf{B}$ and all cyclic permutations of this.

\subsection{Boundaries in the MSC}\label{subsec:Boundaries_MSC}

Topological systems sometimes admit gapped boundaries, with a classification that is possible purely by considering the model's anyon content (see Appendix~\ref{app:Lagrangian_subgroups} for details).
A system is gapped if the gap in energy between ground and lowest-energy excited states is bounded from below by a constant independent of the system size. 
Gapped boundaries of topological models are not only important from a condensed matter perspective \cite{Kitaev_Kong, Protected_Edge_Modes}, but also play an important role in quantum computation \cite{Bravyi_Kitaev_Codes_with_Bdry, BBrown_Twists_CC, QC_with_Gapped_Boundaries2017}. 

\begin{figure}
    \centering
    \includegraphics[width=0.98\linewidth]{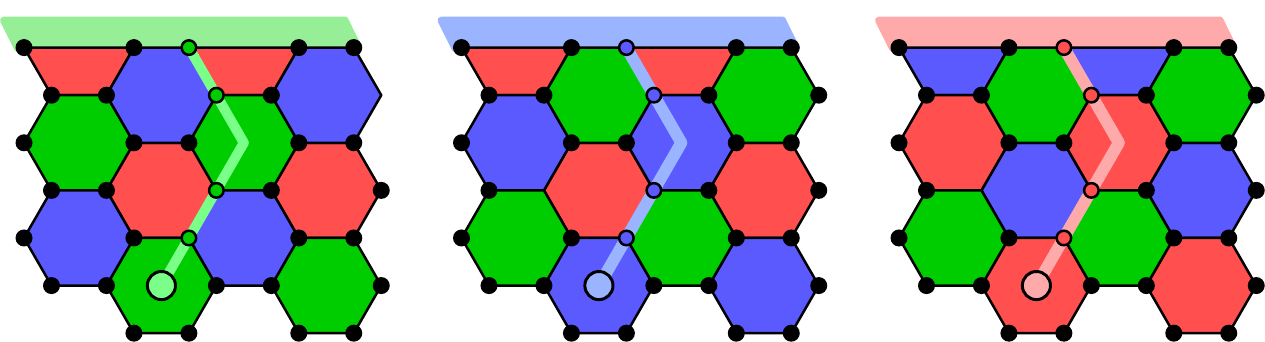}
    \caption{Three boundary types of the MSC and their anyon condensation processes. Each boundary can be colored green (left), blue (middle) or red (right) depending on the anyons that may be condensed at that boundary.}
    \label{fig:MSC_bdrys}
\end{figure}

In the MSC there exist three topologically-distinct types of gapped boundary, which can be colored red, green and blue respectively.
These are shown in Figure~\ref{fig:MSC_bdrys}.
The three boundaries are distinguished from one another by the color of string operator that can terminate on that boundary. 
(Equivalently by the plaquette color not appearing on the boundary.)
For example, in Figure~\ref{fig:Patch}, the top and bottom boundaries are green, since the green string operator $\bar{Z}$ terminates on both of them. 
Equivalently, any green anyons present in the system may be brought into proximity of a green boundary, where local, fermion-parity-preserving processes (e.g., the application of a green bond operator) can result in the anyon being absorbed by the boundary: we say that the anyon ``condenses" at that boundary.
$\mathsf{C}$-boundaries (for $\mathsf{C}=\mathsf{R},\mathsf{G},\mathsf{B}$) can condense only $\mathsf{C}$-anyons.
Anyon condensation processes are shown in Figure~\ref{fig:MSC_bdrys}.
Beyond this classification, fermionic systems have an extra $Z_2$ fermion parity grading along their one-dimensional boundaries.
We will explore this further in the next section.

\section{Twist Defects of the Majorana Surface Code}\label{sec:Domain_Walls_Twists}

If $\varphi$ is a permutation of the anyon labels, we say that $\varphi$ is a symmetry if it preserves the fusion rules and statistics (see Appendix~\ref{app:Anyon_Statistics}) of the anyons~\cite{Barkeshli_defects_gauging2019}.
In two-dimensional topologically ordered systems, (transparent) domain walls are one-dimensional defects that are associated with symmetries of the anyon labels \cite{Kitaev_Kong,Twist_Symmetry_Review_Teo_2016,Barkeshli_defects_gauging2019}.
An anyon $a$ crossing the wall (in a given direction) is transformed into $\varphi(a)$ where $\varphi$ is the symmetry associated with the domain wall.
If $a$ crosses the wall in the opposite direction, it is transformed into $\varphi^{-1}(a)$.

Domain walls can terminate at point-like twist defects (``twists" for short)~\cite{Kitaev_Kong,Twist_Symmetry_Review_Teo_2016,Barkeshli_defects_gauging2019,Wen_Twists2013,Barkeshli_Nematic_States2012,Defects_Abelian_states,Twists_genons,Twist_Liquid}. 
It has been shown that these twist defects are closely related to non-Abelian anyons \cite{Bombin_Twists, TEE_with_twist, Defects_Abelian_states,Dua_Twists_Majorana_stats}, and can be used for topological quantum computation in surface codes \cite{Holes_Twists_SurfaceCode, Bombin_Twists, Bombin_twists_code_deformation,Twists_genons,Webster_Bartlett_Defects2020}. 
Twists can be classified by the permutation that is enacted on an anyon encircling the twist in a clockwise direction. 
If a domain wall associated with symmetry permutation $\varphi$ terminates at two twists, one will be a $\varphi$ twist, and the other will be a $\varphi^{-1}$ twist.

We now introduce the twist defects appearing in the MSC.
While some examples of MSC twists have previously appeared in the literature~\cite{Maj_Ferm_Codes,QC_with_MFCs}, we establish here their fundamental features and provide their exhaustive categorization\footnote{Note that twists have appeared in the MSC context for codes conceptually equivalent to bosonic codes. In Ref.~\cite{QC_with_MFCs}, one of the MSC stabilizer colors is implemented as a parity constraint (i.e. the Majoranas are located on ``hexons" or ``tetrons") and the others are actively measured. This yields a code which is a Majorana realization of a bosonic surface code and hence the twists considered are those of the surface code. This corresponds to a restricted class of MSC twists: only bosonic $Z_2$ twists of a single type (e.g. $T_\textsf{B}$) can be included in such setups.}.
In this model we have three non-vacuum anyon labels (either $\lbrace \textsf{R},\textsf{G},\textsf{B}\rbrace $ or $\lbrace \textsf{RG},\textsf{GB},\textsf{BR}\rbrace$) that can be permuted in any way while still preserving the anyonic data, and hence the anyons possess an $S_3$ symmetry. 
Any non-identity element of the permutation group $S_3$ either swaps a pair of labels, generating a cyclic group $Z_2$, or it is a length-3 cycle, generating the subgroup $Z_3$.
Thus domain walls and twists in the MSC are either of type $Z_2$ or $Z_3$.

\begin{figure*}
    \null\hfill%
    \subfloat[\label{subfig:Z2_Z3_Twists_bosonic}]{%
      \includegraphics[width=0.45\textwidth]{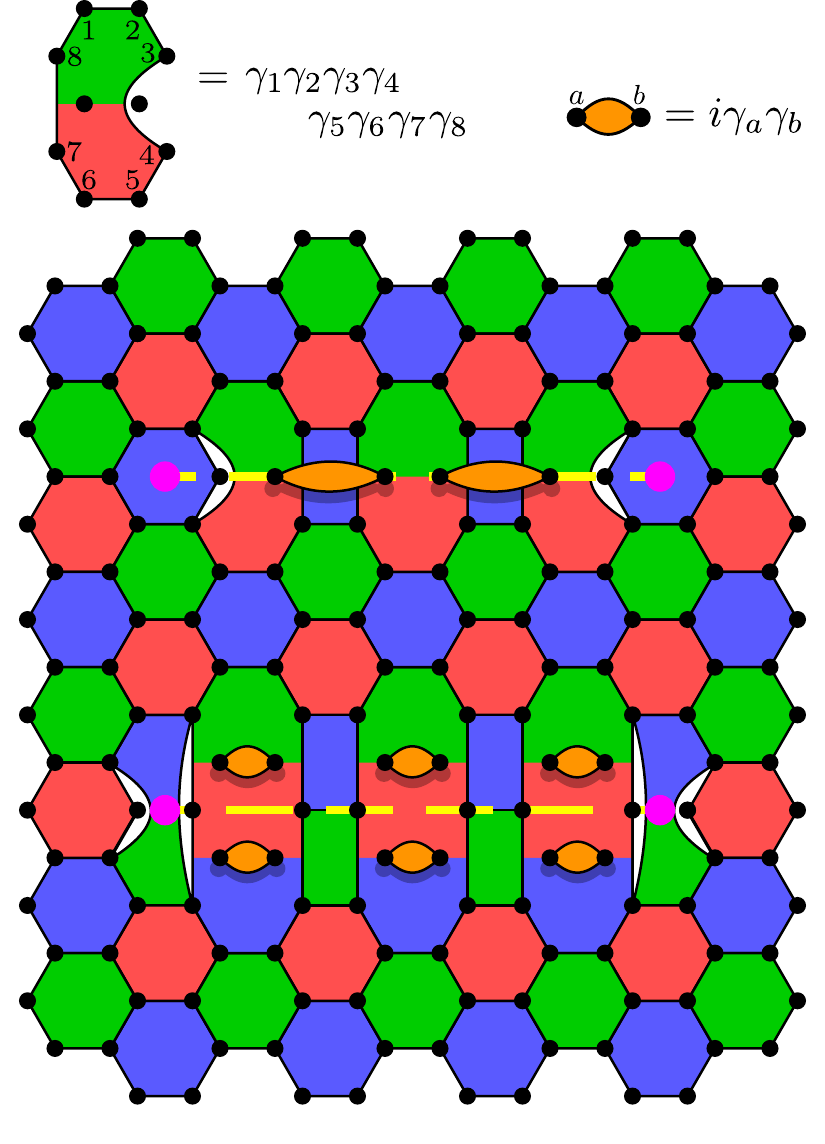}
    }
    \hfill
    \subfloat[\label{subfig:Z2_Z3_Twists_fermionic}]{%
      \includegraphics[width=0.45\textwidth]{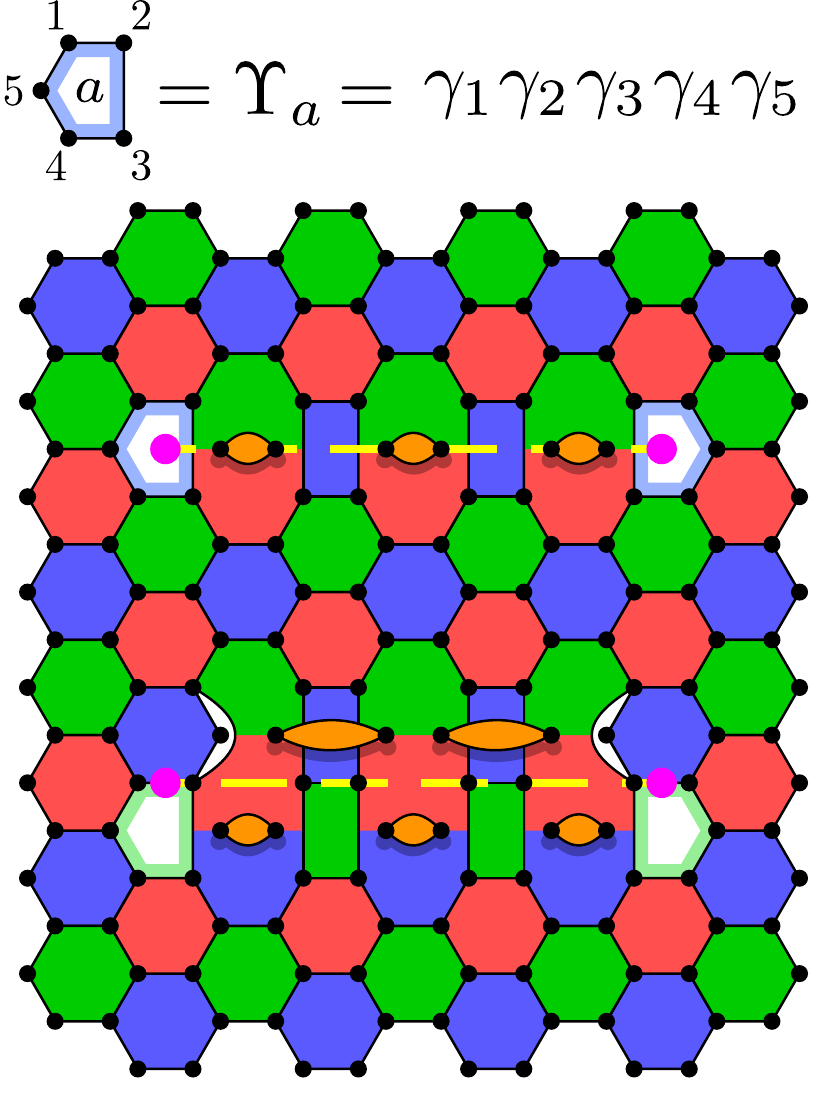}
    }
    \hfill\null\\
    \subfloat[\label{subfig:Ferm_Corners_a}]{\includegraphics[width=0.49\textwidth]{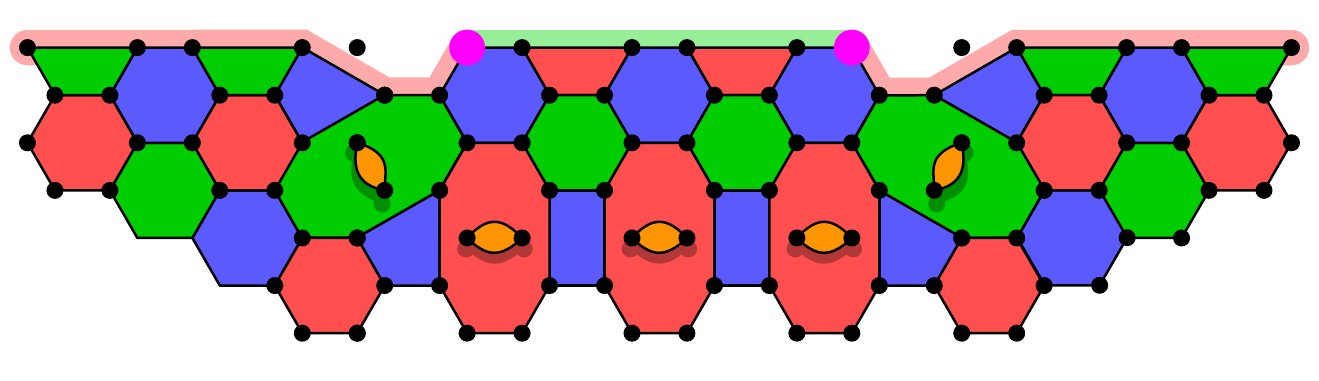}}
    \hfill
    \subfloat[\label{subfig:Ferm_Corners_b}]{\includegraphics[width=0.49\textwidth]{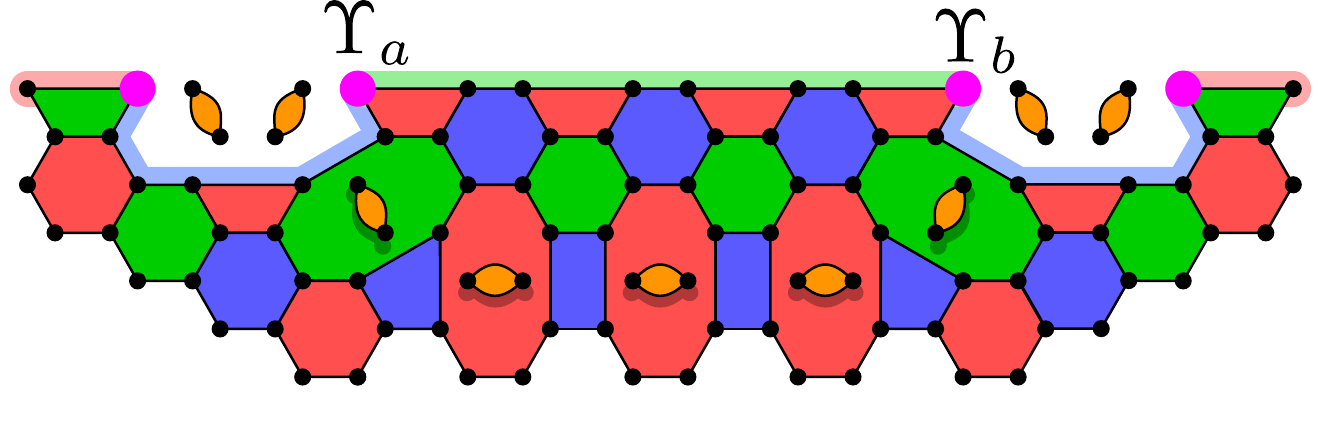}}
    \caption{Lattice realizations of twists and logical MZMs in the MSC. In (a) and (b), twists and domain walls are indicated by pink dots and yellow dashed lines respectively. The Majoranas around the boundary of floating orange, pill-shape faces belong to the original honeycomb lattice but, after the introduction of twists, are not included in any plaquette operators. We introduce additional stabilizers of the form $i\gamma_a\gamma_b$, represented by the orange faces, to remove the degeneracy from these Majorana operators. In (a) we display an example of bosonic $Z_2$ twists, specifically of type $T_\mathsf{B}$ (top), and of $Z_3$ twists (bottom) and the domain walls connecting them. Examples of the new stabilizers introduced are shown at the top. In (b) we display the fermionic counterparts of each of these twists and the resulting change to the orange stabilizers. This results in logical MZMs at twist locations, defined at the top. In (c), a fermionic stretch of boundary is displayed, with pink dots indicating corners. Unpaired MZMs exist at the ends of the green boundary. Local changes to the stabilizer group result in the lattice in (d), which expands the unpaired MZMs into logical MZMs $\Upsilon_a$ and $\Upsilon_b$ hosted on the odd-length blue boundaries.}\label{fig:Twists_and_Corners}
\end{figure*}

Beyond this classification, we can further identify sub-types. 
There are three types of $Z_2$ twists, which we label $T_{\textsf{R}}$, $T_{\textsf{B}}$ and $T_\textsf{G}$, according to the color of the anyon that is preserved when wound completely around the twist.
There are two types of $Z_3$ twists, labeled $T_\mathsf{RGB}$ and $T_\mathsf{RGB}^{-1}$.
The former enacts the cycle $\mathsf{R}\rightarrow \mathsf{G}\rightarrow \mathsf{B}\rightarrow \mathsf{R}$ on anyons wound clockwise around the twist, and the latter enacts the inverse cycle.
We provide lattice-based realizations of $Z_2$ and $Z_3$ twists in Figure~\ref{fig:Twists_and_Corners}.
Discussion of these sub-types is convenient, but note that we can inter-convert twists between sub-types by relabeling the plaquettes.
Indeed, a relabeling of an area $A$ of code in this way amounts to having introduced a closed loop of domain wall at its boundary $\partial A$. 
Similarly, by relabeling an area adjacent to the domain wall, we can, effectively, move the path of the wall. This shows that domain walls -- but not the locations of their endpoints -- are gauge dependent. 
Furthermore, if the loop of domain wall encircles a twist, it can change a $Z_2$ twist's sub-type to any other within the $Z_2$ class ($T_\mathsf{B}$, $T_\mathsf{G}$ or $T_\mathsf{R}$) and, similarly, can map between the two $Z_3$ twist sub-types ($T_\mathsf{RGB}$ and $T_\mathsf{RGB}^{-1}$).
See Appendix~\ref{app:Gauge_Dependence} for further details of such gauge dependencies.

Twists in the MSC differ from those in bosonic models in an important way: they can be further subdivided into ``bosonic" and ``fermionic" types of twist. 
These differ by a single MZM.
However, rather than an extra MZM simply furnishing the twists, as might be expected, in Figure~\ref{fig:Twists_and_Corners} we show that in fact the result can be a \emph{logical} MZM located at the twist (cf. Section~\ref{subsec:MFCs}).
In Figure~\ref{subfig:Z2_Z3_Twists_bosonic} we provide a lattice realization of bosonic $Z_2$ and $Z_3$ twists and the domain walls connecting them, while in Figure~\ref{subfig:Z2_Z3_Twists_fermionic} we show $Z_2$ and $Z_3$ fermionic twists.
In the latter, the lattice contains pentagonal holes. 
The (Hermitian) product of all Majorana operators around the boundary of this hole is a fermion-parity-odd operator that, by Equation~\eqref{eqn:MajModeCommutation}, commutes with all plaquette operators.
These operators cannot be included in the stabilizer group, owing to their odd fermion parity. 
Instead they are logical MZMs. 
They generalize the key features of MZMs, including their commutation with the Hamiltonian. 
We discuss these logical MZMs appearing at fermionic twists below, in Section~\ref{subsec:LogMZM_fermion_twist}.

A domain wall lying along a boundary can change the color of that boundary, since it alters the color of anyons condensing there.
Thus, there is an association between twists and corners, or interfaces between boundaries of different types (see Appendix~\ref{app:Twists_Corners})~\cite{BBrown_Twists_CC}.
In Figure~\ref{fig:Patch}, we have indicated the corners between $\textsf{R}$ and $\mathsf{G}$ boundaries in the MSC using pink dots.
When such corners are fermionic, they host an MZM (see Figure~~\ref{subfig:Ferm_Corners_a}).

Twists can also be fused or split.
For example, a $Z_3$ twist can be split into two $Z_2$ twists of different sub-types; this follows from $S_3$ being generated by transpositions (i.e., $Z_2$ operations)~\cite{rotman2000first}. 
However, the splitting or fusion of twists must preserve their overall fermionic or bosonic character: it cannot change the number of logical MZMs (mod 2). 
When a fermionic corner is split, logical MZMs appear, associated with boundaries of odd-length (see Figure~~\ref{subfig:Ferm_Corners_b}).
Appendix~\ref{app:Twists_fusion} has further twist fusion details and examples.

When the MSC is defined on the honeycomb lattice, some Majoranas along domain walls are not included in any plaquettes.
In Figure~\ref{subfig:Z2_Z3_Twists_bosonic} and~\ref{subfig:Z2_Z3_Twists_fermionic}, we draw orange, pill-shape faces between pairs of Majoranas left out of the lattice in this way.
These Majorana operators furnish extraneous fermion modes (cf.~Section~\ref{subsec:MFCs}), so we introduce additional bilinear stabilizers ($i\gamma_a\gamma_b$, for left-out operators $\gamma_a$, $\gamma_b$), represented by the orange faces in Figure~\ref{fig:Twists_and_Corners},  to remove the ground-state degeneracy associated with these modes, i.e., to remove these states from our code space.
Those Majorana operators included in orange stabilizers are not included in any other stabilizer of the code.

These extraneous Majorana operators, along with the logical MZMs at the twists, form a system similar to the Kitaev chain~\cite{Kitaev_chain2001}: a one-dimensional string of adjacent, nearest-neighbor Majorana bilinears. 
Specifically, we replace the first and last of the $2n$ Majoranas of a Kitaev chain with logical MZMs $\Upsilon_a$ and $\Upsilon_b$. 
The Hamiltonian~\cite{Kitaev_chain2001} therefore has two fixed points, one in a topological phase in which unpaired logical MZMs exist at the endpoints (the situation of Figure~\ref{subfig:Z2_Z3_Twists_fermionic}), and one in a trivial phase without unpaired logical MZMs.
In the latter phase, $\Upsilon_a$ and $\Upsilon_b$ combine with neighboring
Majoranas to form parity-even plaquette operators (e.g., the blue plaquettes at $Z_2$ twists in Figure~\ref{subfig:Z2_Z3_Twists_bosonic}).
A transition between fermionic and bosonic twists therefore requires the chain to go through a topological phase transition. 
This corresponds to a change of the chain Hamiltonian (i.e., domain wall stabilizer set) along its entire length, so that the bilinears change from one assignment of adjacent nearest neighbors to the other. 
The two sets can be seen in Figures~\ref{subfig:Z2_Z3_Twists_bosonic} and~\ref{subfig:Z2_Z3_Twists_fermionic}. 
The existence of these two types of twists thus corresponds to the existence of a topologically non-trivial phase for one-dimensional fermionic systems, as modeled by the Kitaev chain~\cite{Turner_Fermion_Phases_1D2011,Kitaev_Fermion_Phases_1D2011,Defects_Abelian_states}.
The fermionic or bosonic nature of a twist is as robust as this topological phase; it is immutable under local, fermion-parity-conserving processes.

\subsection{Quantum Dimensions of Twists}\label{subsec:Quantum_Dimension}

The ground state degeneracy of a code with twist defects (or equivalently the dimension of the code space) is governed by the quantum dimension of those twists. 
Given a code with $N$ twists of quantum dimension $d$, the dimension of the code's ground space grows as $d^N$ for large $N$.
The quantum dimensions of twists in the MSC differ between bosonic and fermionic types, due to fermionic twists also contributing to the degeneracy via their logical MZMs.

We can calculate these quantum dimensions by counting the number of anyons that can ``localize" at a given twist.
An anyon is localized at a twist if it is brought close to that twist and absorbed through local processes.
For example, if anyon $a$ obeys $a = b\times c$ and twist $\varphi$ is such that $\varphi(b) = \bar{c}$ (the anti-particle of $c$, such that $c\times \bar{c} = \mathbf{1}$), then anyon $a$ can be dragged to the vicinity of a twist, then split into anyons $b$ and $c$. 
$b$ is then wound clockwise around the twist and thus transformed to $\bar{c}$ which can then annihilate with $c$. 
This process results in anyon $a$ being localized at the twist.
The quantum dimension of a twist $d_\varphi$ is given by $d_\varphi^2 = \sum_{a\in\mathcal{A}} N_{a,\varphi}$, where $\mathcal{A}$ is the collection of anyons the system permits, and $N_{a,\varphi}$ is $1$ if anyon $a$ can be localized at $\varphi$, and $0$ otherwise~\cite{BBrown_Twists_CC}.

\begin{figure}
    \centering
    \includegraphics[width=0.98\linewidth]{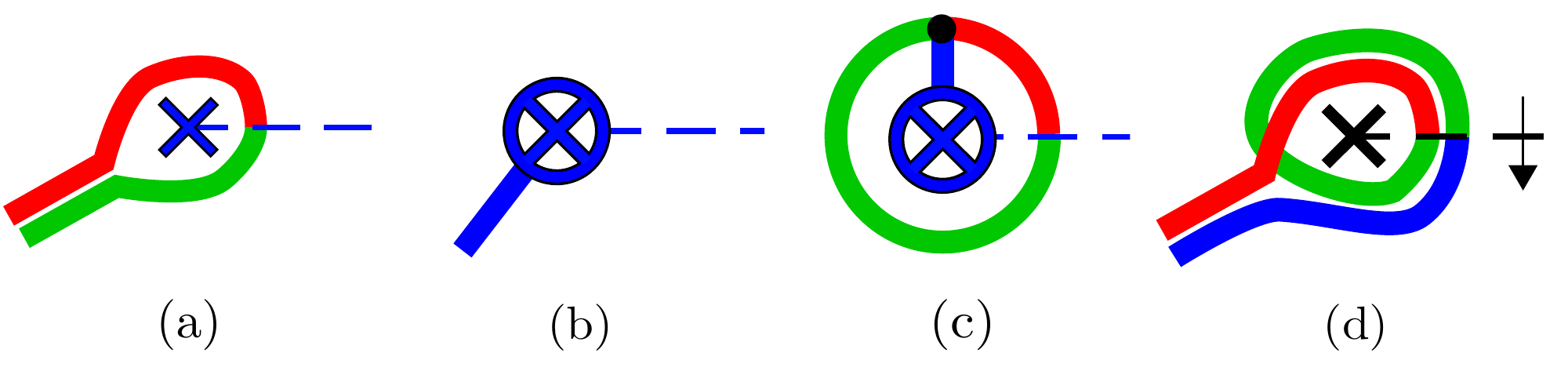}
    \caption{Schematic illustration
    of examples of anyon localization. Bosonic twists are indicated by crosses, and fermionic twists by circles containing crosses. Domain walls are indicated by dashed lines. Solid blue/red/green lines are string operators. (a) $\mathsf{RG}$ anyon localized at a bosonic $T_\mathsf{B}$ twist indicated in blue. (b) $\mathsf{B}$ anyon localized at a fermionic $T_\mathsf{B}$ twist indicated in blue. (c) $\mathsf{RGB}$ particle localized at a fermionic $T_\mathsf{B}$ twist. String operators of three different colors meet at a single Majorana operator (black dot). (d) $\mathsf{BR}$ anyon is localized at a bosonic $T_\mathsf{RGB}$ twist indicated in black. The arrow on the domain wall indicates the direction in which an anyon crossing the wall undergoes the permutation $T_\mathsf{RGB}$.}
    \label{fig:anyon_localization}
\end{figure}

Let us now count the anyons that can localize at each type of twist.
Some of these localization processes are shown in Figure~\ref{fig:anyon_localization}.
Trivially, all twists localize the vacuum particle $\mathbf{1}$.
Additionally, for bosonic $Z_2$ twists, anyon $\mathsf{RG}$ can localize at $T_\mathsf{B}$, $\mathsf{GB}$ at $T_\mathsf{R}$ and $\mathsf{BR}$ at $T_\mathsf{G}$.
We do not count $\mathsf{B}$ as localizing at $T_\mathsf{B}$, nor similarly for the other twists, because in order to split $\mathsf{B}$ to $\mathsf{R}\times \mathsf{G}$, we need to introduce an $\mathsf{RGB}$ anyon to the system, and hence the twist is localizing $\mathsf{RGB}\times \mathsf{B} = \mathsf{RG}$, rather than just the $\mathsf{B}$ anyon.
However we can localize $\mathsf{B}$ at a fermionic $T_\mathsf{B}$ twist, because a blue string operator can directly terminate at the odd-length, blue boundary associated with this twist (see Figure~\ref{subfig:Z2_Z3_Twists_fermionic}), and thereby commute with all stabilizer generators in the vicinity of the twist.
Fermionic twists also localize $\mathsf{RGB}$ anyons since, for example, $\mathsf{RGB} = \mathsf{B} \times \mathsf{RG}$, so one can split $\mathsf{RGB}$ into these two anyons. Then both $\mathsf{B}$ and $\mathsf{RG}$ can localize at $T_\mathsf{B}$.
Thus bosonic $Z_2$ twists localize two types of anyons (including $\mathbf{1}$), while fermionic $Z_2$ twists localize four. 

Bosonic $Z_3$ twists can localize $\mathsf{RG}$, $\mathsf{GB}$ and $\mathsf{BR}$ (and $\mathbf{1}$), since these particles can be split into their constituents (e.g., $\mathsf{BR} = \mathsf{B}\times \mathsf{R}$), one of which can be wound either once or twice around the twist until it is transformed into the same type as its partner (e.g., $\mathsf{R}$ transformed into $\mathsf{B}$, as shown in Figure~\ref{fig:anyon_localization}). 
Fermionic $Z_3$ twists also have an odd-length boundary of a single color and hence can localize either $\mathsf{R}$, $\mathsf{G}$ or $\mathsf{B}$ in the same manner as fermionic $Z_2$ twists.
However, $\mathsf{R}$, $\mathsf{G}$ and $\mathsf{B}$ can be transformed into one another by encircling the twist, and hence fermionic $Z_3$ twists can localize all three of these. 
Finally, they also localize $\mathsf{RGB}$, similarly to fermionic $Z_2$ twists.
Thus in total, fermionic $Z_3$ twists can localize all eight anyons.

From the above, we can see that bosonic $Z_2$ twists have quantum dimension $d_2^B = \sqrt{2}$. 
This signals their similarity to Ising anyons~\cite{Bombin_Twists}.
Fermionic $Z_2$ twists have quantum dimension $d_2^F = 2$.
Bosonic $Z_3$ twists have quantum dimension $d_3^B = 2$, whereas fermionic $Z_3$ twists localize all anyons and hence have quantum dimension $d_3^F = 2\sqrt{2}$. 
These quantum dimensions are verified using a different method in Appendix~\ref{app:Quantum_Dimensions}.

MZMs are also similar to Ising anyons~\cite{MZM_TQC_Review2015}; hence they also have quantum dimension $\sqrt{2}$. 
This precisely equals the ratio between fermionic and bosonic twist quantum dimensions.
Thus going from bosonic to fermionic twists can be seen as attaching an MZM, or a logical MZM, to the defect, as discussed previously.

\subsection{Logical Qubits from Twists}

\begin{figure}
    \subfloat[\label{subfig:Z2_Twist_logical_ops}]{%
      \includegraphics[width=0.45\linewidth]{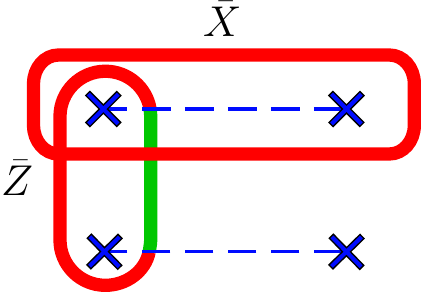}
    }\hspace{10pt}
    \subfloat[\label{subfig:Z3_Twist_logical_ops}]{%
      \includegraphics[width=0.45\linewidth]{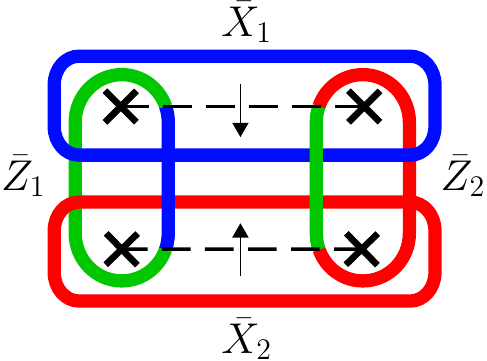}
    }
    \caption{Schematic illustration of logical operators for examples of $Z_2$ (a) and $Z_3$ (b) twist defects (either bosonic or fermionic; we depict bosonic twists here). Twists are shown as crosses, domain walls as dashed lines, and string operators as solid lines, colored according to their type. The $Z_2$ twists and domain walls shown are of type $T_\mathsf{B}$. For the $Z_3$ twists and domain walls, the arrows indicate the direction in which a particle crossing the wall undergoes the permutation $T_\mathsf{RGB}$.}\label{fig:Twists_Logical_ops}
  \end{figure}

The degeneracy associated with twist defects discussed in the preceding section can be exploited for quantum computation.
Introducing $k$ pairs of bosonic $Z_2$ twists of a single type (e.g., $T_\mathsf{B}$) into the MSC (without boundaries), results in $2^{k-1}$ additional ground states and hence $k-1$ additional logical qubits.
(The $-1$ is because it takes at least four twists to get a set of logical operators. This is analogous to at least four Ising anyons being needed to furnish a qubit~\cite{Non-Abelian_TQC_Review}.)
For reasons discussed below, we will refer to this type of logical qubit as a ``bosonic twist (BT) qubit" and the information it stores as ``bosonic logical information."

Logical operators for these BT qubits
correspond to string operators encircling pairs of twists. 
Specifically, if $l$ is a set of red bonds forming a closed loop around a pair of $T_\mathsf{B}$ twists, then the red string loop $\prod_{(j,k)\in l}i\gamma_j\gamma_k$ is a logical operator. 
The fact that it is not in the stabilizer group results from the inability to deform it past any of the twists.
If one could do this, it would change the number of $\mathsf{RG}$ anyons localized at that twist, and hence violate the total color conservation of anyons in the system.
The %
conjugate logical operator is given by a red/green string operator encircling a pair of twists and crossing two domain walls (see Figure~\ref{subfig:Z2_Twist_logical_ops}).
Blue string operators encircling $T_\mathsf{B}$ twists are stabilizer operators.

We schematically illustrate the logical operators associated with four $Z_2$ twists in Figure~\ref{subfig:Z2_Twist_logical_ops}, using colored lines to indicate string operators.
The operators $\bar{X}$ and $\bar{Z}$ shown anti-commute, since different-colored string operators that cross each other once intersect at a vertex, and hence anti-commute.
These logical operators generate the full algebra of operators acting on the logical qubit created by the addition of four twists to the system.

In Figure~\ref{subfig:Z3_Twist_logical_ops} we illustrate the logical operators for four $Z_3$ twists. 
In the figure, an anyon crossing a domain wall in the direction of the arrows is transformed according to the permutation $T_\mathsf{RGB}$, while those crossing in the opposite direction are transformed according to the inverse permutation. 
As can be seen, four $Z_3$ twists permit twice the number of logical operators as four $Z_2$ twists, owing to the extra degeneracy they afford (cf. Section~\ref{subsec:Quantum_Dimension}).

There is an arbitrariness to the logical operators shown. 
In Figure~\ref{subfig:Z2_Twist_logical_ops}, one may wrap blue string operators around the two upper-most or left-most twists (these operators are stabilizer elements).
The blue strings can be fused with the logical operators indicated, in the sense outlined in Section~\ref{subsec:Anyons_MSC}, thereby changing the red string operators to green ones and vice versa. 
If the code is embedded on a sphere, red and green string operators encircling all four twists are stabilizers, since they may be deformed around to the other side of the sphere, where they are not encircling any twists.
With different boundary conditions, this becomes complicated, and we deal with these cases in the following section. For now, we continue with the simple case of a sphere.
Operators $\bar{X}$ and $\bar{Z}$ can then be deformed into operators that instead circle the two bottom and right-most twists respectively.
In Figure~\ref{subfig:Z3_Twist_logical_ops}, one cannot change the colors of the strings, since any string operator encircling a pair of $Z_3$ twists is not a stabilizer.
But on a sphere, all strings may be deformed to encircle the other two twists in the set of four.
Twists can admit either sparse or dense encodings of logical qubits. 
This is analogous to encodings with MZMs~\cite{MZM_TQC_Review2015}.
In a dense encoding, $2k+2$ twists encode $k$ logical qubits. 
In sparse encodings, certain logical operators from the dense encoding become stabilizers, thereby lowering the number of logical qubits.
For example, in a collection of $4k$ $T_\mathsf{B}$ twists, we can promote to stabilizers all red (and hence also green) string operators surrounding each group of four twists. 
This results in $k$ logical qubits, each stored within one group of four twists, and having the logical operators shown in Figure~\ref{subfig:Z2_Twist_logical_ops}.

\subsection{Logical MZMs from Fermionic Twists}
\label{subsec:LogMZM_fermion_twist}

\begin{figure}
    \centering
    \subfloat[\label{subfig:Fermionic_logical_ops}]{%
      \includegraphics[width=0.49\linewidth]{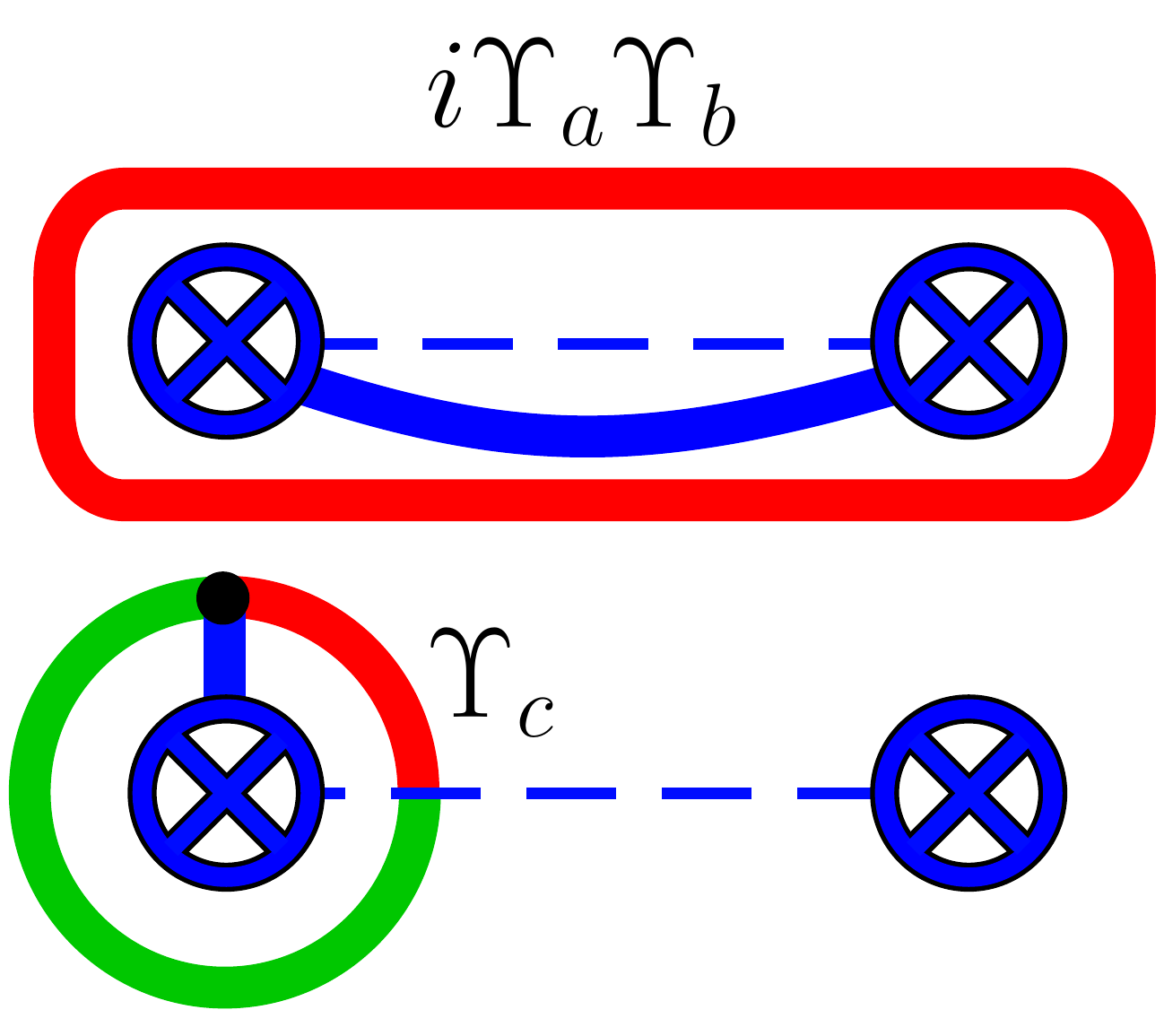}
    }\hfill
    \subfloat[\label{subfig:Odd_Boundary}]{\includegraphics[width=0.45\linewidth]{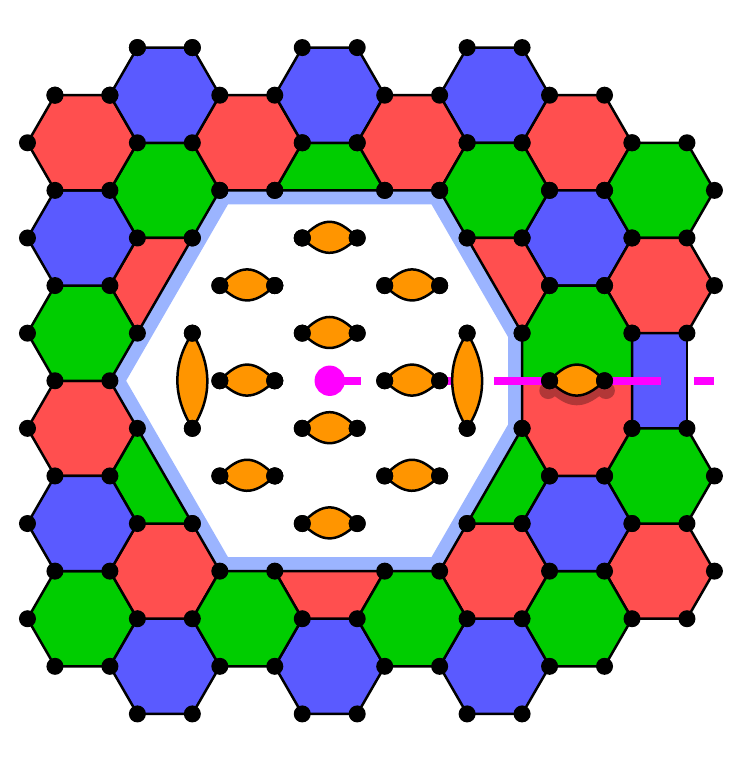}}
    \caption{Logical MZMs in the MSC with fermionic twists (blue circles containing crosses). In (a) a logical MZM $\Upsilon_c$ and a logical MZM bilinear $i\Upsilon_a\Upsilon_b$ are represented as string operators, the latter being the product of the red string encircling the two twists, and the blue string terminating on the twists. The holes associated with fermionic twists may be enlarged, so long as the resulting boundary has odd length. An example is shown in (b), with the location of the twist and domain wall indicated by a pink dot and dashed line respectively.}
\end{figure}

Both bosonic and fermionic twists in the MSC host the BT qubits discussed above. 
This same type of topological degeneracy is seen in twists of qubit-based codes, which is why we refer to it as ``bosonic."
For fermionic twists in the MSC we will label the corresponding bosonic logical operators as $\bar{X}_B$ and $\bar{Z}_B$.
In addition, fermionic twists also host logical MZMs.
In the same way as four physical MZMs can store a single qubit in a fermionic parity sector~\cite{MZM_TQC_Review2015}, so too can four logical MZMs $\Upsilon_a, \Upsilon_b, \Upsilon_c, \Upsilon_d$. 
We can also define encoded Pauli operators acting on the qubits stored by these logical MZMs.
For example, for four logical MZMs, we may have ``fermionic" logical operators $\bar{X}_F = i\Upsilon_a\Upsilon_b$ and $\bar{Z}_F = i\Upsilon_a\Upsilon_c$. 
We will refer to qubits stored in this way as ``fermionic twist (FT) qubits."
Figure~\ref{subfig:Fermionic_logical_ops} shows the operator $i\Upsilon_a\Upsilon_b$ for two twists, deformed such that it is the product of a red string operator encircling the twists, and a blue string operator terminating on the two twists.
It also shows the representation in terms of string operators of a single logical Majorana $\Upsilon_c$, which is equivalent to the localization of an $\mathsf{RGB}$ anyon at that twist (cf. Section~\ref{subsec:Quantum_Dimension}).

The logical operators $\bar{X}_F$ and $\bar{Z}_F$ have low minimum weights but large diameters, if the twists are far-separated (cf. Section~\ref{subsec:MFCs}). 
If only local, fermion parity conserving (FPC, i.e., even-weight) errors can occur, FT qubits are protected by the twists' separation~\cite{Maj_Ferm_Codes}.
(Here we assume local coupling to the environment so that elementary FPC error operators are locally supported; all possible FPC error operators are products of these elementary FPC operators.)
In realistic setups, however, we also have to contend with quasi-particle poisoning (QPP) resulting in non-FPC errors.
The simplest such error operator is a single Majorana operator.
To protect FT qubits from QPP, we can grow the size of the odd-length boundaries associated with twists and hence grow the weights of logical MZMs.
As such (assuming we also increase twist separation), the FT qubits will have larger distances and so an increased protection against QPP noise.
An example of such an enlarged hole is shown in Figure~\ref{subfig:Odd_Boundary}.

Note that in systems where FPC errors are more common than QPP events, and if the former have below-threshold error rates, we can reduce the logical error rate to a low value set by QPP by increasing only the twist separation (i.e., increasing the code diameter but not the distance). 
This contrasts to MSC strategies without logical MZMs, where the code distance must be increased to reduce the logical error rate.
Using fermionic twists in systems with low QPP probability, we can store double the number of logical qubits with only a minor increase in overheads, resulting from the size of the odd-boundary holes (see Figure~\ref{subfig:Odd_Boundary}).
While for low enough QPP probability microscopic MZMs could be used to store topological qubits, without error-correcting codes such a strategy is not fault-tolerant as any finite QPP rate will inevitably limit the coherence times of the topological qubit. 
Meanwhile, using logical MZMs, logical error rates resulting from QPP can be made arbitrarily small by increasing the code distance, provided the QPP rate is below the corresponding error threshold.

Logical MZMs can also appear in systems with only bosonic twists, when a boundary is introduced to the system.
To our knowledge, this type of logical MZM has not been identified previously in the literature. 
We introduce it in Appendix~\ref{app:Logical_Maj_Boundary}.

\section{Fault-Tolerant Computation With Twists}\label{sec:FT_Comp_with_Twists}

In this section we detail how to perform fault-tolerant quantum computation with FT and BT qubits.
These procedures will require the ability to move twists via ``code deformation" (Section~\ref{subsec:Code_Deformation}).
In Section~\ref{subsec:Logical_qubit_braids} we discuss how to prepare logical Pauli eigenstates and apply Clifford gates to BT qubits.
In Sections~\ref{subsec:Fermion_Parity_Mmt} and \ref{subsec:Hybrid_Approaches} we discuss performing measurements of BT and FT logical operators.
The above, combined with the preparation of magic states (Section~\ref{subsec:Magic_State}), is sufficient for universal computation with both types of qubit.

\subsection{Code Deformation}\label{subsec:Code_Deformation}

\begin{figure*}[t]
\subfloat[\label{subfig:Bosonic_Z2_Code_Def}]{\includegraphics[width=0.38\textwidth]{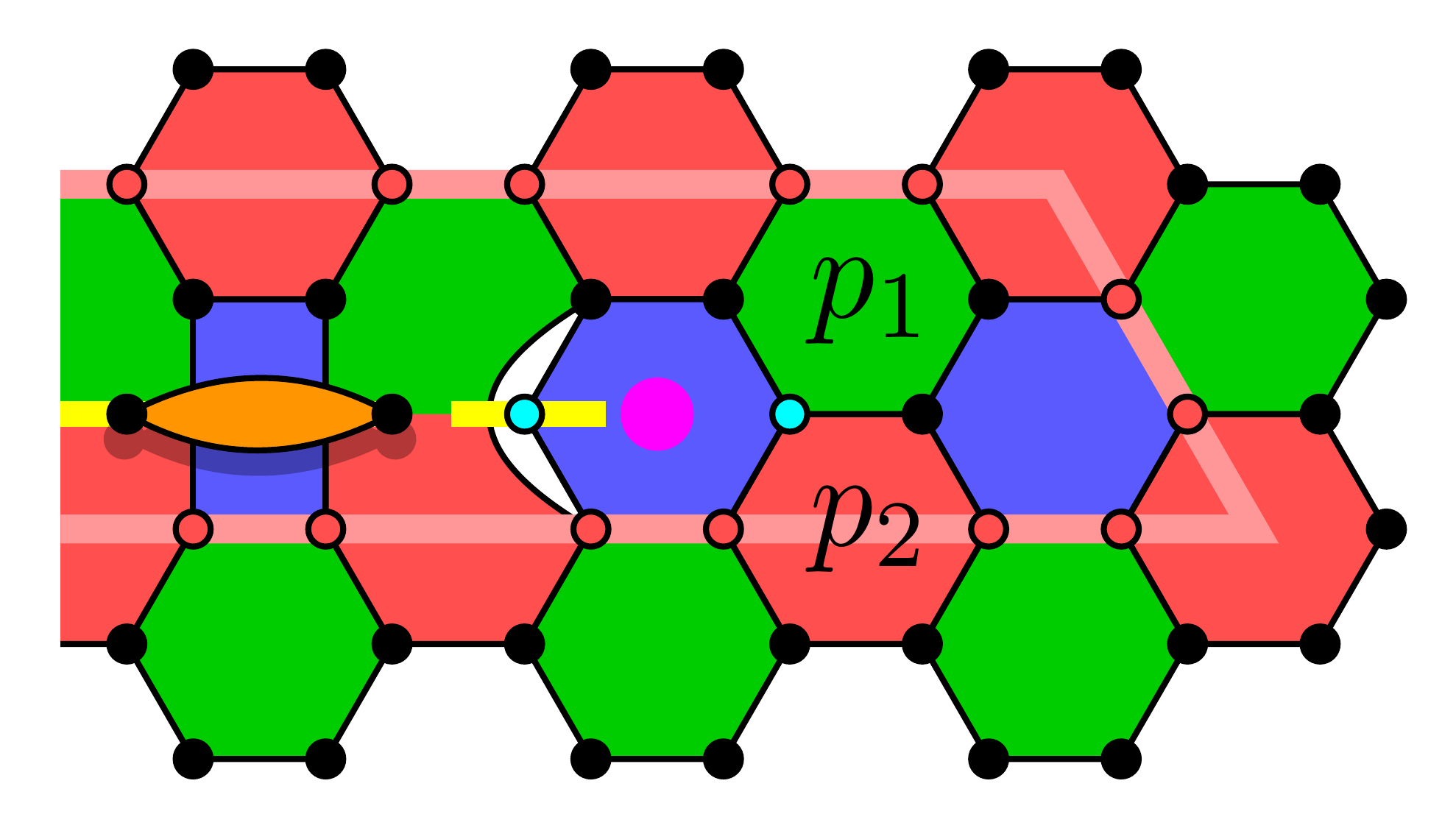}}
\hfill
\subfloat[\label{subfig:Bosonic_Z3_Code_Def}]{\includegraphics[width=0.46\textwidth]{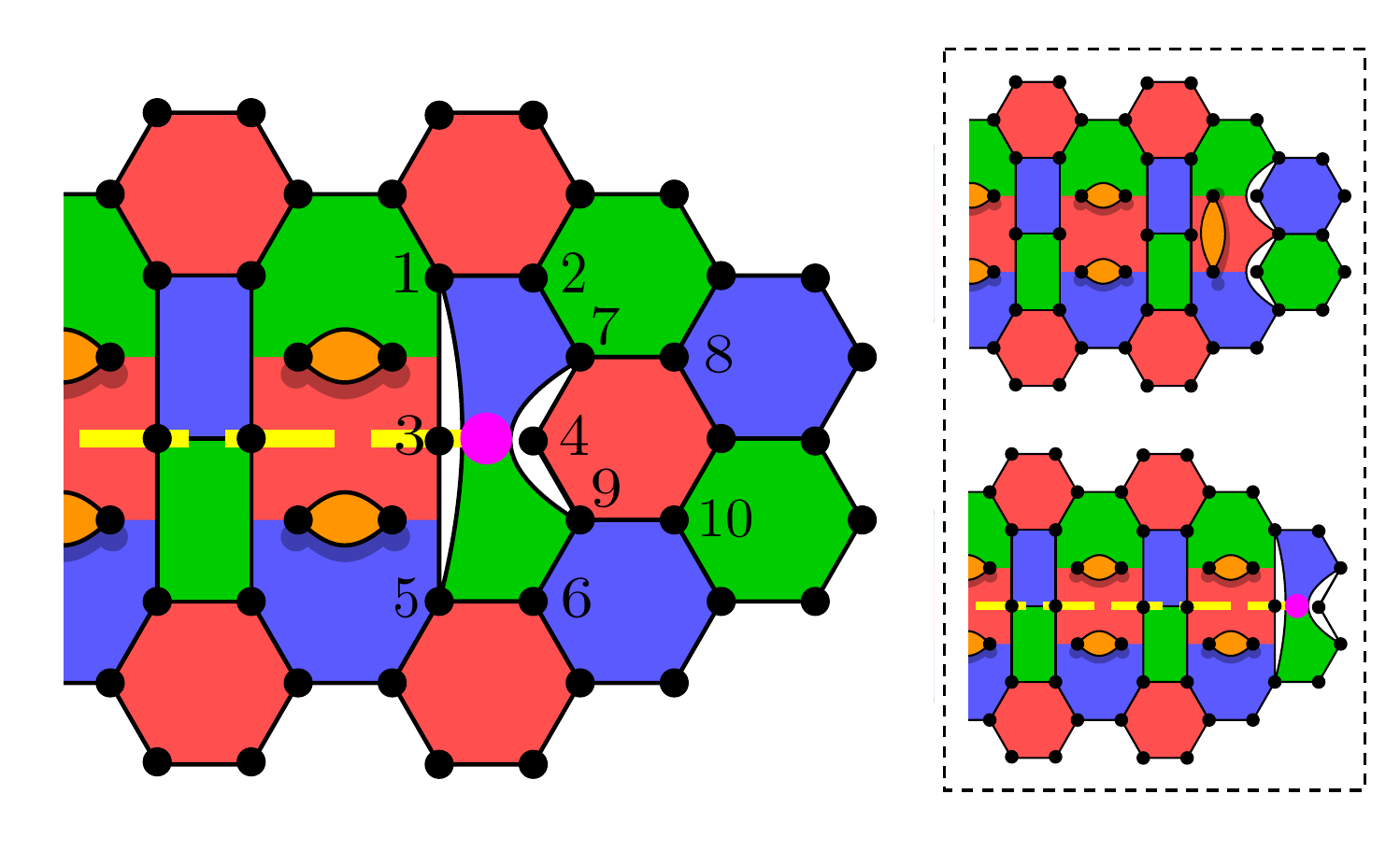}}\\
\subfloat[\label{subfig:Fermionic_Z2_Enlarge}]{\includegraphics[width=0.43\textwidth]{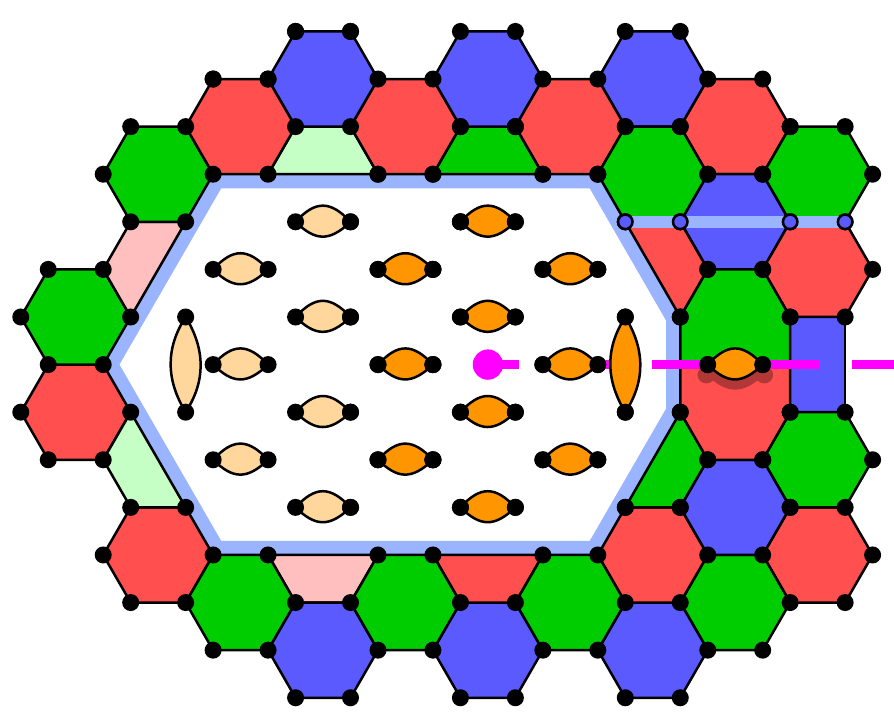}}
\hfill
\subfloat[\label{subfig:Fermionic_Z2_Shrunk}]{\includegraphics[width=0.43\textwidth]{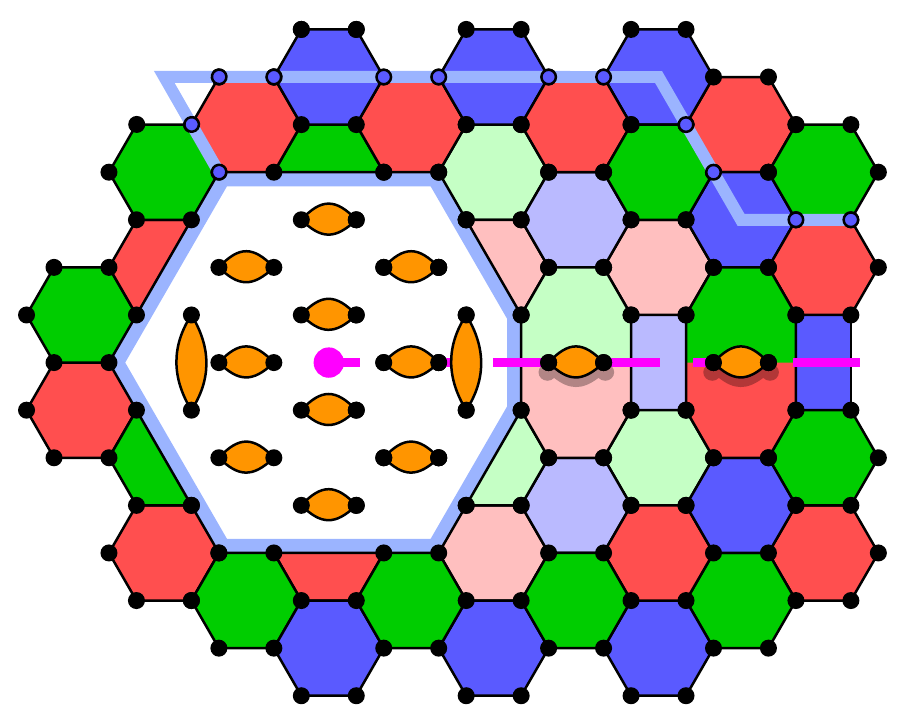}}
\caption{Code deformation to move bosonic (a)-(b) and fermionic (c)-(d) twists (pink dots). Domain walls are yellow/pink dashed lines. (a) A $Z_2$ bosonic twist along with part of a logical operator (red string), deformed away from the twist. To move the twist one plaquette to the right, we measure the bilinear with the light blue Majorana operators and merge the stabilizers at plaquettes $p_{1,2}$. (b) A $Z_3$ bosonic twist. After measuring the new stabilizers $\gamma_1\gamma_2\gamma_3\gamma_4$ and $\gamma_3\gamma_4\gamma_5\gamma_6$, the stabilizer generators appear as shown in the top inset. Performing two more bilinear measurements produces the stabilizers shown in the bottom inset. The result of these measurements is that the twist is moved to the right. (c) The hole from Figure~\ref{subfig:Odd_Boundary}, associated with a fermionic twist, is enlarged to the left by measuring all light-colored operators. The blue string operator indicated commutes with all these measurements. (d) The hole is shrunk from the right by measuring all light-colored operators. The same blue string operator is now deformed to meet the hole on the left-hand side, where it commutes with all of these measurements.}
\label{fig:Patch_twists}
\end{figure*}

For fault-tolerant computation using twists, we need to understand how to move them. 
We can do this with code deformation~\cite{Bombin_twists_code_deformation, Bombin_Code_Deformation, Punctures_Raussendorf_2006, Raussendorf_2007, Gauge_Color_Code}:
a process where a code with stabilizer group $\mathcal{S}$ is mapped to another code with stabilizer group $\mathcal{S}^\prime$. 
To project to a code state of $\mathcal{S}^\prime$, we perform measurements of its stabilizer generators. 
In particular, in the new code those previous $\mathcal{S}$ stabilizer elements that anti-commute with any $s^\prime \in \mathcal{S}^\prime$ are replaced.
While this changes the code's stabilizer group, it preserves the logical state if each logical operator equivalence class $[L]$ of the original code has members that are also logical operators of the new code. 

Here we introduce a novel code deformation procedure for the MSC that allows us to move twists.
The procedure differs between fermionic and bosonic twists.

\subsubsection{Moving Bosonic Twists}

The code deformation procedure for bosonic $Z_2$ twists is illustrated in Figure~\ref{subfig:Bosonic_Z2_Code_Def}.
In the figure, a $T_\mathsf{B}$ twist along with part of a logical operator (a red string operator encircling this and another twist) are shown. 
To move the twist to the blue plaquette to its right, we measure the bilinear $\upsilon$ involving Majorana operators at the two light-blue sites. 
This then becomes an orange stabilizer along the newly enlarged domain wall. 
Since $\upsilon$ anticommutes with the plaquette operators $\mathcal{O}_{p_1}$ and $\mathcal{O}_{p_2}$ (indicated in the figure), these two operators are not included in $\mathcal{S}^\prime$. 
However their product $\mathcal{O}_{p_1}\mathcal{O}_{p_2}$ commutes with $\upsilon$ and hence is included as an 8-body stabilizer element. 
Note that measuring $\upsilon$ means that the blue hexagonal plaquette formerly hosting the twist can be replaced by a rectangular (4-body) plaquette: the new stabilizer generator is the product of $\upsilon$ and the original hexagonal plaquette operator.

The logical operator in the figure commutes with the updated stabilizer group.
In general, we can always deform these string-like logical operators away from twists~\cite{bravyi2009no,Maj_Ferm_Codes},
so that they commute with the code deformation measurements performed.
Thus, twists can be moved via series of small steps like the above, without destroying the information stored in logical qubits.

In Figure~\ref{subfig:Bosonic_Z3_Code_Def}, a $Z_3$ bosonic twist is illustrated. 
If we again wish to move this twist to the right, we first measure the rectangular operators $\gamma_1\gamma_2\gamma_3\gamma_4$ and $\gamma_3\gamma_4\gamma_5\gamma_6$, with the vertex numbering indicated in the figure.
Again, these operators anti-commute with certain members of the original stabilizer group $\mathcal{S}$. 
After measuring both rectangular operators, the adjacent red, green and blue hexagonal operators to their right are replaced by a 10-body operator (see upper inset of Figure~\ref{subfig:Bosonic_Z3_Code_Def}).
The rectangular operators both commute with the operator defined on the twist's original location, $\mathcal{O}_T = i\gamma_1\gamma_2\gamma_5\gamma_6\gamma_7\gamma_9$.
The product of the two rectangular operators is $-\gamma_1\gamma_2\gamma_5\gamma_6$ and multiplying this with $\mathcal{O}_T$, we find that the bilinear $i\gamma_9\gamma_7$ is a stabilizer element. 
This is shown in orange in the top inset of Figure~\ref{subfig:Bosonic_Z3_Code_Def}.
We then measure operators $i\gamma_7\gamma_8$ and $i\gamma_9\gamma_{10}$. 
This produces the same orange operators as in other locations along the domain wall.
By following how the stabilizer group is updated after each of these two measurements, we see that they result in the stabilizer generators indicated in the lower inset of Figure~\ref{subfig:Bosonic_Z3_Code_Def}.

We can similarly move twists in different directions or generate a pair of twists in the lattice by measuring the stabilizer generators of the new code lattice. 
If any of these measurements return a $-1$ result, we can simply define the new stabilizer generators to be the negative of all measured operators whose eigenvalues are $-1$.

Moving the twists a short distance (compared to the twist separation) does not change the encoded states of the BT qubits, since the changes are restricted to be far from the paths of a suitable set of logical operators. 
However we shall see that performing a large sequence of small code deformation steps can result in unitary transformations of the code states.

\subsubsection{Moving Fermionic Twists}

This procedure must preserve the information stored both by BT qubits and by logical MZMs.
As above, the former can be ensured by deforming bosonic logical operators away from the twists. 
However, logical MZMs cannot be deformed in such a way: their support always has an odd-weight intersecton with the boundary associated with the twist (cf.~Figure~\ref{subfig:Fermionic_logical_ops}).

Code deformation can most easily be seen to preserve information in MZMs by considering moving an enlarged hole, such as the one in Figure~\ref{subfig:Odd_Boundary}. 
We move this hole through a series of enlarging and contracting procedures. 
In Figure~\ref{subfig:Fermionic_Z2_Enlarge}, the hole from Figure~\ref{subfig:Odd_Boundary} is enlarged to the left by measuring the bilinear operators indicated in light-orange in Figure~\ref{subfig:Fermionic_Z2_Enlarge}, along with the four-body plaquette operators colored light-green and light-red in the figure.
Figure~\ref{subfig:Fermionic_Z2_Shrunk} then shows the same hole shrunk from the right by measuring the light-colored plaquette operators indicated.
The final result is that the location of the hole, and hence the twist, has moved to the left.

To see the preservation of the logical state, consider the logical operators in Figure~\ref{subfig:Fermionic_logical_ops}.
The logical operator $i\Upsilon_a\Upsilon_b$ is the product of a string operator encircling twists $a$ and $b$, and a string operator connecting twists $a$ and $b$. 
Similarly, any parity-even product of logical MZMs can be deformed into an operator that intersects each of the twists' boundaries at a single site (e.g., operator $i\Upsilon_a\Upsilon_b$ in Figure~\ref{subfig:Fermionic_logical_ops}).
Other sections of the fermionic logical operator can be deformed away from the twists during code deformation, but the string operator(s) terminating on the twist boundaries cannot.
The procedure will preserve fermionic information if we can deform these operators so that their support intersects none of the sites involved in the measurements to be performed. 
In Figure~\ref{subfig:Fermionic_Z2_Enlarge} we indicate such a string operator terminating on the right-hand side of the boundary of a $T_\mathsf{B}$ twist - this is unaffected by the measurements that enlarge the hole.
In Figure~\ref{subfig:Fermionic_Z2_Shrunk} we indicate the same string operator, deformed to terminate on the left-hand side of the boundary. 
This operator is now unaffected by the measurements that shrink the hole. 
Hence the information stored in these logical MZMs is protected during code deformation.

\subsection{State Preparation of BT Qubits and Clifford Gates via Braiding Twists}\label{subsec:Logical_qubit_braids}

\begin{figure*}
\centering
\includegraphics[width=0.75\textwidth]{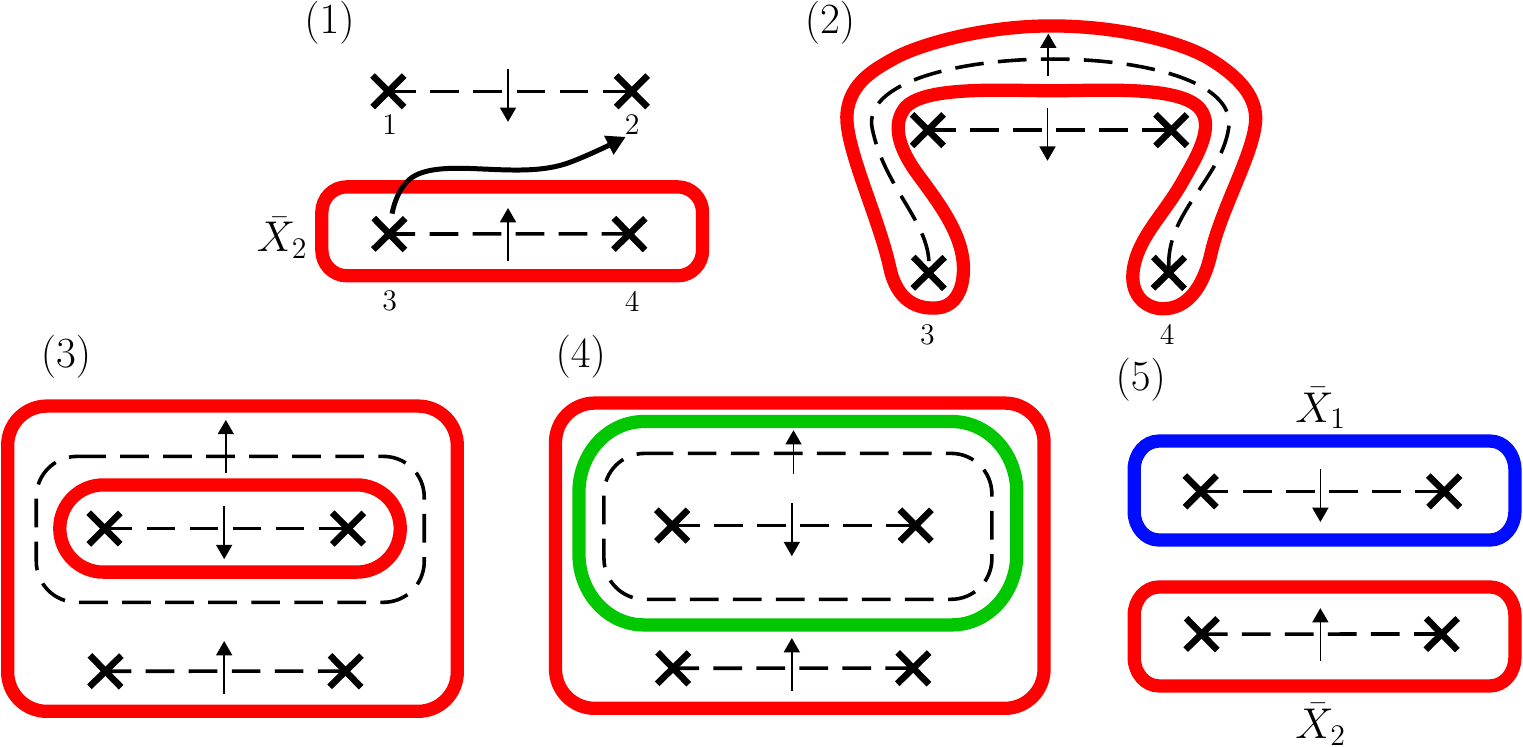}
\caption{A braid of $Z_3$ twist $T_3$ around the pair $T_1$ and $T_2$ results in an entangling gate between the logical qubits. We show how this braid affects the logical operator $\bar{X}_2$. In panel (3) we have altered the location of the domain wall between $T_3$ and $T_4$ with a relabeling of plaquettes (see Appendix~\ref{app:Gauge_Dependence}). In panel (5) we have deleted the closed loop of domain wall with another relabeling.}\label{fig:Z3_CNOT_Braid}
\end{figure*}

We will now discuss how to prepare logical Pauli eigenstates and perform fault-tolerant Clifford gates using BT qubits. 
The latter will be achieved purely by braiding $Z_2$ and $Z_3$ twists, in analogy to Clifford gate procedures in the bosonic Color Code~\cite{Bombin_twists_code_deformation} (though Clifford gates may also be achieved through lattice surgery; see Section~\ref{subsec:Hybrid_Approaches}).
Our MSC scheme requires no ancilla logical qubits for the implementation of single-qubit Clifford gates, unlike other approaches~\cite{MjFerm_Surface_Code,Roadmap_to_MSCs}.
We shall consider BT qubits stored sparsely in quartets of either $Z_2$ (storing one qubit in four twists) or $Z_3$ (two qubits in four twists) bosonic twists.
We will see that we need only consider two sub-types of $Z_2$ twist, which we take (without loss of generality) to be $T_\mathsf{B}$ and $T_\mathsf{R}$.

To prepare four $Z_2$ twists in an eigenstate of a given logical Pauli operator, we start with a MSC patch hosting these twists, such as the one of Figure~\ref{fig:Patch}.
For example, to prepare this patch in the $+1$ eigenstate of $\bar{X}$ (a red string operator), we initialize the patch in the $+1$ eigenstate of all red bond operators. 
This prepares $\bar{X}$ and green and blue plaquette operators in their $+1$ eigenstates since they are all products of red bond operators. 
To prepare red plaquette operators in their $+1$ eigenstates, we measure them and apply  red string operators that take the state to the code space given the observed syndrome (in practice, we would not apply this operator and instead, redefine the stabilizers).
Since $\bar{X}$ and the green/blue plaquette operators commute with the red plaquette operators, these measurements do not change their eigenvalues from the preceding step. 
The inequivalent choices of string in the last step differ by $\bar{X}$; this is again immaterial given we already had a $+1$ eigenstate of $\bar{X}$.
This procedure thus prepares the patch in the $+1$ eigenstate of $\bar{X}$ in the code space.
We can then move the twists away from the boundaries of the patch using code deformation (cf. Section~\ref{subsec:Code_Deformation}).
Finally, we can fuse the patch with the rest of the lattice hosting logical information by adding stabilizer operators that bridge the gap between the boundaries of the two lattices.

As is possible in the bosonic surface code~\cite{Holes_Twists_SurfaceCode}, we can braid bosonic twists in the MSC to perform certain encoded Clifford gates on logical qubits. 
This procedure is achieved with code deformation via sequences of small steps.
It is a naturally fault-tolerant procedure, since the code distance (and the diameters of logical operators) can be kept large throughout the whole braiding process.
If we were restricted to using $Z_2$ twists of a single type (e.g., $T_\mathsf{B}$) then, just as in the bosonic surface code, braiding could not generate the entire Clifford group for the logical qubits.
Braiding $T_\mathsf{B}$ twists can generate only the single-qubit Clifford gates, as explained in Appendix~\ref{app:Braiding}.
This is due to the similarity between bosonic $T_\mathsf{B}$ twists and Ising anyons~\cite{Non-Abelian_TQC_Review}, as is also the case in the surface code~\cite{Bombin_Twists,TEE_with_twist}.

However, unlike the bosonic surface code, the MSC also has $Z_3$ twist defects and hence allows for a larger set of gates to be implemented by braids. This is similar to the case of the bosonic Color Code~\cite{Bombin_twists_code_deformation}.
We can enact an entangling gate between the two qubits stored in a quartet of $Z_3$ twists by using the braid shown in Figure~\ref{fig:Z3_CNOT_Braid}. 
If we call this braiding operation $B$, it can be seen from the figure that the operator $\bar{X}_2$ is mapped to $B\bar{X}_2B^\dagger = \bar{X}_1\bar{X}_2$ by this braid.
Similarly (see Appendix~\ref{app:Braiding}), the other logical Pauli operators are transformed as: $\bar{X}_1\rightarrow \bar{X}_2$, $\bar{Z}_1 \rightarrow \bar{Y}_1 \bar{Y}_2$ and $\bar{Z}_2\rightarrow \bar{Z}_1\bar{X}_2$. 
We can fix the signs in these mappings by carefully defining the logical operators and stabilizer generators.
It can be verified that $B$ is equivalent to the following circuit between the two logical qubits:
\[\Qcircuit @C=1em @R=1.5em {
\lstick{1} & \gate{S} & \gate{H} & \gate{S} & \ctrl{1} & \gate{S} & \gate{H} & \gate{S} & \targ & \qw \\
\lstick{2} & \qw & \gate{X} & \qw &\targ & \gate{S} & \gate{H} & \gate{S} & \ctrl{-1} & \qw
}\]

A $Z_3$ twist can be split into two differently-colored $Z_2$ twists (cf. Appendix~\ref{app:Twists_fusion}).
Splitting all four $Z_3$ twists into $T_\mathsf{B}$ and $T_\mathsf{R}$ twists results in qubit 1 being stored between the $T_\mathsf{R}$ twists and qubit 2 between the $T_\mathsf{B}$ twists.
Thus, through this splitting, we can enact all single-qubit Clifford gates on qubits $1$ and $2$ with $Z_2$ twist braids.
We can therefore achieve the full set of Clifford gates on qubits 1 and 2 via twist braids (note that a CNOT gate controlled on qubit 1 and targeted on qubit 2 can be achieved with the gate $\bar{Z}_1 \bar{H}_1 B^3 \bar{H}_1$, where $\bar{H}_1$ is the Hadamard gate acting on qubit 1).

To enact a Clifford circuit on $n$ BT qubits, we can store some of them in  $T_\mathsf{B}$ twists and the rest in $T_\mathsf{R}$ twists. 
This allows single-qubit Cliffords to be implemented within same-color quartets and entangling gates to be implemented between $T_\mathsf{B}$-stored and $T_\mathsf{R}$-stored qubits with the above procedures.
We can also entangle $T_\mathsf{B}$-stored (equivalently $T_\mathsf{R}$-stored) qubits by introducing a quartet of $T_\mathsf{G}$ twists as an ancilla qubit.
Entangling gates between the ancilla and the two logical qubits (which are possible due to the above discussion) are all that is required to perform an entangling gate on the two logical qubits.

\subsection{State Preparation of FT Qubits and Logical Majorana Parity Measurements via Lattice Surgery}\label{subsec:Fermion_Parity_Mmt}

To compute with the logical MZMs of fermionic twists, we need the ability to prepare eigenstates, perform computational basis measurements, and perform logical $B_4$ and $T_2$ gates,
which have the form $B_{4,abcd} = \exp(i\frac{\pi}{4}\Upsilon_a\Upsilon_b\Upsilon_c\Upsilon_d)$ and $T_{2,ab} = \exp(\frac{\pi}{8}\Upsilon_a \Upsilon_b)$ respectively.
In general we may also wish to perform $B_{2k} = \exp{(i^{k+1}\frac{\pi}{4}\Upsilon_{j_1}\ldots \Upsilon_{j_{2k}})}$ gates as these too are fermionic variants of Clifford gates acting on FT qubits. 
While these could be synthesized from $B_4$ gates~\cite{FPBC}, it may be preferable to perform them directly.
We also want to perform all of these fermionic gates while preserving the bosonic information stored in the twists.
We will, similarly to the previous section, detail how to achieve this with $T_\mathsf{B}$ and $T_\mathsf{R}$ twists as this will be sufficient for universal computation.

We can perform $B_{2k}$ gates if we can prepare ancilla logical MZMs furnishing known eigenstates and measure arbitrary logical MZM products, i.e., logical fermion parity operators~\cite{Bravyi_Kitaev_Fermionic_QC2002}.
The ability to perform these measurements also allows us to prepare eigenstates of operators $i\Upsilon_a\Upsilon_b$ for arbitrary logical MZMs $a,b$.
In this section, we detail how to prepare patches of code that host ancilla logical MZMs, and how to perform logical fermion parity measurements.
In the next section, we detail how to perform $T_2$ gates.
This, combined with the above, allows us to perform universal computation with logical MZMs.

\subsubsection{Bilinear Measurement and Eigenstate Preparation}\label{subsec:Bilinear_Mmt}

\begin{figure}[t]
\centering
\includegraphics[width=0.95\linewidth]{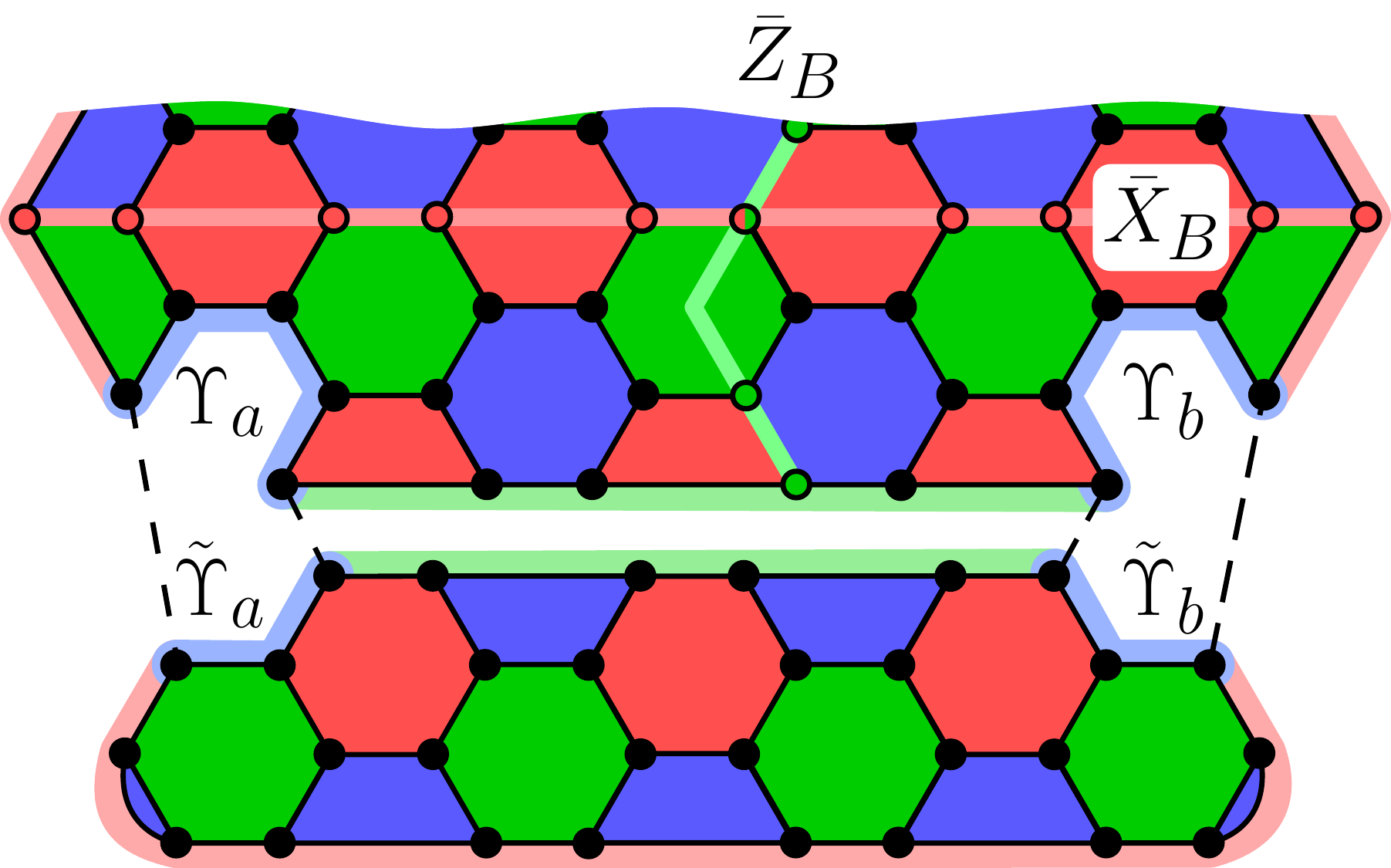}
\caption{Fault-tolerant measurement
of $i\Upsilon_a\Upsilon_b$ through lattice surgery with a neighboring MSC patch  containing ancillary logical MZMs $\tilde{\Upsilon}_a$ and $\tilde{\Upsilon}_b$. A small MSC section is shown in the top part of the figure. This patch contains the logical MZMs to be measured ($\Upsilon_a$ and $\Upsilon_b$), and some BT qubits. Some examples of bosonic logical operators ($\bar{X}_B$ and $\bar{Z}_B$) are shown. 
The ancilla MSC patch, shown in the bottom of the figure, is prepared in the $+1$ eigenstate of $i\tilde{\Upsilon}_a\tilde{\Upsilon}_b$. Operators $\tilde{\Upsilon}_a$ and $\tilde{\Upsilon}_b$ are associated with the length-$3$ blue boundary sections. 
$i\Upsilon_a\tilde{\Upsilon}_a$ and $i\Upsilon_b\tilde{\Upsilon}_b$ are eight-body plaquette operators indicated by dashed lines. The measurement of these two operators fuses the two patches together along their blue boundary sections.}\label{fig:Bilinear_mmt}
\end{figure}

The simplest measurement to consider is that of logical MZM bilinears, i.e., operators of the form $i\Upsilon_a\Upsilon_b$.
This type of measurement may appear at the end of a fermionic circuit (sampling the output distribution) and can also be used to prepare computational basis states for the start of a circuit.
To perform such a measurement in a non-destructive, fault-tolerant manner, we first prepare an ancillary pair of logical MZMs ($\tilde{\Upsilon}_a$ and $\tilde{\Upsilon}_b$) in a separate patch of code located close to the first.
This ancilla patch is shown in Figure~\ref{fig:Bilinear_mmt}. 
Note that this patch contains no BT qubits, owing to its boundary conditions. 
Comparing this patch with the one in Figure~\ref{fig:Patch}, we see that while in the latter, there exist multiple inequivalent logical operators running between separated boundaries of the same color ($\bar{X}$ and $\bar{Z}$ in Figure~\ref{fig:Patch}), the ancilla patch in this case has only one such string operator, a blue string terminating on the two blue boundary sections.
This is equivalent to the operator $i\tilde{\Upsilon}_a\tilde{\Upsilon}_b$.
The patch can be prepared in the $+1$ eigenstate of $i\tilde{\Upsilon}_a\tilde{\Upsilon}_b$ by, similarly to the preparation of BT qubits, preparing the $+1$ eigenstate of all blue bond operators, then measuring all blue stabilizers, and applying a correction string if needed.

To measure $i\Upsilon_a\Upsilon_b$ on the neighboring patch of code, we merge the twists supporting $\Upsilon_a$ and $\Upsilon_b$ with a boundary colored differently from the twists, so that the logical Majoranas are located along the exterior boundary of the lattice (see Figure~\ref{subfig:Ferm_Corners_b} and~\ref{fig:Bilinear_mmt}).
For $T_\mathsf{B}$ and $T_\mathsf{R}$ twists we merge the twists with a green-colored boundary.
In Figure~\ref{fig:Bilinear_mmt} we show an example in which $\Upsilon_a$ and $\Upsilon_b$ are both supported along blue boundaries. 
See Appendix~\ref{app:Lattice_Surg} for a more general example.

We can measure operators $i\Upsilon_a\tilde{\Upsilon}_a$ and $i\Upsilon_b\tilde{\Upsilon}_b$ by fusing the blue boundaries associated with the logical Majoranas, thus forming an extra blue plaquette operator (see Figure~\ref{fig:Bilinear_mmt}).
We begin by measuring the operator $i\Upsilon_a\tilde{\Upsilon}_a$, obtaining the result $\eta_{a}$.
This operator anticommutes with both $i\Upsilon_a\Upsilon_b$ and $i\tilde{\Upsilon}_a\tilde{\Upsilon}_b$, but the effect of the measurement is simply to logical-braid
$\Upsilon_a$ with $\tilde{\Upsilon}_b$~\cite{Bravyi_PBC2016,FPBC}.
To see this, note that %
if the pre-measurement state is $\ket{
\psi}$ [which by construction satisfies $(i\tilde{\Upsilon}_a\tilde{\Upsilon}_b)\ket{\psi}=\ket{\psi}$], then the post-measurement state is $\frac{1}{\sqrt{2}}[1+\eta_{a}(i\Upsilon_a\tilde{\Upsilon}_a)]\ket{
\psi}$ which equals
\begin{equation}
B_{2,ba}\ket{\psi} = \frac{1}{\sqrt{2}}\left[1 + \eta_{a}(i\Upsilon_a\tilde{\Upsilon}_a)(i\tilde{\Upsilon}_a\tilde{\Upsilon}_b)\right]\ket{\psi}
\end{equation}
with the logical braid $B_{2,ba} = \exp(\eta_{a}\frac{\pi}{4}\tilde{\Upsilon}_b\Upsilon_a)$. 

$B_{2,ba}$ transforms logical Majoranas as: $B_{2,ba} \Upsilon_a B_{2,ba}^\dagger = \eta_{a}\tilde{\Upsilon}_b$ and $B_{2,ba}\tilde{\Upsilon}_b B_{2,ba}^\dagger = -\eta_{a}\Upsilon_a$.
The operator we wish to measure, $i\Upsilon_a\Upsilon_b$, is thus mapped to $\eta_{a}i\tilde{\Upsilon}_b\Upsilon_b$ through the above measurement.
Hence we must now measure the operator $i\tilde{\Upsilon}_b\Upsilon_b$, another blue plaquette operator fused from the corresponding blue boundaries. 
We denote the outcome by $\eta_{b}$.
The desired measurement outcome of $i\Upsilon_a\Upsilon_b$ is thus $\eta_{a}\eta_{b}$.
After these measurements, $-\eta_{a}\eta_{b}\Upsilon_a\Upsilon_b\tilde{\Upsilon}_a\tilde{\Upsilon}_b$ is a stabilizer of the resulting state of the full system.
Finally, we measure $i\tilde{\Upsilon}_a\tilde{\Upsilon}_b$ destructively (i.e., via its constituents~\cite{Bravyi_Kitaev_Fermionic_QC2002})
to 
decouple the ancilla patch: we measure all blue bond operators of the patch, yielding the eigenvalue of $i\tilde{\Upsilon}_a\tilde{\Upsilon}_b\prod_{p\in B} \mathcal{O}_p$ with the blue stabilizers $\mathcal{O}_{p
\in B}$ of the patch. We then measure these $\mathcal{O}_{p
\in B}$ (or use previous syndrome measurement outcomes) and thereby infer the outcome $\tilde{\eta}_{ab}$ for $i\tilde{\Upsilon}_a\tilde{\Upsilon}_b$. 
This yields a $+1$ eigenstate of $\tilde{\eta}_{ab}\eta_{a}\eta_{b}i\Upsilon_a\Upsilon_b$.
If $\tilde{\eta}_{ab}= -1$, this state is not the one corresponding to the obtained $i\Upsilon_a\Upsilon_b$ outcome  $\eta_{a}\eta_{b}$.
But this error can simply be classically tracked, and subsequent computations can be updated.

We can generalize this procedure to cases in which $\Upsilon_a$ and $\Upsilon_b$ have larger weights, and are hosted along boundaries of different colors. 
We outline the more involved setups required in Appendix~\ref{app:Lattice_Surg}.

The bosonic information in the twists is again preserved by these lattice surgery procedures, since bosonic logical operators (such as $\bar{Z}_B$ and $\bar{X}_B$ of Figure~\ref{fig:Bilinear_mmt}) can be deformed away from all of the lattice surgery measurements.

We need not move twists to a distant boundary if they are deep within the code bulk.
Instead, we can create a hole within the lattice by ceasing the stabilizer measurements within and altering the stabilizers along the boundary to ensure that it has a single color. 
We then introduce the MSC patch with the ancilla logical MZMs into this hole and proceed  as above.
Afterwards we fill in the hole by measuring the original plaquette operators.
Such a hole may add an extra logical qubit, but if so, filling in the hole simply discards the qubit again. 
While the introduction of the hole will affect the weights of certain logical operators, and hence may reduce the code distance, this distance can still be kept arbitrarily large by increasing the separation of the code's twists.

\begin{figure*}[ht]
\centering
\subfloat[\label{subfig:Four_body_1}]{\includegraphics[width=0.49\textwidth]{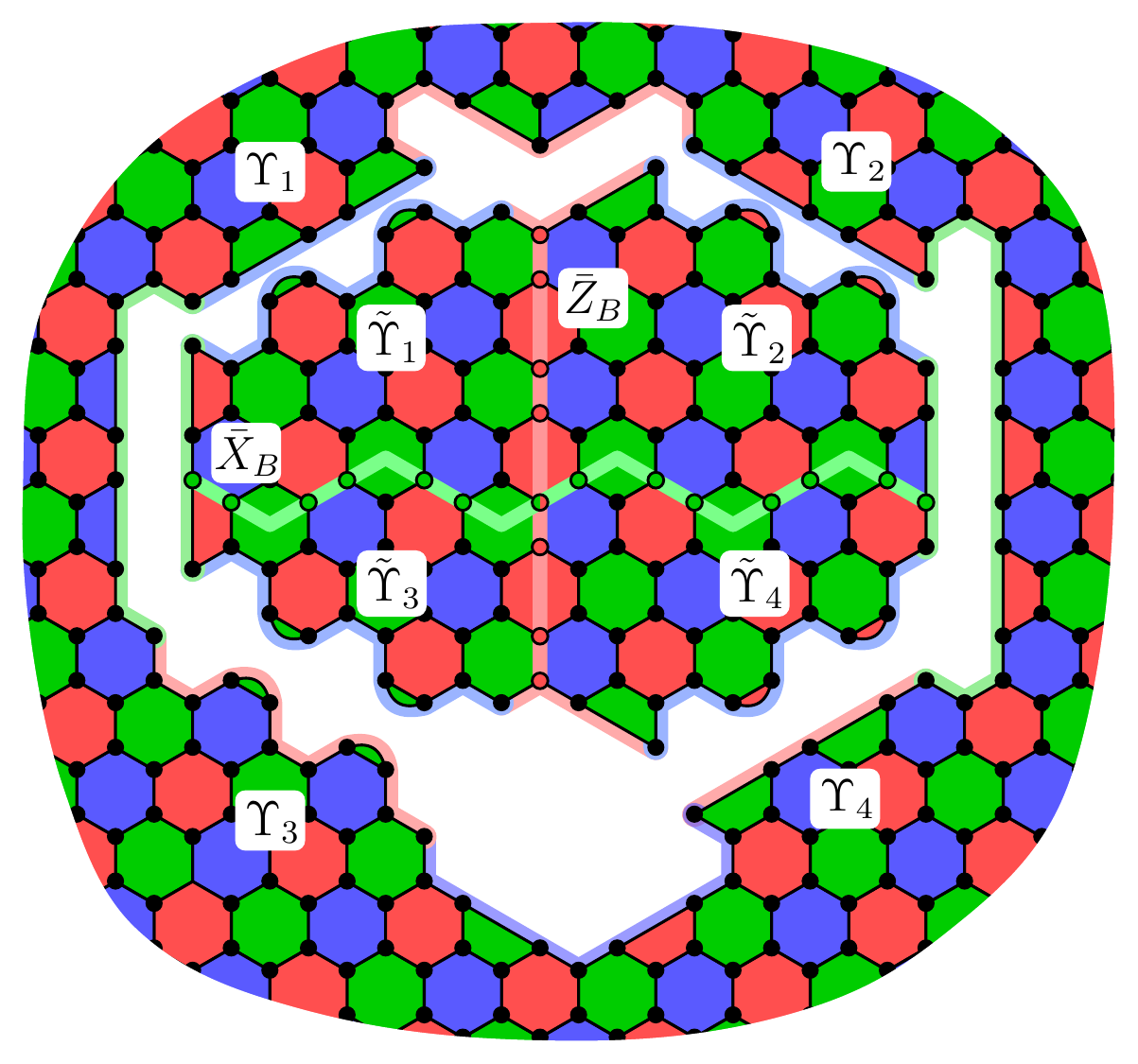}}\hfill
\subfloat[\label{subfig:Four_body_2}]{\includegraphics[width=0.49\textwidth]{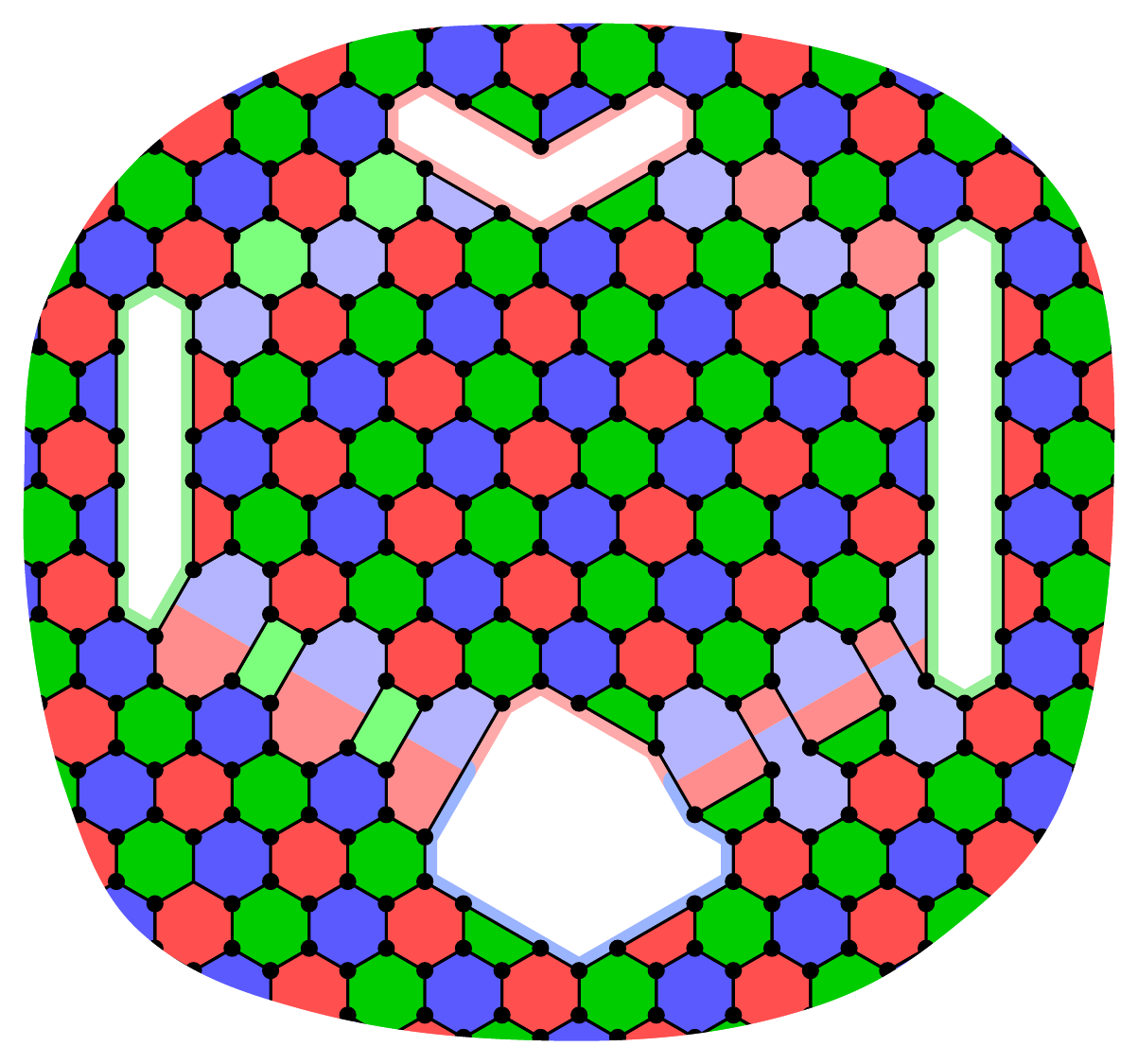}}
\caption{Lattice surgery procedure for measuring logical parity operator 
$-\Upsilon_1\Upsilon_2\Upsilon_3\Upsilon_4$. (a) Logical MZMs are supported on blue (for $\Upsilon_1$ and $\Upsilon_2$) or red (for $\Upsilon_3$ and $\Upsilon_4$) sections of odd-length boundaries on the patch surrounding the central hole. The patch inside the hole has ancilla logical MZMs ($\tilde{\Upsilon}_1,\ldots,\tilde{\Upsilon}_4$) supported on blue, odd-length boundaries. It also encodes a BT qubit, with logical operators $\bar{X}_B$ (green string) and $\bar{Z}_B$ (red string) indicated. The lattice surgery involves measuring the plaquette operators connecting logical MZMs $\Upsilon_j$ and $\tilde{\Upsilon}_j$, for $j=1,2,3,4$. These plaquette operators are shown lightly colored in (b).}\label{fig:Four_body_mmt}
\end{figure*}

\renewcommand{\arraystretch}{1.5}
\begin{table*}[ht]
\centering
\resizebox{\textwidth}{!}{%
\begin{tabular}{ccccc}
     & \hspace{5pt} Pre-measurement \hspace{5pt} & $i\Upsilon_1\tilde{\Upsilon}_1 $ & $i\Upsilon_2\tilde{\Upsilon}_2$ & $i\Upsilon_3\tilde{\Upsilon}_3$ \\%
     \Xhline{2\arrayrulewidth}
     \begin{tabular}{@{}c@{}}
       Stabilizing\\
       Logical Operators
             \end{tabular}
     & \begin{tabular}{@{}c@{}}$\lbrace \Gamma, \tilde{\Gamma}_4, i\tilde{\Upsilon}_1 \tilde{\Upsilon}_2 \bar{X}_B$\\
     $i\tilde{\Upsilon}_2\tilde{\Upsilon}_4 \bar{Z}_B\rbrace $
     \end{tabular}  
     &\begin{tabular}{@{}c@{}}$\lbrace \lambda_1 i\Upsilon_1\tilde{\Upsilon}_1 ,\Gamma \tilde{\Gamma}_4,$\\
     $i\tilde{\Upsilon}_3\tilde{\Upsilon}_4 \bar{X}_B, i\tilde{\Upsilon}_2\tilde{\Upsilon}_4 \bar{Z}_B\rbrace $ \end{tabular} 
     &\begin{tabular}{@{}c@{}}$\lbrace \lambda_1 i\Upsilon_1\tilde{\Upsilon}_1 , \lambda_2 i\Upsilon_2\tilde{\Upsilon}_2 ,$\\
     $\Gamma \tilde{\Gamma}_4, i\tilde{\Upsilon}_3\tilde{\Upsilon}_4 \bar{X}_B \rbrace $ \end{tabular} 
     &\begin{tabular}{@{}c@{}}$\lbrace \lambda_1 i\Upsilon_1\tilde{\Upsilon}_1 , \lambda_2 i\Upsilon_2\tilde{\Upsilon}_2 ,$\\
     $\lambda_3 i\Upsilon_3\tilde{\Upsilon}_3, \Gamma \tilde{\Gamma}_4 \rbrace $ \end{tabular} \\
     \hline
     & $\Upsilon_1$ 
     & $i\lambda_1 \tilde{\Upsilon}_2\tilde{\Upsilon}_3 \tilde{\Upsilon}_4$ 
     & $-\lambda_1 \tilde{\Upsilon}_3 \bar{Z}_B $ 
     & $\lambda_1 i \tilde{\Upsilon}_4 \bar{X}_B \bar{Z}_B $ \\
     \cline{2-5} 
     Logical MZMs
     & $\Upsilon_2$ 
     & $\Upsilon_2$ 
     & $\lambda_2\tilde{\Upsilon}_4\bar{Z}_B $ 
     & $\lambda_2\tilde{\Upsilon}_4 \bar{Z}_B $ \\
     \cline{2-5} 
     & $\Upsilon_3$ & $\Upsilon_3$ & $\Upsilon_3$ 
     & $\lambda_3 \tilde{\Upsilon}_4 \bar{X}_B$ \\
     \cline{2-5}
     & $\Upsilon_4$ & $\Upsilon_4$ & $\Upsilon_4$ & $\Upsilon_4$ \\
     \hline 
     \hspace{5pt} \begin{tabular}{@{}c@{}}
          Measurement\\
          Operator
     \end{tabular} \hspace{5pt} 
     & $\Gamma_4$ 
     & $-i\lambda_1 \tilde{\Upsilon}_2\tilde{\Upsilon}_3\tilde{\Upsilon}_4 \Upsilon_2\Upsilon_3\Upsilon_4$ 
     & $\lambda_1\lambda_2 \tilde{\Upsilon}_3 \tilde{\Upsilon}_4 \Upsilon_3\Upsilon_4$ 
     & $i\lambda_1\lambda_2\lambda_3 \Upsilon_4 \tilde{\Upsilon}_4$ \\
     \hline
\end{tabular}}
\caption{Operators and their updates after each set of lattice surgery measurements. Top row: the operator measured by lattice surgery. First column: we track updates to the set of logical operators stabilizing the state, the logical MZMs $\Upsilon_1,\ldots,\Upsilon_4$, and measurement operator $\Gamma_4$. The updates to these after measuring  $i\Upsilon_j\tilde{\Upsilon}_j$ are shown in the respective columns. The final measurement $i\Upsilon_4\tilde{\Upsilon}_4$ is not considered, as this, unlike the preceding $i\Upsilon_j\tilde{\Upsilon}_j$, commutes with all stabilizing logical operators.}\label{table:Four_body_mmt_ops}
\end{table*}

\subsubsection{Arbitrary Fermion Parity Measurement}\label{subsec:Arb_Ferm_Parity_Mmt}

We can measure arbitrary products $\Gamma_{2k} = i^{k} \prod_{j=1}^{2k}  \Upsilon_j$ of logical MZMs similarly.
We again treat the case in which all twists hosting the $\Upsilon_j$ are of types $T_\mathsf{B}$ or $T_\mathsf{R}$.
We move the twists to a boundary of color different to all the twists (i.e., green), using code deformation. 
Similarly to above, this can either be an exterior boundary or we can introduce a hole in the center of the lattice.
Now the logical MZMs are supported on $2k$ sections of odd-length boundary.
A $k=2$ example where these boundary sections  lie around a hole is shown in Figure~\ref{fig:Four_body_mmt}.

An ancilla patch of code is then prepared close to the boundary that now hosts all $\Upsilon_j$.
This patch will also host $2k$ logical MZMs labeled $\tilde{\Upsilon}_j$ for $j=1,\ldots, 2k$, supported along the boundary of the patch. 
Each $\tilde{\Upsilon}_j$ is located close to $\Upsilon_j$ on the main lattice.
The ancilla patch will have the $2k$-MZM parity operator $\tilde{\Gamma}_{2k} =i^{k} \prod_{j=1}^{2k}\tilde{\Upsilon}_j$.
We choose these logical MZMs to be supported on blue sections of boundary.
This patch will contain BT qubits as well as logical MZMs. Specifically, it encodes $k-1$ BT qubits.
In Figure~\ref{subfig:Four_body_1} we show the logical operators for the patch in the example considered. 

We prepare the ancilla patch by first preparing all blue bonds within the patch in their $+1$ eigenstates and then measuring all blue plaquette operators.
In the example of Figure~\ref{fig:Four_body_mmt}, there are three independent logical operators for the ancilla patch that are made up only of blue string operators, as seen from representing logical MZM bilinears similarly to Figure~\ref{subfig:Fermionic_logical_ops} (but with the green $\bar{X}_B$ replacing the red loop of Figure~\ref{subfig:Fermionic_logical_ops} for $i\tilde{\Upsilon}_1 \tilde{\Upsilon}_2$ and $i\tilde{\Upsilon}_3\tilde{\Upsilon}_4$): $i\tilde{\Upsilon}_1 \tilde{\Upsilon}_2 \, \bar{X}_B$, $i\tilde{\Upsilon}_2 \tilde{\Upsilon}_4 \,\bar{Z}_B$, and $-\tilde{\Upsilon}_1 \tilde{\Upsilon}_2 \tilde{\Upsilon}_3 \tilde{\Upsilon}_4 = \tilde{\Gamma}_4$.
After preparation, the system will be in the $+1$ eigenstate of all three of these operators.

We now measure $i\Upsilon_j\tilde{\Upsilon}_j$, obtaining outcome $\lambda_j$, for all $j= 1,\ldots, 2k$, using lattice surgery.
This involves measuring plaquette operators that bridge the gaps between logical MZMs $\Upsilon_j$ and $\tilde{\Upsilon}_j$. 
The product of these plaquette operators is equivalent to the operator $i\Upsilon_j\tilde{\Upsilon}_j$.
Examples of these plaquette operators are shown in Figure~\ref{subfig:Four_body_2}.
Notice that, in this example, the product of blue plaquette operators connecting $\Upsilon_1$ and $\tilde{\Upsilon}_1$ give the outcome for $i\Upsilon_1\tilde{\Upsilon}_1$, since the green plaquette operators between these two boundaries are simply products of original four-body and two-body stabilizer generators.
Similar situations apply to the measurement of $i\Upsilon_2\tilde{\Upsilon}_2$, $i\Upsilon_3\tilde{\Upsilon}_3$ and $i\Upsilon_4\tilde{\Upsilon}_4$.

To explain the effect of these $i\Upsilon_j\tilde{\Upsilon}_j$ measurements, we consider the example of Figure~\ref{fig:Four_body_mmt} more closely.
Here we will use $\Gamma$ to denote the total fermion parity operator of the main lattice (containing $\Upsilon_1,\ldots, \Upsilon_4$).
We can assume the system is initially in a $\Gamma = +1$ state (this just sets the parity sector for the logical MZMs).
Before any lattice surgery measurements, there exist four independent logical operators that stabilize the full system (i.e., the system is in a $+1$ eigenstate of all these operators): $\Gamma$, $\tilde{\Gamma}_4$, $i\tilde{\Upsilon}_1\tilde{\Upsilon}_2\, \bar{X}_B $ and $i\tilde{\Upsilon}_2\tilde{\Upsilon}_4\, \bar{Z}_B$.
The first measurement operator, $i\Upsilon_1\tilde{\Upsilon}_1$, anti-commutes with $\Gamma$, $\tilde{\Gamma}_4$ and $i\tilde{\Upsilon}_1\tilde{\Upsilon}_2\, \bar{X}_B $, and hence only products of pairs of these three operators continue to stabilize the state of the system.
The full list of such (independent) stabilizing logical operators is: $\lambda_1 i\Upsilon_1\tilde{\Upsilon}_1$, $\Gamma\tilde{\Gamma}_4$, $i\tilde{\Upsilon}_2 \tilde{\Upsilon}_4\, \bar{Z}_B$ and $i\tilde{\Upsilon}_1 \tilde{\Upsilon}_2\, \bar{X}_B \tilde{\Gamma}_4 = i\tilde{\Upsilon}_3\tilde{\Upsilon}_4\, \bar{X}_B$.
Similarly to Section~\ref{subsec:Bilinear_Mmt}, we may replace the projector associated with the logical measurement with a logical braid: $\exp(\frac{\pi}{4}\lambda_1 i\Upsilon_1\tilde{\Upsilon}_1 \tilde{\Gamma}_4) = \exp(-\frac{\pi}{4}\lambda_1 i\Upsilon_1 \tilde{\Upsilon}_2\tilde{\Upsilon}_3\tilde{\Upsilon}_4)$.
This produces the correct post-measurement state.
This operator can be thought of as a braid between $\Upsilon_1$ and $i\lambda_1\tilde{\Upsilon}_2 \tilde{\Upsilon}_3\tilde{\Upsilon}_4$: it maps $\Upsilon_1 \mapsto i\lambda_1\tilde{\Upsilon}_2\tilde{\Upsilon}_3\tilde{\Upsilon}_4 $.
Thus the operator $\Gamma_4$ we wish to measure is mapped to operator $-i\lambda_1 \tilde{\Upsilon}_2\tilde{\Upsilon}_3\tilde{\Upsilon}_4 \Upsilon_2\Upsilon_3\Upsilon_4$.

The subsequent measurements of $i\Upsilon_j\tilde{\Upsilon}_j$ (for $j=2,3,4$, each with outcome $\lambda_j$) commute with this and, hence, the lattice surgery measurements can be seen to result in the measurement of $\Gamma_4$: it has outcome $\lambda_1\lambda_2\lambda_3\lambda_4$.
This is explained in detail in Table~\ref{table:Four_body_mmt_ops} by tracking updates to the stabilising logical operators, logical MZMs and $\Gamma_4$ after each measurement.

$\lambda_1\lambda_2\lambda_3\lambda_4$ is also the eigenvalue of the operator $\Gamma_4 \tilde{\Gamma}_4 = \prod_{j=1}^4 i\Upsilon_j\tilde{\Upsilon}_j$. 
To disentangle ancilla and logical degrees of freedom, we measure all blue bond operators on the ancilla patch, from which we can infer the eigenvalues of $\tilde{\Gamma}_4$ (call it $\eta_1$), $i\tilde{\Upsilon}_1\tilde{\Upsilon}_2\, \bar{X}_B$ (call it $\eta_2$) and $i\tilde{\Upsilon}_2\tilde{\Upsilon}_4\, \bar{Z}_B$ (call it $\eta_3$). 
These operators do not commute with the operators $i\Upsilon_j\tilde{\Upsilon}_j$ but they do commute with  $\Gamma_4\tilde{\Gamma}_4$. 
Hence from the previous measurement outcomes and the measured eigenvalue of $\tilde{\Gamma}_4$, we can find the $\Gamma_4$ eigenvalue to be $\eta_1 \lambda_1\lambda_2\lambda_3\lambda_4$. 
If $\eta_1 = -1$, we need to apply a correction to map the state to an eigenstate of $\Gamma_4$ with the measured eigenvalue $\lambda_1\lambda_2\lambda_3\lambda_4$. 
This correction operation can simply be kept track of classically.
There are multiple such possible correction operations owing to the degeneracy of the $\Gamma_4 = \pm 1 $ eigenspaces. 
To determine which we should apply, we need to determine if all logical bilinears ($i\Upsilon_1\Upsilon_2$, $i\Upsilon_3\Upsilon_4$, etc.) have been correctly preserved by the entire series of measurements above.
We discuss the details of this in Appendix~\ref{app:Correction_Operation}.

Note that, similarly to the case of Section~\ref{subsec:Bilinear_Mmt}, this measurement procedure does not directly affect the bosonic information stored between the twists, as all bosonic logical operators can be deformed away from the locations of the lattice surgery measurements. 
The introduction of the hole again reduces the distance of some logical qubits, but this distance can be preserved by further separating the twists in the code.
Hence all encoded information can remain topologically protected throughout the entire procedure.

\subsection{Magic State Injection}\label{subsec:Magic_State}

Combining the above operations with the injection of magic states into MSC patches allows one to perform universal quantum computation fault-tolerantly with both bosonic logical qubits and logical MZMs. 
Here we detail how to achieve magic state injection in two ways: the first is within an existing patch of code, while the second is in a separate patch of code (e.g., in an ancilla register).

Several noisy copies of this magic state can be purified to a single copy using Clifford gates and measurements~\cite{Magic_State_Dist}.
``Magic state gadgets," also composed of Clifford gates and measurements~\cite{nielsen_chuang}, can be used to apply $T$ gates on BT or FT qubits in the code, using these purified magic states as a resource.

\begin{figure*}
\centering
\subfloat[\label{subfig:On_Patch_Inj}]{\includegraphics[width=0.4\linewidth]{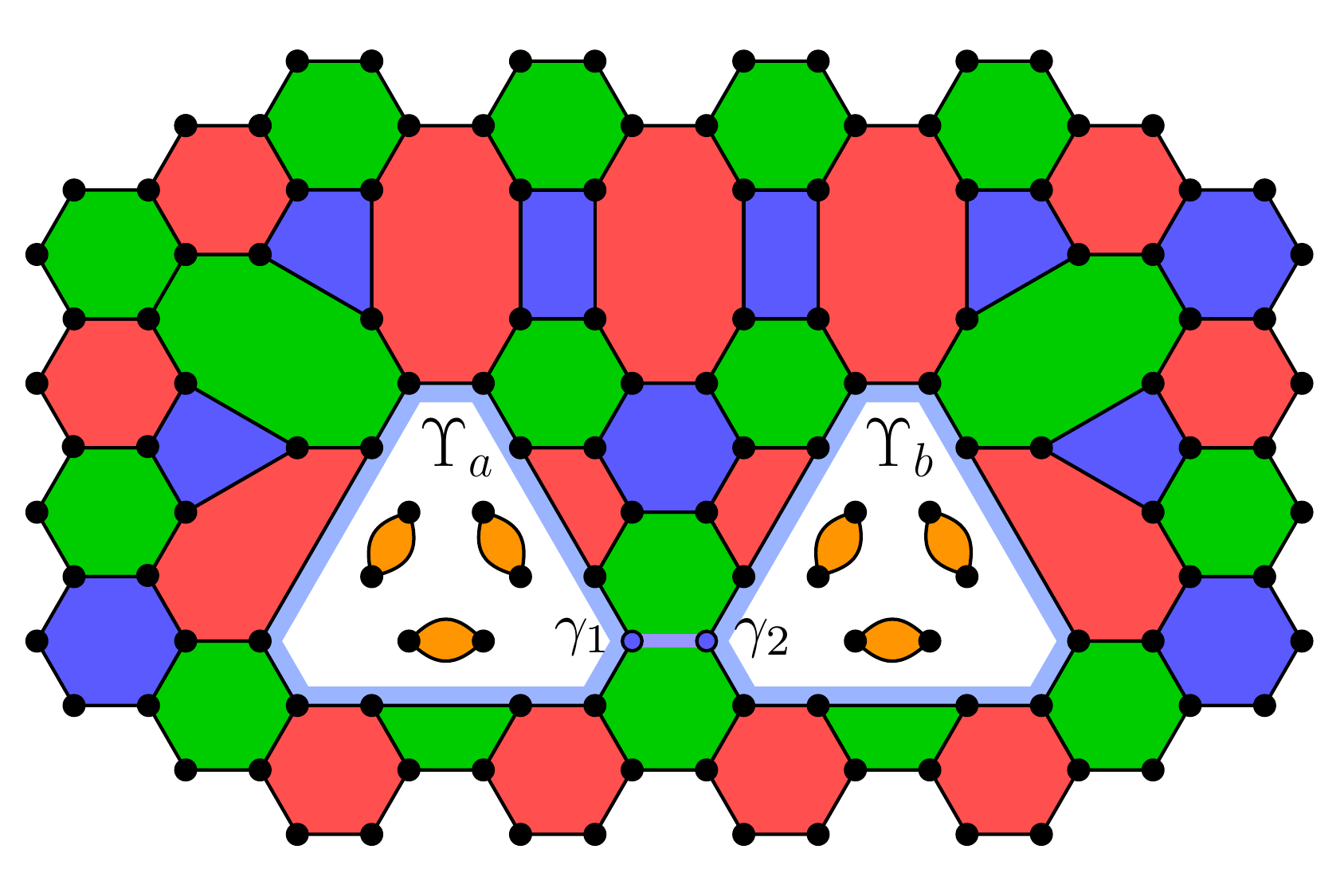}}\hspace{20pt}
\subfloat[\label{subfig:Off_Patch_Inj}]{\includegraphics[width=0.5\linewidth]{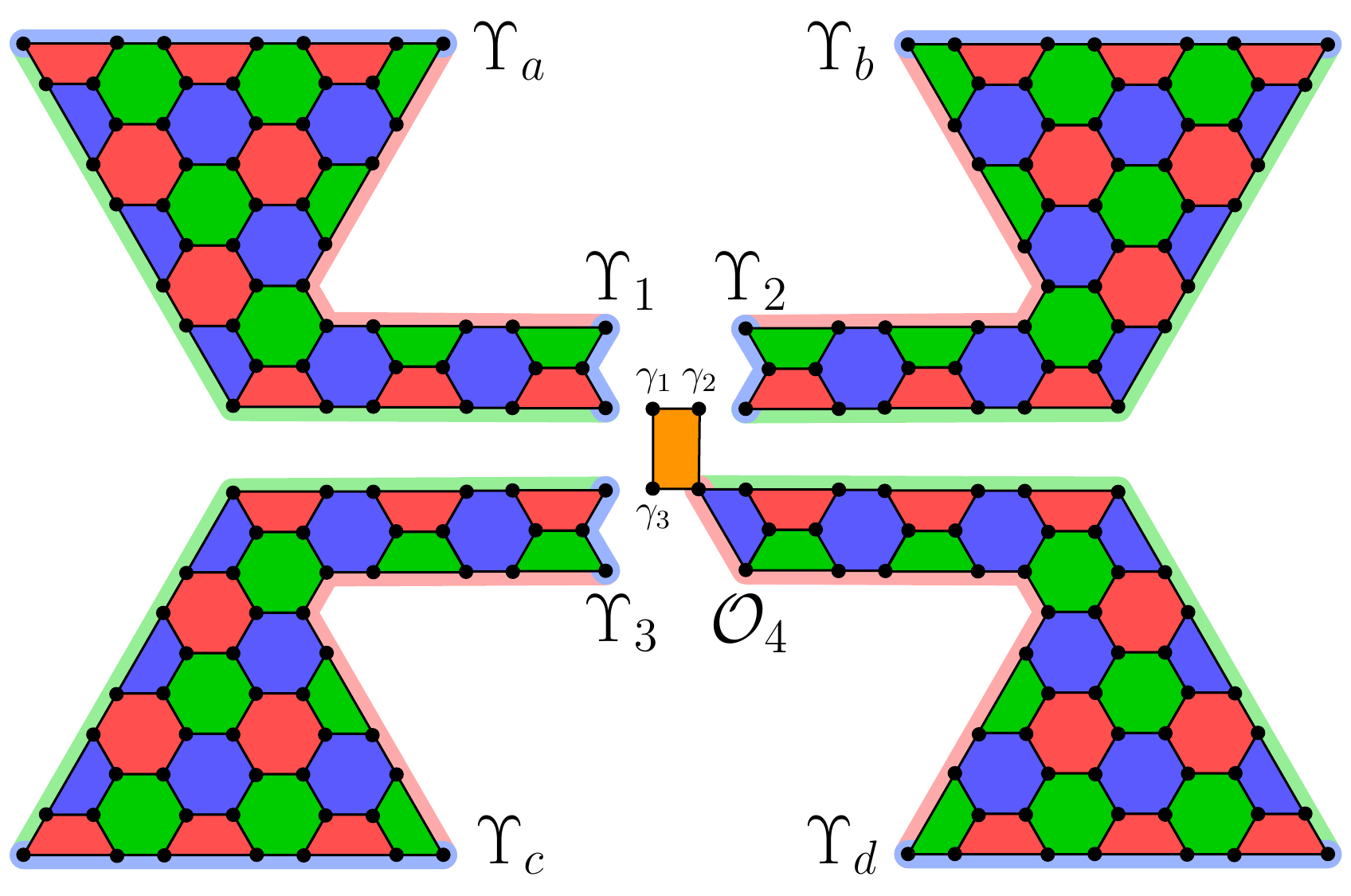}}
\caption{Code configurations for magic state injection. (a) On-patch injection. Two fermionic $T_\mathsf{B}$ twists are at minimal separation, with another two twists located far away (not shown). A magic state can be injected by preparing a suitable initial state and applying the gate $\exp(\frac{\pi}{8}\gamma_1\gamma_2)$. Logical MZMs $\Upsilon_a$ and $\Upsilon_b$ reside on the blue boundaries shown. The blue bond operator $i\gamma_1\gamma_2$ is the logical $i\Upsilon_a\Upsilon_b \bar{Z}_B$, acting on both FT and BT qubits stored between the four twists. (b) Off-patch injection. Four MSC patches are prepared, each with two logical MZMs along blue boundaries (e.g., $\Upsilon_a$ and $\Upsilon_1$ indicated).
A qubit with Majorana operators $\gamma_1,\gamma_2,\gamma_3,\gamma_4$ (orange rectangle) is prepared in a magic state. Plaquette operators $\mathcal{O}_j = i\Upsilon_j\gamma_j$ (for $j=1,2,3,4$, with $\mathcal{O}_4$  shown) are measured to inject this magic state into the four-patch qubit furnished by $\Upsilon_a$, $\Upsilon_b$, $\Upsilon_c$, $\Upsilon_d$.}\label{fig:State_Injection}
\end{figure*}

\subsubsection{On-Patch Magic State Injection}

Figure~\ref{subfig:On_Patch_Inj} shows two fermionic twists (hosting logical MZMs $\Upsilon_a$, $\Upsilon_b$) at minimum separation. Suppose there exist another two twists at large separation from this pair (hosting logical MZMs $\Upsilon_c$, $\Upsilon_d$).
For this setup, the logical operator $i\gamma_1\gamma_2$, shown in the figure, equals $i\Upsilon_a\Upsilon_b\bar{Z}_B$, where $\bar{Z}_B$ is a logical Pauli operator (green string loop encircling the two logical MZMs, not shown) for the BT qubit stored between the four twists.
Hence, $i\gamma_1\gamma_2$ acts on both FT and BT qubits stored in the twist.
To prepare a magic state, we first prepare the state $\ket{+}_F\ket{0}_B$ (cf. Sections~\ref{subsec:Logical_qubit_braids} and~\ref{subsec:Bilinear_Mmt}), where $\ket{0}_B = \bar{Z}_B\ket{0}_B$ is a computational basis state of the BT qubit and $\ket{+}_F = i\Upsilon_a\Upsilon_c\ket{+}_F$ is a basis state of the FT qubit stored between $\Upsilon_a,\Upsilon_b,\Upsilon_c,\Upsilon_d$.
We then perform the rotation $\exp{(\frac{\pi}{8}\gamma_1\gamma_2)}$; we assume we can perform at least a noisy version of this gate using established methods~\cite{Karzig_geom_magic_2016,Karzig_geom_magic_meas2019,Karzig_Majorana2017}. 
This approximately yields
\begin{align}
    \frac{1}{\sqrt{2}}\left(\ket{0}_F + e^{i\pi/4}\ket{1}_F\right)\ket{0}_B
\end{align}
where $\ket{0}_F$ and $\ket{1}_F$ are the $+1$ and $-1$ eigenstates of operator $i\Upsilon_a\Upsilon_b$, respectively.

Throughout this process, twists $a$ and $b$ are far-separated from twists $c$ and $d$, and hence $\bar{X}_B$ and $i\Upsilon_a\Upsilon_c$ logical errors are suppressed by this large separation, and the weights of the logical MZMs, which can be made arbitrarily large.
So, provided error rates are below the error threshold, the BT qubit remains in the $\ket{0}_B$ state throughout the process with arbitrarily high probability, while the FT qubit is prepared in a noisy magic state.
Alternatively, we could prepare the BT qubit in a magic state by instead first preparing the state $\ket{0}_F \ket{+}_B$. 
The same gate as above then approximately prepares the state:
\begin{align}
    \frac{1}{\sqrt{2}}\ket{0}_F \left( \ket{0}_B + e^{i\pi/4}\ket{1}_B\right).
\end{align}
We can also prepare magic states in both FT and BT qubits by enacting the following operators on the state $\ket{+}_F\ket{+}_B$:
\begin{align}
    \text{CX}_{F,B} \exp{\left(\frac{\pi}{8}\Upsilon_a\Upsilon_b \bar{Z}_B\right)} \text{CX}_{F,B} = \exp{\left(-i\frac{\pi}{8}\bar{Z}_B\right)},\\
    \text{CX}_{B,F} \exp{\left(\frac{\pi}{8}\Upsilon_a\Upsilon_b \bar{Z}_B\right)} \text{CX}_{B,F} = \exp{\left(\frac{\pi}{8}\Upsilon_a\Upsilon_b\right)},
\end{align}
where $\text{CX}_{J,K}$ is the CNOT gate controlled on qubit $J$ and targeted on qubit $K$.

\subsubsection{Off-Patch Magic State Injection}\label{subsec:Off-Patch_Magic_state_inj}

In Figure~\ref{subfig:Off_Patch_Inj} we illustrate injecting magic states into 
separate MSC patches via measurements.
Four MSC patches are prepared, each with two logical MZMs. 
On each patch, one of these logical MZMs has low weight  and the other has high weight.
The low-weight logical MZMs are labeled $\Upsilon_1$, $\Upsilon_2$, $\Upsilon_3$, and $\Upsilon_4$ (cf. the weight-three logical MZMs in Figure~\ref{subfig:Off_Patch_Inj}), and the large-weight logical MZMs are labeled $\Upsilon_a$, $\Upsilon_b$, $\Upsilon_c$, and $\Upsilon_d$, respectively (cf. the weight-seven logical MZMs of Figure~\ref{subfig:Off_Patch_Inj}).
These patches do not contain any BT qubits, owing to their boundary conditions. 
Our goal will be to inject a magic state into the four-patch FT qubit furnished by $\Upsilon_a$, $\Upsilon_b$, $\Upsilon_c$, and $\Upsilon_d$, i.e., using the large-weight MZMs. 

We start by preparing the patches in the $+1$ eigenstates of $i\Upsilon_a\Upsilon_1$, $i\Upsilon_b\Upsilon_2$, $i\Upsilon_c\Upsilon_3$ and $i\Upsilon_d\Upsilon_4$ (cf. Section~\ref{subsec:Fermion_Parity_Mmt}).
We then prepare a patch with four Majorana operators (orange rectangle in Figure~\ref{subfig:Off_Patch_Inj} and labeled $\gamma_1,\gamma_2,\gamma_3,\gamma_4$) in a possibly noisy magic state encoded in subspace stabilized by the patch parity $ -\gamma_1\gamma_2\gamma_3\gamma_4$. 
We inject this magic state into an FT qubit furnished by $\Upsilon_a,\ldots, \Upsilon_d$ by measuring operators $\mathcal{O}_j = i\Upsilon_j\gamma_j$ for $j=1,2,3,4$. 
The example of $\mathcal{O}_4$ is represented as a plaquette operator in Figure~\ref{subfig:Off_Patch_Inj}.
This measurement maps $\gamma_1$ to $\Upsilon_a$ (and similarly for the other measurements),
by the same logic that we have used repeatedly: due to the anticommutation of $\mathcal{O}_1$ and $i\Upsilon_a\Upsilon_1$, the measurement, with outcome $\lambda_1$ effectively implements the logical braid $\exp(\frac{\pi}{4}\lambda_1\Upsilon_a \gamma_1)$ and similarly for $j=2,3,4$.
In this way, the magic state furnished by $\gamma_1$, $\gamma_2$, $\gamma_3$, $\gamma_4$ is transferred to the FT qubit furnished  by logical MZMs $\Upsilon_a,\Upsilon_b,\Upsilon_c,\Upsilon_d$ (specifically in the subspace stabilized by $-\Upsilon_a\Upsilon_b\Upsilon_c\Upsilon_d$), up to a correction operation (if $
\lambda_1 \lambda_2 \lambda_3 \lambda_4 = -1$) which can be tracked. 
The resource cost of this magic state preparation method is compared with alternatives in Section~\ref{sec:Low_Over_Ferm_QC}.

\subsection{Arbitrary String Operator Measurements}\label{subsec:Hybrid_Approaches}

We now discuss measuring various string operators. The following considerations make this particularly interesting.
The twist framework established above allows for not only fermionic and bosonic computation to be done side-by-side, but also for hybrid computing schemes involving both FT and BT qubits. 
To complete the set of universal gates for both types of logical qubit, we require a gate that entangles the two.
This would, for example, allow us to use magic states stored in FT qubits (cf. Section~\ref{subsec:Magic_State}) to apply $T$ gates to BT qubits.
Moreover, the measurement of BT qubit logical operators opens up an additional route for implementing Clifford gates on top of braiding twists, which could result in reduced overheads~\cite{Lattice_surg_low_overhead}.

We can achieve the effect of entangling gates between BT and FT qubits by fault-tolerantly measuring $\Gamma_F P_B$, where $\Gamma_F$ is a logical MZM parity and $P_B$ is a bosonic logical Pauli operator.
For example, take an FT, a BT, and an ancilla twist qubit, with logical operators $\lbrace \bar{Z}_F,\bar{X}_F \rbrace$, $\lbrace \bar{Z}_B,\bar{X}_B \rbrace$ and $\lbrace \bar{Z}_A,\bar{X}_A \rbrace$, respectively.
To implement a CNOT gate controlled on qubit $F$ and targeted on $B$, we can prepare the ancilla qubit in the $\ket{+}_A$ state, then measure $\bar{Z}_F \bar{Z}_A$, then $\bar{X}_A\bar{X}_B$, and finally $\bar{Z}_A$.
This results in the desired gate, up to a logical Pauli operator which can be tracked.
We can avoid introducing these ancilla qubits by instead classically tracking Clifford gates and updating circuit measurements accordingly (see also Section~\ref{sec:Low_Over_Ferm_QC})~\cite{Bravyi_PBC2016,Tracking_Color_Codes,QC_with_MFCs}.

Here, we detail how to measure arbitrary $\Gamma_F P_B$, all of which will be some product of string operators.
We distinguish between two types of string operator, with arbitrary $\Gamma_F P_B$ operators being products of multiple instances of these two types.
Define fermionic string operators as those that terminate on the odd-length boundaries associated with two fermionic twists, and bosonic string operators as simply products of bosonic logical operators (thus they are simply of the form $P_B$).
The latter will either encircle some number of twists or alternatively terminate at a boundary of even length.
Hence only fermionic string operators have a non-trivial action on FT qubits.
We describe the measurement of fermionic string operators first, then move on to bosonic string operators. 
We describe how to measure arbitrary products of these two types in Appendix~\ref{app:Measurement_Details}.

\begin{figure}
\centering
\subfloat[\label{subfig:ZB_ZF_lattice_surg_a}]{\includegraphics[width=0.48\linewidth]{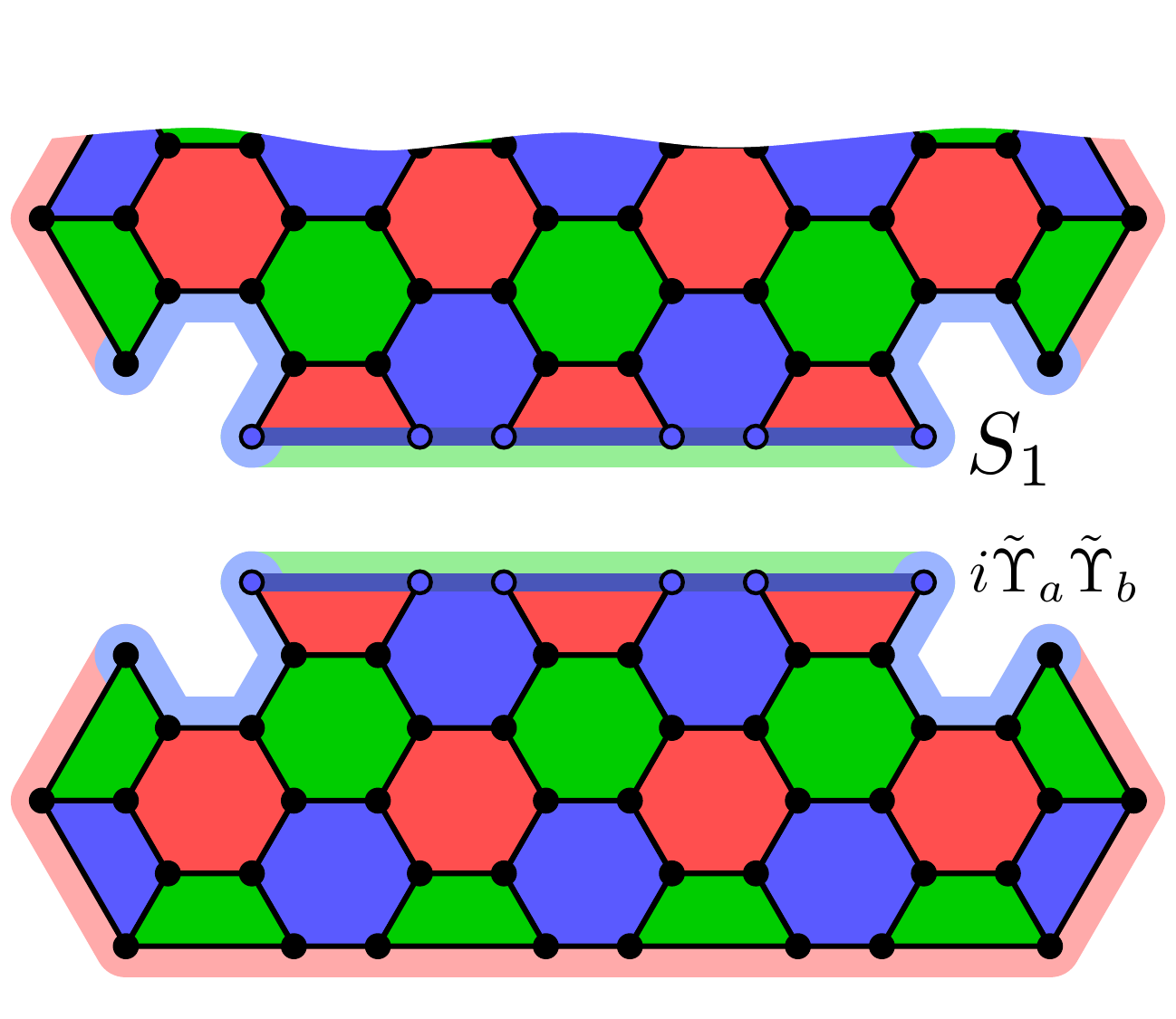}}\hfill 
\subfloat[\label{subfig:ZB_ZF_lattice_surg_b}]{\includegraphics[width=0.48\linewidth]{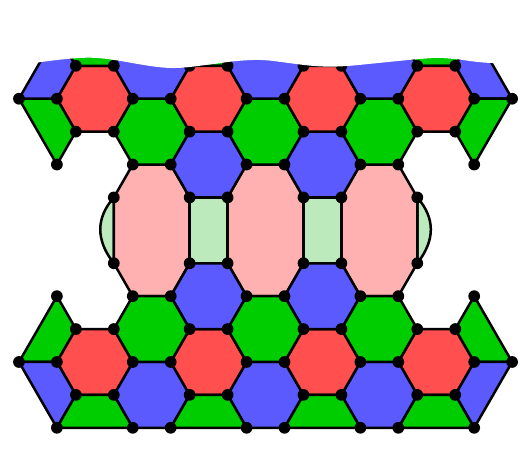}}
\caption{Lattice surgery procedure for measuring fermionic string operator $S_1$. The lower ancilla patch contains logical MZMs $\tilde{\Upsilon}_a$ and $\tilde{\Upsilon}_b$, while the upper patch contains logical information. (a) The setup before lattice surgery, with the ancilla patch prepared in the $i\tilde{\Upsilon}_a\tilde{\Upsilon}_b=+1$ state. Blue, red and green boundaries on both patches are indicated; blue boundaries host logical MZMs.
Also indicated are blue string operators $S_1$ and $i\tilde{\Upsilon}_a\tilde{\Upsilon}_b$
running between blue boundaries on the logical and ancilla patches respectively. (b) Lattice surgery is performed, with extra operators fusing the two patches indicated with lightly colored plaquettes.}\label{fig:Open_String_Parity_Mmts}
\end{figure}

The procedures for measuring string operators use lattice surgery and are very similar to those for measuring fermion parity operators, as described in Section~\ref{subsec:Fermion_Parity_Mmt}.
We begin with the case of a single fermionic string operator, $S_1$.
For concreteness, we assume that $S_1$ terminates on the boundaries of two (fermionic) $T_\mathsf{B}$ twists.
As in Section~\ref{subsec:Bilinear_Mmt}, we move these two twists to a (green- or red-colored) boundary of the lattice, and introduce an ancilla patch of code nearby to this boundary, hosting two logical MZMs $\tilde{\Upsilon}_a$ and $\tilde{\Upsilon}_b$ (hosted on blue boundaries).
Alternatively, a hole can be created in the lattice and the ancilla patch prepared in this hole, as described in previous sections.
An example of the resulting setup is shown in Figure~\ref{subfig:ZB_ZF_lattice_surg_a}.
The ancilla patch is initialized in the $+1$ eigenstate of $i\tilde{\Upsilon}_a\tilde{\Upsilon}_b$ (cf. Section~\ref{subsec:Bilinear_Mmt}), which is a blue string operator.
(The red string operator part of $i\tilde{\Upsilon}_a\tilde{\Upsilon}_b$, akin to that in Figure~\ref{subfig:Fermionic_logical_ops}, would now connect the red boundaries and hence can be shrunk: it is a stabilizer element.)

To measure the operator $S_1$, we perform lattice surgery between the logical patch and the ancilla patch. 
For the example of Figure~\ref{fig:Open_String_Parity_Mmts}, we perform measurements of the green plaquette operators bridging the gap between the two patches. 
Notice that the product of these green plaquettes is equal to $\pm S_1(i\tilde{\Upsilon}_a\tilde{\Upsilon}_b)$.
This procedure differs from that of Section~\ref{subsec:Bilinear_Mmt} since there we only measure two plaquette operators bridging the gap between the patches.
Those plaquette operator measurements fused pairs of logical MZMs on the ancilla and code patches. 
In the present case we fuse the patches along the two string operators shown in Figure~\ref{fig:Open_String_Parity_Mmts} resulting in different logical information being extracted.
Call the product of all measurement outcomes of these green plaquette operators $\lambda$.
Similarly to the cases examined in Section~\ref{subsec:Fermion_Parity_Mmt}, $\lambda$ is the measurement outcome for $S_1$ that we desire (up to a known sign defined by the ordering of the operators).
To decouple the ancilla patch, we then destructively measure $i\tilde{\Upsilon}_a\tilde{\Upsilon}_b$ (cf. Section~\ref{subsec:Bilinear_Mmt}).
This produces the result $i\tilde{\Upsilon}_a\tilde{\Upsilon}_b$ with certainty (assuming perfect measurements) since we can find a representative of this logical operator that commutes with all lattice surgery measurements.
Hence we need not track any correction operation, as we had to do in previously-described lattice surgery procedures.

\begin{figure}
    \centering
    \subfloat[\label{subfig:Bosonic_string_mmt_1}]{\includegraphics[width=0.48\linewidth]{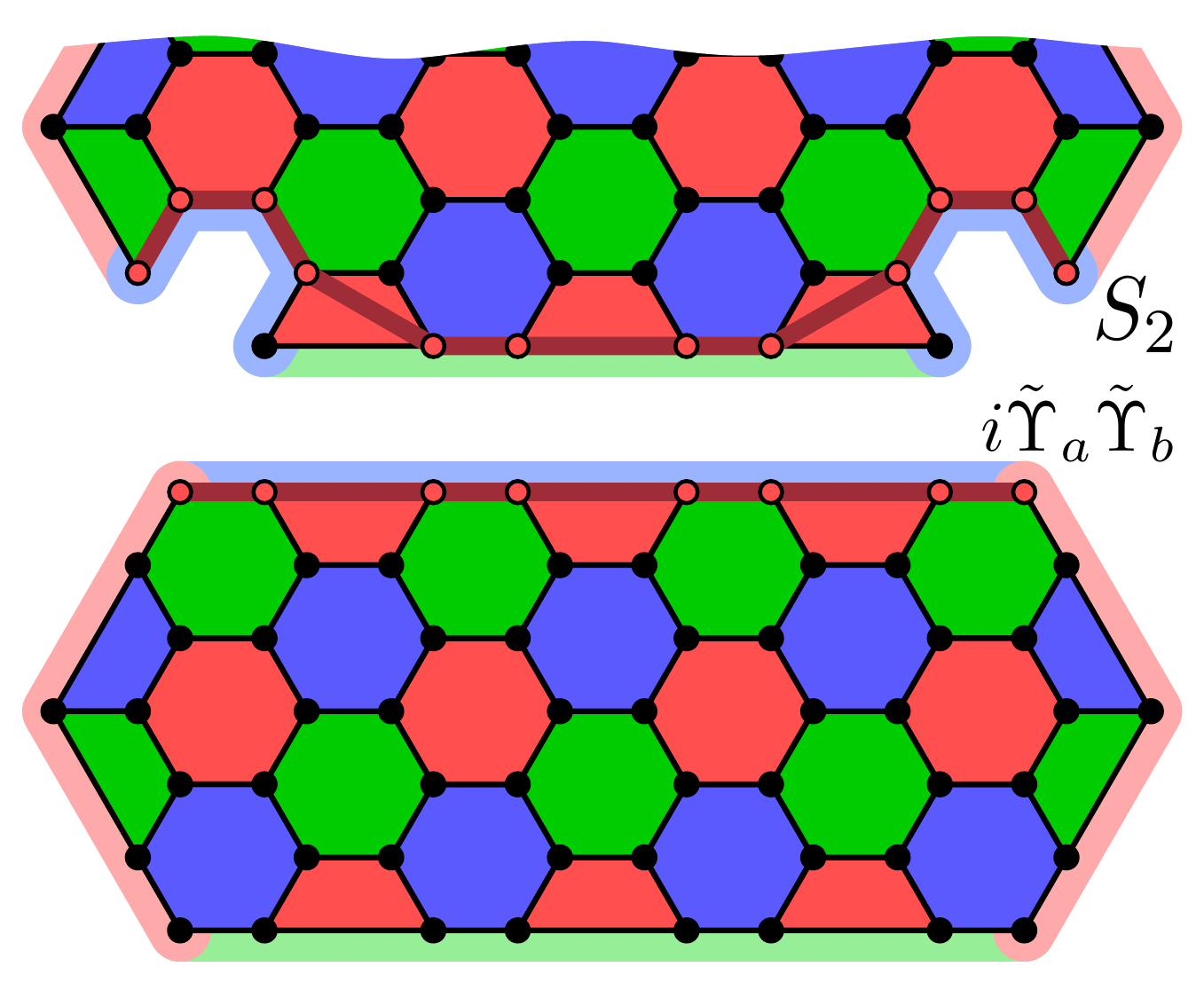}}\hfill 
    \subfloat[\label{subfig:Bosonic_string_mmt_2}]{\includegraphics[width=0.48\linewidth]{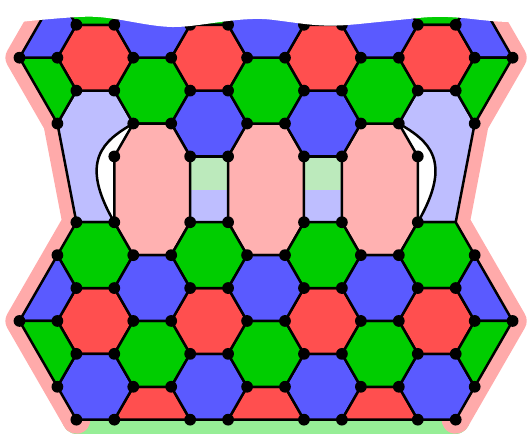}}
    \caption{Lattice surgery procedure for measuring a bosonic string operator. 
    Lattices before being fused are shown in (a):  $S_2$ (red) is the bosonic string to be measured; the ancilla patch has logical MZMs $\tilde{\Upsilon}_a$ and $\tilde{\Upsilon}_b$ (red boundaries) with $i\tilde{\Upsilon}_a\tilde{\Upsilon}_b$ realized by the red string connecting them.
     The lattice after fusion is shown in (b), with the extra plaquettes connecting the two patches shown lightly colored.}
    \label{fig:bosonic_string_op_mmt}
\end{figure}

We now explain the measurement of a bosonic string operator using lattice surgery. Without loss of generality we assume this is a red string operator, which we call $S_2$.
If $S_2$ encircles some twists, we move these to one of the boundaries of the lattice, as before. 
Thus $S_2$ now terminates on even boundaries.
An example involving two twists is shown in Figure~\ref{subfig:Bosonic_string_mmt_1}. 
The lattice hosting $S_2$ in this figure is the same as that hosting $S_1$ in Figure~\ref{subfig:ZB_ZF_lattice_surg_a}.
While the ancilla patch looks slightly different, it still only hosts two logical Majoranas $\tilde{\Upsilon}_a$ and $\tilde{\Upsilon}_b$.
The patch is prepared in the $+1$ eigenstate of $i\tilde{\Upsilon}_a\tilde{\Upsilon}_b$, a red string operator shown in Figure~\ref{subfig:Bosonic_string_mmt_1}.
Measuring the blue and green/blue plaquettes that fuse the two patches (shown lightly shaded in Figure~\ref{subfig:Bosonic_string_mmt_2}) measures the operator $\pm S_2 i\tilde{\Upsilon}_a\tilde{\Upsilon}_b$.
These measurements commute with the fermionic logical operators on the upper patch and, moreover, do not reveal any information about them. 
Finally, the ancilla patch is destructively measured leaving the upper patch in the measured eigenstate of $S_2$. 

In Appendix~\ref{app:Measurement_Details}, we extend these procedures to the measurement of arbitrary products of bosonic and fermionic string operators.
Thus, any logical operator in the code can be measured using lattice surgery.
We can therefore achieve universal quantum computation with BT and FT qubits side-by-side, by combining these measurements with the procedures of Sections~\ref{subsec:Logical_qubit_braids}-\ref{subsec:Magic_State}.

\begin{figure*}
\captionsetup[subfigure]{
  position=top,
  captionskip=1pt,
  singlelinecheck=false,
  margin={-17cm,0cm},
  font=Large
}
\centering
\subfloat[\label{subfig:Large_Parity_Mmt_FPBC_Example}]{\includegraphics[width = 0.95\linewidth]{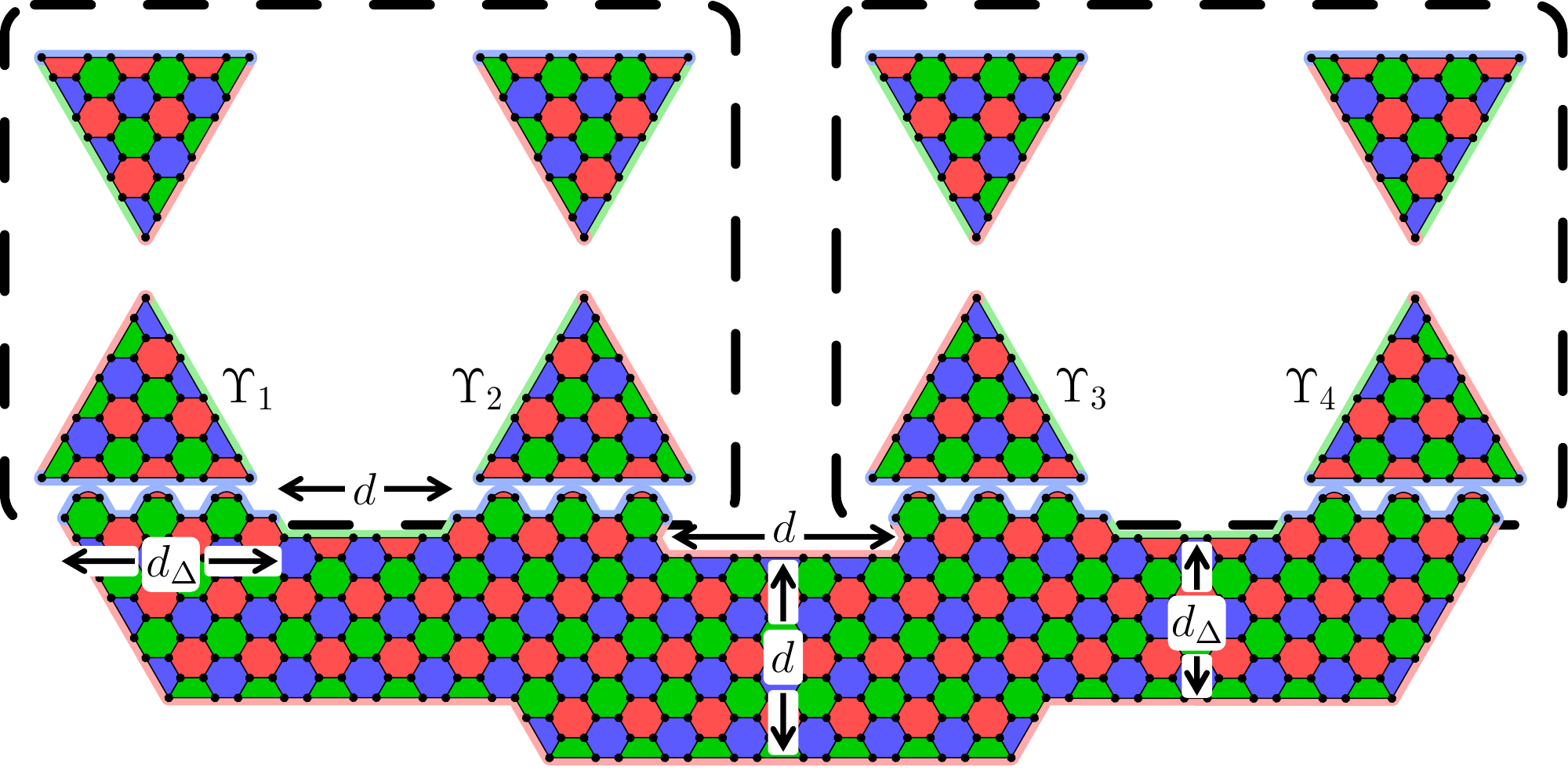}}\\
\subfloat[\label{subfig:Ancilla_Patch_Long_range_mmt}]{\includegraphics[width=0.8\linewidth]{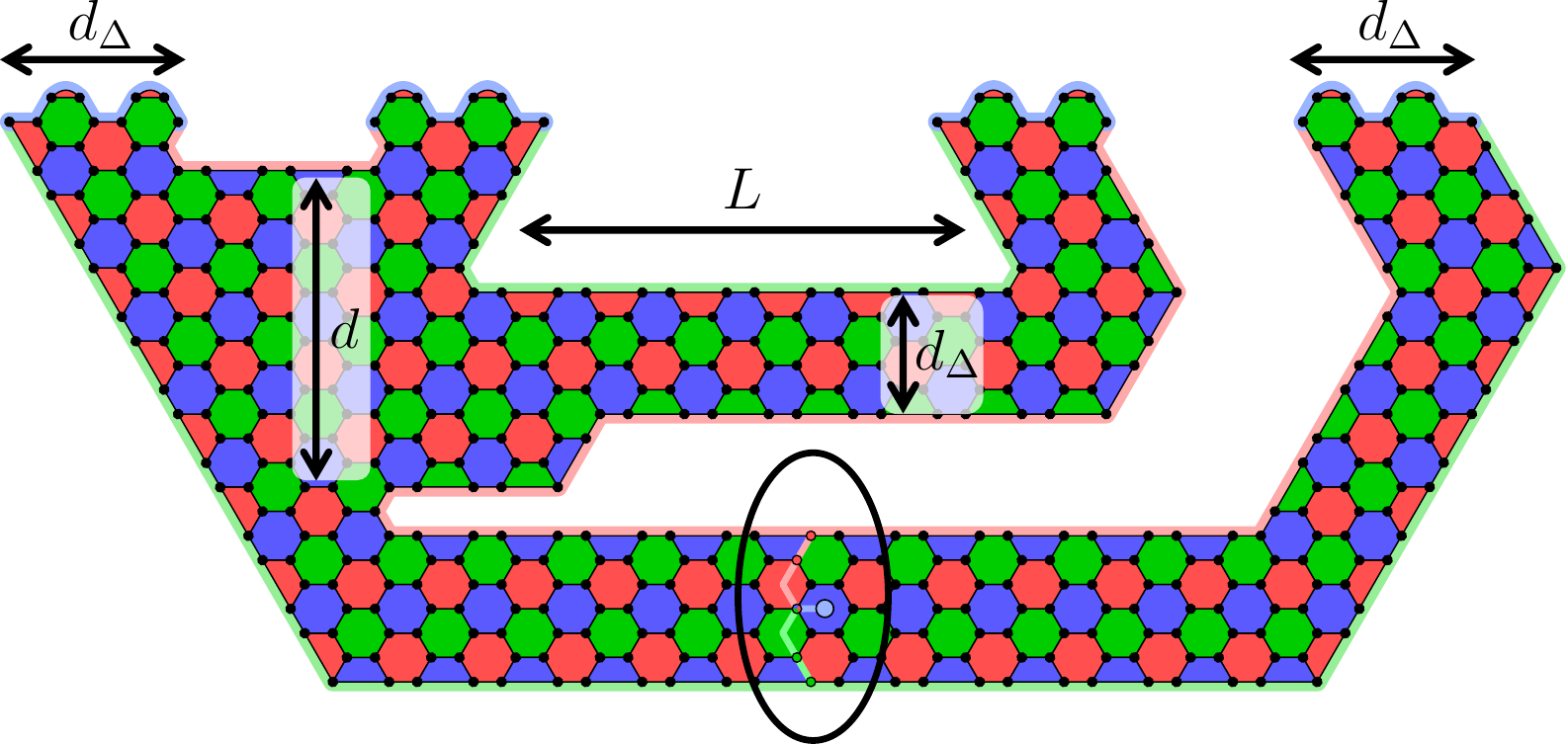}}
\caption{
(a) Setup for measuring a two-qubit logical Pauli operator, with logical qubits encoded 
in Majorana triangle code quartets (in dashed boxes). The operator being measured is $\Upsilon_1\Upsilon_2\Upsilon_3\Upsilon_4$, with logical MZMs on triangle codes indicated. An ancilla patch hosts logical MZMs on blue boundaries of weight $d_\Delta$, where $d_\Delta$ is the minimum weight of logical MZMs on triangle code patches. This patch has alternating thickness $\sim d_\Delta$ and $\sim d$ as shown, to protect from the error mechanism from (b) and FPC errors respectively. The separation of logical MZMs on the patch is everywhere $\sim d$.
(b) Ancilla patch altered such that $O(d_\Delta)$-thickness sections run between far-separated logical qubits, located distance $L$ apart. The width of the patch is $\sim d + d_\Delta$ and $\sim 2 d_\Delta$ in alternating sections (in this example, $d=10$, $d_\Delta = 5$). A potentially undetectable
error string is shown, circled. A parity-odd string flips the blue plaquette indicated. Preparation of this patch involves preparing eigenstates of blue bond operators, followed by the measurement of blue plaquettes (with random outcomes). Plaquette measurement errors can result in this flipped plaquette not being detected.}\label{fig:Large_FPBC_Mmt}
\end{figure*}

\section{Overheads for Universal Fault-Tolerant Computation}\label{sec:Low_Over_Ferm_QC}

\bgroup
\def\arraystretch{1.5}%
\begin{table*}
    \centering
    \begin{tabular}{cc|c|c}
         & \textbf{Qubit Storage} & \textbf{State Preparation} & \textbf{Long-range Measurement}\\
        \hline
        \textbf{All bosonic} & $\frac{3}{2}d^2 - 2d + 2$ & $0$
        & $dL + O(d^2)$\\
        \hline
        \begin{tabular}{@{}c@{}}\textbf{Fermionic qubits and} \\ \textbf{bosonic ancillas}\end{tabular}  & $3 d_\Delta^2 + 1$ & $O(d^2)$ 
        & \text{N/A}\\
        \hline
        \textbf{All fermionic} & $3 d_\Delta^2 + 1$ & $O(1)$ 
        & $2 d_\Delta L + O(d^2)$,\\
        \hline
    \end{tabular}
    \caption{Number
    of Majoranas required for Pauli-based computing with various schemes, utilizing either bosonic or fermionic aspects of the MSC. $d$ is the code distance chosen for bosonic patches, while $d_\Delta \lesssim d$, is half the distance of the triangle code patches (i.e., $d_\Delta$ is the minimum weight of the logical Majoranas hosted therein), and the minimum thickness of ancilla patches.
    The chosen $d_\Delta$ will depend on the QPP probability. %
    We compare the requirements
    for storing a logical qubit in a 6.6.6 lattice (1$^\text{st}$ column), and the
    ancilla costs of preparing a magic state (2$^\text{nd}$ column) and measuring a long-range multi-qubit logical Pauli operator with an ancilla patch of length roughly $L$ (3$^\text{rd}$ column). 1$^\text{st}$ row: Majorana requirements for storing magic states in BT qubits and using the same ``bosonic" MSC patches (with no logical MZMs) for logical Pauli operator measurements~\cite{QC_with_MFCs}. For state preparation, only those MZMs in the storage patch are required and hence the ancilla cost is zero~\cite{QC_with_MFCs}. 2$^\text{nd}$ row: requirements for storing qubits in Majorana triangle codes and using bosonic ancilla patches where possible~\cite{Maj_Triangle_Code2018}. 3$^\text{rd}$ row: requirements for storing qubits in Majorana triangle codes and using ancilla patches that contain logical MZMs. %
    }
    \label{tab:Majorana_overhead}
\end{table*}

Universal quantum computation can be achieved via the measurement of arbitrary Pauli operators on (distilled) magic states~\cite{Bravyi_PBC2016,Mithuna_Magic_State_PBC2019}. 
Universal fermionic quantum computation, which is more suitable for fermionic simulation, can be performed in a similar manner~\cite{FPBC}. 
Here, we consider the resource costs for performing such computations using the MSC, in setups in which QPP is suitably rare.
We consider encoding logical MZMs in ``Majorana triangle codes"~\cite{Maj_Triangle_Code2018} and performing Pauli measurements on magic states encoded using these. %
That is, we perform ``Pauli-based computation" with the MSC, taking full advantage of the Gottesman-Knill theorem to reduce the quantum resources required.
Our method is able to find potential avenues for increased resource savings by exploiting the fermionic aspects of the MSC. 

A Majorana triangle code is a patch of MSC encoding a single logical MZM.
It has three odd-length boundaries of different colors (see Figure~\ref{subfig:Large_Parity_Mmt_FPBC_Example}).
We begin by preparing $n$ magic states in $4n$ Majorana triangle codes, using for each quartet the ``off-patch'' method outlined in Section~\ref{subsec:Off-Patch_Magic_state_inj}, followed by additional measurements
to sever the ``tails" in Figure~\ref{subfig:Off_Patch_Inj}.
(We assume having high-fidelity magic states, otherwise we can employ magic state distillation to improve the magic-state quality.)
Let the distance of this encoding be $2d_\Delta$; that is, the minimum weights of the logical MZMs hosted on the four triangle codes are all $d_\Delta$.
Meanwhile, let the distance of any bosonic MSC patch be $d$.
We will assume we are in a parameter regime in which $d_\Delta < d$ and discuss the possibility of this below.

Our state-injection procedure 
takes only a $d_\Delta$-independent time
(that of the measurements) to increase the distance of the logical qubit from $2$ to $2d_\Delta$.
However, the code patches for magic state injection (cf. Figure~\ref{subfig:Off_Patch_Inj}) include low-weight logical MZMs, along with thin-width patches of code vulnerable to the type of error process described below.
These two features, respectively, act in favor of and against the fidelity of the injected state.
The overall magic state preparation process, however, is fault-tolerant because of magic state distillation used.

Note that the ``tails'' of the initial state-injection patches (Figure~\ref{subfig:Off_Patch_Inj}) may be maintained at a constant length, independent of the code distance.
Increasing the tail length beyond a certain length simply increases the diameter of operators $i\Upsilon_a\Upsilon_1$, $i\Upsilon_b\Upsilon_2$, etc., not the minimum weight $d_\Delta+3$ of these operators.
While this decreases the probability of parity-conserving processes resulting in logical errors such as $i\Upsilon_a\Upsilon_1$, these logical ``errors" act as $+1$ on the initial state of the patch and hence are inoperative. %

Hence, since we use tails of constant width and length, the total number of constituent Majoranas involved in the preparation of a single (potentially noisy) magic state is $4N_\Delta(d_\Delta) + O(1)$, where $N_\Delta(d_\Delta)$ is the number of Majoranas in a single side-length-$d_\Delta$ triangle code.
After the magic states are prepared, the tails can be deleted, resulting simply in four triangle codes (see Figure~\ref{subfig:Large_Parity_Mmt_FPBC_Example}).
For triangle codes based on hexagonal, 6.6.6 lattices (i.e., those with red, green and blue plaquettes having support on 6 Majoranas each) $N_\Delta(d_\Delta) = \frac{3}{4}d_\Delta^2 + \frac{1}{4}$.
Meanwhile, for triangle codes based on 4.8.8 lattices (i.e., those with square and octagonal plaquettes~\cite{QC_with_MFCs}) $N_\Delta (d_\Delta) = \frac{1}{2}(d_\Delta^2 + 2d_\Delta -1)$.
We will continue to analyse the resources of 6.6.6 codes, but we can obtain lower asymptotic overheads with 4.8.8 codes, at the cost of higher-weight stabilizer measurements.
The Majorana overhead for qubit storage using triangle codes and bosonic MSC patches are presented in Table~\ref{tab:Majorana_overhead}.

Previous work demonstrated computation with Majorana triangle codes but utilized only bosonic aspects of the MSC for the ancilla patches containing magic states~\cite{Maj_Triangle_Code2018}.
This yielded a scheme where
magic state injection required an $O(d^2)$ number of ancilla Majoranas in addition to four triangle codes: %
The ancilla cost is at least that
of a distance-$d$ patch (e.g., that of Figure~\ref{fig:Patch}).
This compares to the $O(1)$ number of ancilla Majoranas [namely the 4 MZMs used to prepare the noisy state for injection, and the $O(1)$-length tails] 
required in the off-patch method from Section~\ref{subsec:Off-Patch_Magic_state_inj}. 
Thus, for large $d$, utilizing fermionic aspects of the MSC in magic state preparation and storage becomes more efficient.
(Note that it is not possible to use bosonic ancilla patches for measurement as we require odd-length boundaries for lattice surgery with triangle codes; see Table~\ref{tab:Majorana_overhead}.)

We will now consider the overheads for information storage and ancilla patches more closely. 
To make our estimates more precise, suppose that between error-correction rounds, $q$ is the QPP probability (for any given Majorana), $p$ the FPC error probability (along any given bond) and $p_m$ is the measurement error probability.
For small values of $q$, $p$ and $p_m$, a useful proxy for the logical error probability on any given code patch is the probability $p_\text{LWL}$ of the lowest-weight logical (LWL) error.
We will assume $q^2 \ll p$, so that we can restrict our attention to logical errors that have only one QPP component.
For example, on a triangle code patch at some time long after state preparation, the largest term in $p_\text{LWL}$ will be proportional to $qp^{(d_\Delta-1)/2}$ (where we consider an undetectable fermion-parity-odd error on a triangle patch to be a logical error, even though technically such an error is not a logical operator, as it is parity-odd).
Meanwhile, the $p_\text{LWL} \propto p^{d/2}$ for a bosonic code patch.
The proportionality refers to the multiplicities of each LWL error (which will be functions of $d$ or $d_\Delta$).

We choose $d$ and $d_\Delta$ such that the same $p_\text{LWL}$ for bosonic and fermionic patches is achieved, which will set
the logical error probability for the entire computation (in practice, one would choose $d$ and $d_\Delta$ with a target maximum logical error probability for the computation in mind).
These distances would depend on $p$ and $q$.
As can be seen from Table~\ref{tab:Majorana_overhead}, if we can achieve a parameter regime such that $d_\Delta < d/\sqrt{2}$ is permissible, using Majorana triangle codes for logical qubit storage becomes asymptotically favourable for large $d$ or $d_\Delta$.

To consider the overheads associated with fermionic ancilla patches for logical measurement, we first
discuss error processes that we must contend with in order for our procedures to be fault-tolerant. 
Measurement errors may make thin-width sections of code %
vulnerable to unobservable, low-weight error mechanisms, depending on the patch preparation and readout used.
We illustrate this error through the example shown
in Figure~\ref{subfig:Ancilla_Patch_Long_range_mmt} (black ring).
In this example, ancilla patch preparation starts by preparing blue bond operators in their $+1$ eigenstates, followed by the measurement of blue plaquettes, which all have random outcomes.
A fermion-parity-odd string that terminates on red and green boundaries can flip a blue plaquette, as shown in Figure~\ref{subfig:Ancilla_Patch_Long_range_mmt}. 
However, if such an error string occurs shortly after state preparation along blue bonds, measurement errors can make it unobservable: suppose, for example, a blue plaquette suffers an initial measurement error, and it is then flipped by the parity-odd string described above. 
Subsequent measurements of the blue plaquette cannot detect the error string because of the initial measurement error (see Ref.~\cite{Anyon_condensation_floquet_codes}
for a discussion of analogous error mechanisms in the bosonic Color Code).

This error mechanism occurs with probability proportional to $q p^{(d_\Delta - 1)/2} p_m$, for measurement error probability $p_m$ (the minimum thickness of the ancilla patch in this example is $d_\Delta$).
To suppress the probability of such errors occurring, we let all ancilla patches have a thickness above some minimum value.
For simplicity, and because we assume $q$ is the smallest of the error probabilities, we assume the minimum patch thickness to be $d_\Delta$.

We next show that
the Majorana overhead of ancilla patches 
may be reduced compared with bosonic MSC schemes
if $d_\Delta < d/2$ can be permitted in the parameter regime considered. 
Let us consider the measurement of an arbitrary fermion parity operator $i^k \Upsilon_1\Upsilon_2\ldots \Upsilon_{2k}$, where logical MZMs are stored in triangle code patches.
To measure this operator, we prepare an ancilla patch containing $2k$ logical MZMs along its boundary and perform lattice surgery between the ancilla patch and the triangle codes in the same way as shown in Figure~\ref{fig:Four_body_mmt}.
Several sections of this patch have opposite boundaries that are colored differently to one another.
In Figure~\ref{subfig:Large_Parity_Mmt_FPBC_Example}, we show a $k=2$ example wherein a two-qubit logical Pauli operator is measured. 
As can be seen, two sections of the ancilla patch can have minimal thickness $d_\Delta$, where opposite boundaries are colored red and green, while the middle section has thickness $d$, where opposite boundaries are both colored red.
The $d_\Delta$-thickness of sections of the patch is afforded by the logical MZMs along blue boundaries -- without these, the patch would have boundaries of two colors only.
In Pauli-based computation, we may have to measure logical Pauli operators with support on any logical qubit.
Thus, the ancilla patch may have to connect two qubits stored in far-separated parts of the system.
For these long-range measurements, we can use ancilla patches with $O(d)$ thickness between logical MZMs of the same logical qubit and $O(d_\Delta)$ thickness in every other section, as shown in Figure~\ref{subfig:Ancilla_Patch_Long_range_mmt}.
Note that this patch differs from that of Figure~\ref{subfig:Large_Parity_Mmt_FPBC_Example} only by a change in geometry -- the topology of the patch is unchanged.
In this way, the $O(d_\Delta)$ thickness sections have arbitrarily long length $L$, while the $O(d)$ thickness sections have length only $\sim d + 2d_\Delta$ (sufficient to accommodate two logical MZMs and maintaining their separation $\sim d$).
The width required for the length-$L$ sections is $2d_\Delta$, since one width-$d_\Delta$ ``arm'' of the patch needs to extend its entire length, as shown in Figure~\ref{subfig:Ancilla_Patch_Long_range_mmt} (this is true for higher-weight logical Pauli measurements too).
Assuming $2d_\Delta < d$, we therefore find that this patch improves on the patch geometry from Figure~\ref{subfig:Large_Parity_Mmt_FPBC_Example} and improves on purely length-$L$ bosonic patches: 
we require $\sim 2d_\Delta L + O(d^2)$ Majoranas, compared to $\sim dL + O(d^2)$ for bosonic patches, since the latter require a thickness of $d$ everywhere~\cite{QC_with_MFCs}.
For a two-dimensional system with $n$ logical qubits and $O(d)$ distance for each,
the maximum separation between distant 
logical qubits is $L_\text{max} = O(d\sqrt{n})$. Hence this resource saving can be significant for large $n$ and $d$. 
Table~\ref{tab:Majorana_overhead} summarizes these Majorana overheads for long-range measurement. %

The ability to use ancilla patches with thinner sections is a unique feature of MSCs, owing to the qualitatively different mechanisms of FPC, QPP and measurement errors. 
For $q\ll p$, parameter regimes may exist where $d_\Delta$ is significantly smaller than $d$ (by some additive amount). 
Determining such regimes precisely would require numerics, which we leave for future work. 
This resource saving depends on all error probabilities. 
For example, as $p_m \rightarrow 0$, the resource cost of such a patch becomes $\sim wL + O(d^2)$ with $w=O(1)$.
There may be non-trivial trade-offs between the overheads of qubit storage and ancilla patches that will depend on $q$, $p$ and $p_m$.

\begin{figure*}
\centering
\subfloat[\label{subfig:Fold_Patch}]{\includegraphics[width=0.38\linewidth]{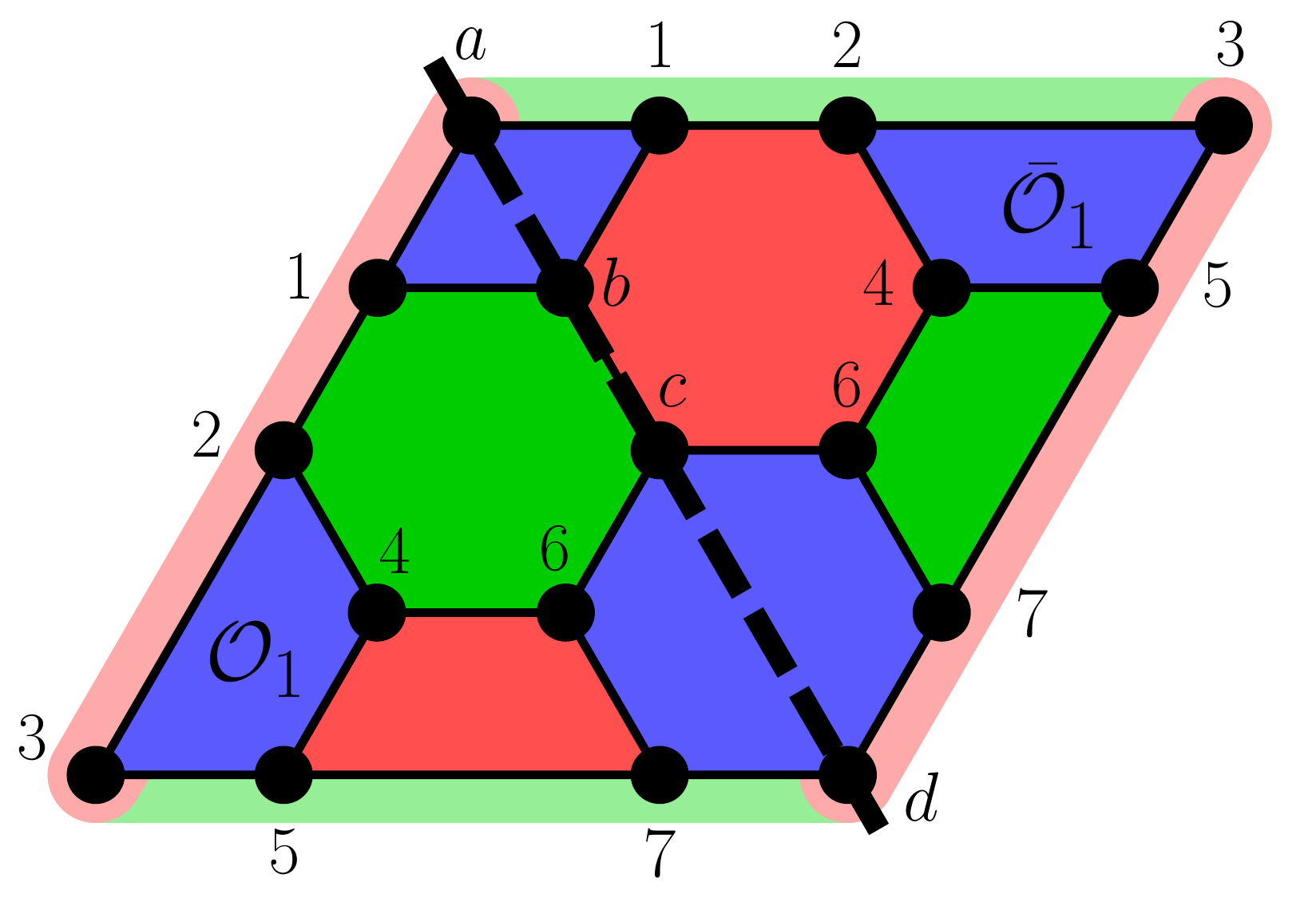}}\hfill
\subfloat[\label{subfig:Pentagon_Patch}]{\includegraphics[width=0.27\linewidth]{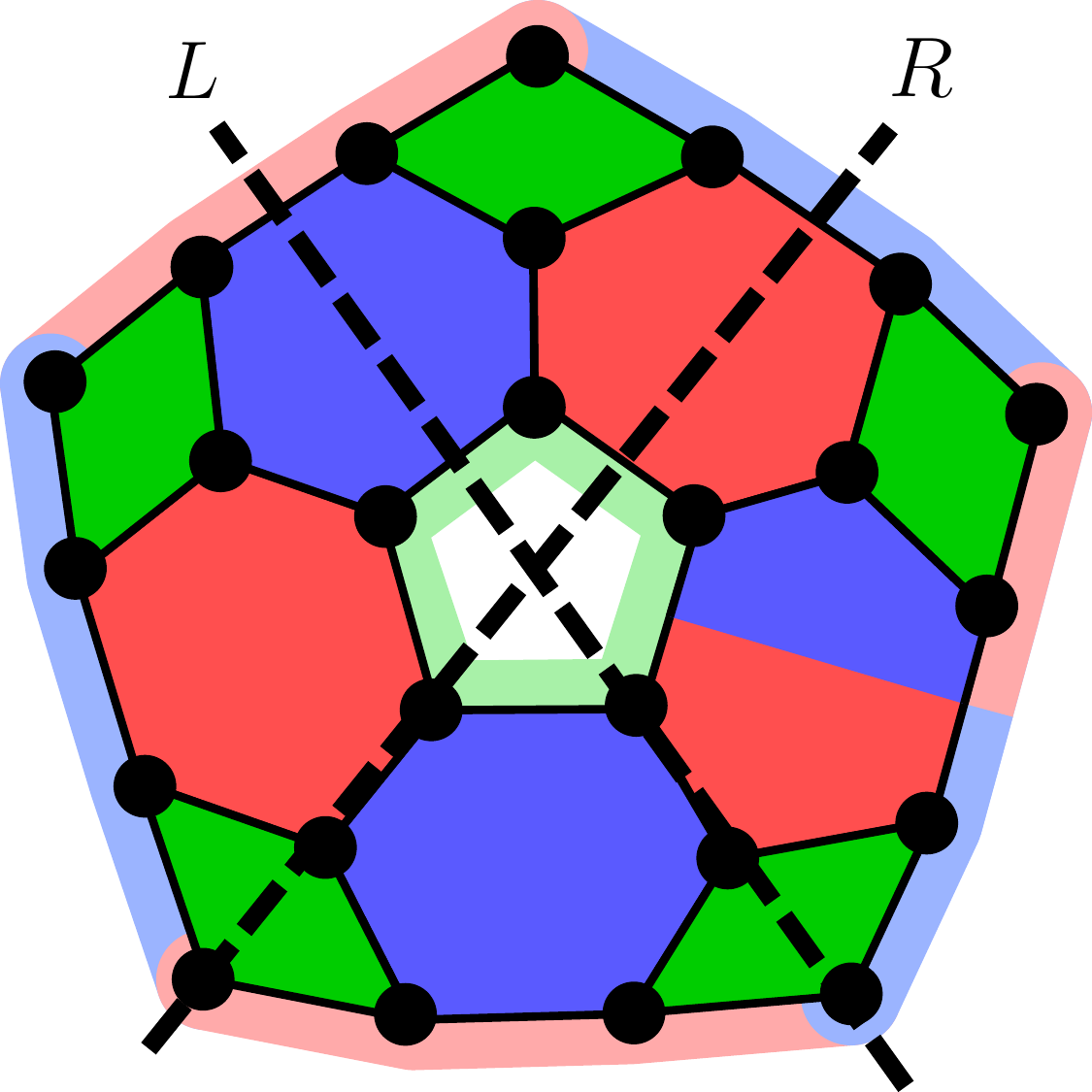}}\hfill
\subfloat[\label{subfig:Pentagon_Logical_Ops}]{\includegraphics[width=0.3\linewidth]{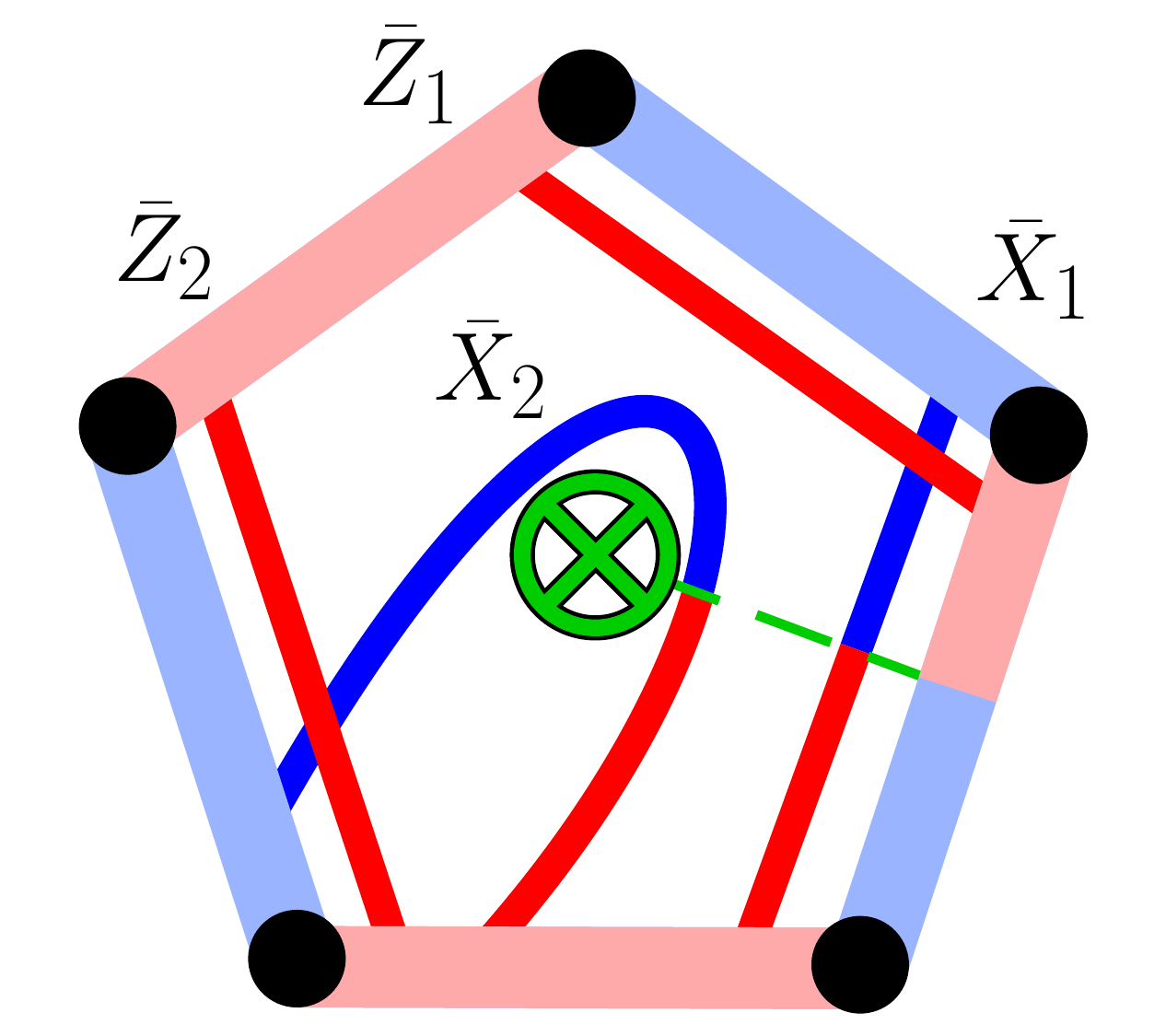}}
\caption{Surface code patches with symmetry axes (some shown dashed) and with boundary colors indicated. Braiding constituent Majoranas enacts Clifford gates on logical qubits.  (a) Patch with one symmetry axis. Braiding equivalently numbered Majoranas across this line results in an encoded Hadamard gate on the logical qubit. (b) Patch with two logical qubits and one logical MZM. This patch has five symmetry axes (the domain wall is a gauge choice, thus does not break this symmetry),
two of which are shown and labeled $L$ and $R$. (c) Logical (string) operators for the patch in (b) are schematically illustrated.}\label{fig:Folded_Codes}
\end{figure*}

\section{Fault-Tolerant Gates on Small Majorana Surface Codes}\label{sec:Braids_Small_Codes}

In this penultimate section, we describe another approach to achieving native Clifford gates in the MSC.
This method has no analogue in bosonic codes, as it arises from the non-Abelian statistics of  constituent Majoranas.
Specifically, braiding these Majoranas in the MSC can implement Clifford gates on the logical qubit(s) stored therein.
The patches of MSC require one or more lines of mirror symmetry to be suitable for this type of gate.
Two examples of these are shown in Figure~\ref{fig:Folded_Codes}, with the symmetry axes indicated by dashed lines.
We require braids to occur across this line of symmetry which, as drawn, are relatively long-range. 
Hence, this technique would be either suitable for small patches of the MSC or patches that are ``folded" along these lines of symmetry to make the braids short-range.

We begin with the code shown in Figure~\ref{subfig:Fold_Patch}, which stores one logical qubit.
We will show that we can enact a single-qubit Clifford gate on this qubit by braiding the constituent Majoranas.
Let $U$ be the unitary that braids all Majoranas on opposite sides of the symmetry axis of the code (i.e., those labeled with the same numbers in Figure~\ref{subfig:Fold_Patch}).
That is, $U = \prod_{j=1}^7 \exp{(\frac{\pi}{4}\bar{\gamma}_j\gamma_j)}$, where $\bar{\gamma}_j$ is in the top-right half of the code and $\gamma_j$ is in the bottom-left. 
To see that $U$ implements an encoded Clifford gate, we firstly show that it preserves the stabilizer group.
We start with the four-body blue plaquette operators in the bottom-left and top-right corners of the code, which we denote $\mathcal{O}_1$ and $\bar{\mathcal{O}}_1$ respectively (see Figure~\ref{subfig:Fold_Patch}).
We choose the ordering $\mathcal{O}_1 = -\gamma_3\gamma_2\gamma_4\gamma_5$ and $\bar{\mathcal{O}}_1 = -\bar{\gamma}_3\bar{\gamma}_2\bar{\gamma}_4\bar{\gamma}_5$.
Then $U\mathcal{O}_1 U^\dagger = -\bar{\gamma}_3\bar{\gamma}_2\bar{\gamma}_4\bar{\gamma}_5 = \bar{\mathcal{O}}_1$,
and similarly $U\bar{\mathcal{O}}_1U^\dagger = \mathcal{O}_1$.

We similarly choose oppositely-oriented plaquette operators between the two halves, so that $U$ maps stabilizer generators to other stabilizer generators under conjugation, without any sign changes.  
The blue plaquette operators that are intersected by the symmetry axis have their orientations reversed by $U$.
In the case of the blue four-body operator in the top-left corner, this results in a change of sign. 
But it picks up another minus sign because, under conjugation, $U:\gamma_1\mapsto \bar{\gamma}_1$ and $\bar{\gamma}_1\mapsto - \gamma_1$. 
It can similarly be seen that the blue hexagonal plaquette operator does not change its sign under conjugation by $U$.
Hence $U$ preserves the whole stabilizer group. 

We also define logical operators for the encoded qubit: $\bar{X} = - \gamma_a\gamma_1\gamma_2\gamma_3$ and $\bar{Z} = -\gamma_3\gamma_5\gamma_7\gamma_d$.
Hence $U\bar{X} U^\dagger = -\gamma_a\bar{\gamma}_1\bar{\gamma}_2\bar{\gamma}_3 = \eta \bar{Z}$
and $U\bar{Z}U^\dagger = \lambda \bar{X}$, where $\eta,\lambda\in \lbrace +1,-1 \rbrace$ depend on the definition of the stabilizer operators. 
Hence $U= \bar{H}$, the encoded Hadamard operation, up to a logical Pauli operator (set by $\eta$ and $\lambda$).

This operation $U$ is fault-tolerant as it retains the locality of errors.
That is, suppose $E\in \text{Maj}(18)$ is an error operator acting on the constituent Majoranas of the code.
Then conjugating $E$ with $U$ results in another error operator $E^\prime$ with the same weight as $E$.
Moreover, we can ensure that $U$ has a very low probability of introducing further errors into the system, since the constituent braids making up $U$ can be made arbitrarily accurate by increasing Majorana separation.
In this way we combine the topological protection of constituent braid operations with the protection afforded by the MSC.

A larger set of encoded gates via braids can be found in codes that have more than one axis of symmetry.
MSCs defined on the same lattice as stellated Color Codes~\cite{BBrown_Twists_CC}, such as the one shown in Figure~\ref{subfig:Pentagon_Patch}, have this property.
They also possess highly favourable encoding rates among planar code architectures~\cite{BBrown_Twists_CC}, making them interesting candidates for near-term devices.
The code of Figure~\ref{subfig:Pentagon_Patch} stores two logical qubits and one logical MZM.
(It contains an odd number of constituent Majoranas; this implies that it is part of a larger Majorana system.) 
The logical operators of the encoded qubits are the string operators indicated in Figure~\ref{subfig:Pentagon_Logical_Ops}.
Let $U_L$ (resp. $U_R$) denote the operation that braids all constituent Majoranas across the first (resp. second) line of symmetry [labeled $L$ (resp. $R$) in Figure~\ref{subfig:Pentagon_Patch}].
One can again show that $U_L$ and $U_R$ both preserve the stabilizer group. 
Furthermore, $U_L\bar{Z}_1 U_L^\dagger = \bar{Z}_2$, $U_L\bar{Z}_2U_L^\dagger = \bar{Z}_1$, $U_L\bar{X}_1 U_L^\dagger = \bar{Z}_1 \bar{Y}_2$, and $U_L\bar{X}_2 U_L^\dagger = \bar{Z}_2 \bar{Y}_1$. 
(Different ordering choices for the logical operators and stabilizers can introduce signs into the above relations.)
These transformations are enacted by the following Clifford circuit:
\begin{align}
    \Qcircuit @C=1em @R=1.5em {
\lstick{1} & \qw      & \qw      & \targ     & \ctrl{1} & \gate{H} & \qw\\
\lstick{2} & \gate{S} & \gate{H} & \ctrl{-1} & \targ    & \gate{S} & \qw 
}
\end{align}
The braid $U_R$ maps $\bar{Z}_1 \mapsto \bar{Z}_1$, $\bar{X}_1 \mapsto \bar{X}_1\bar{X}_2$, $\bar{Z}_2 \mapsto \bar{Z}_1\bar{Y}_2$ and $\bar{X}_2 \mapsto \bar{X}_2$, up to signs. 
The Clifford circuit that enacts this gate is the following:
\begin{align}
    \Qcircuit @C=1em @R=1.5em {
    \lstick{1} & \qw & \ctrl{1} & \qw & \qw & \qw & \qw \\
    \lstick{2} & \qw & \targ    & \gate{H} & \gate{S} & \gate{H} &\qw 
    }
\end{align}

It is interesting to note that $(U_L U_R)^5 = \mathds{1}$ up to a phase, while $U_L^2 = \bar{Z}_1\bar{Z}_2$, $U_R^2 = \bar{X}_2$ so that $U_L^2, U_R^2\in \mathcal{P}_2$, the group of 2-qubit Pauli operators. 
Hence, $U_L$ and $U_R$ do not generate a representation of the dihedral group $D_5$ (the symmetry group of the code lattice) in $\mathcal{C}_2$ (the 2-qubit Clifford group) as one might expect.
Rather they generate a representation of $D_5$ in $\mathcal{C}_2/\mathcal{P}_2$.
We note that $|D_5| = 10$ while $|\mathcal{C}_2/\mathcal{P}_2| = 720$ and hence we can  generate only a small fraction of Clifford gates with these two braiding operations.
By considering stellated codes with larger dihedral symmetries, we can obtain larger numbers of $\mathcal{C}_2/\mathcal{P}_2$ gates with braids. 
But since $|D_n| = 2n$, this is not a very promising avenue for generating the entire $\mathcal{C}_2$.
However, expanding the size of this code so that the corners may be braided via code deformation (cf. Section~\ref{subsec:Logical_qubit_braids}), one can fault-tolerantly attain more Clifford gates.
An interesting direction for further study is to consider higher-dimensional Majorana codes: these may have larger symmetry groups and so could have a larger set of gates implementable  via constituent braids.

\section{Conclusion}\label{sec:Conclusion}

We have presented a comprehensive framework for twist defects in the MSC and have established new methods for fault-tolerant quantum computation with these codes.
A complete characterization of the topological features of the MSC reveals a $Z_2$ fermion-parity grading on boundaries and domain walls.
This indicates that twist defects (and equivalently corners) can possess fermionic features leading to a larger set of topologically-protected code states.
However, rather than this grading allowing one to simply attach extra microscopic MZMs at twists, as might be naively expected, we find that twists host \emph{logical} MZMs: large-weight, odd-parity operators commuting with the stabilizer group. 
Twists in the MSC therefore host not only the same logical qubits seen in bosonic codes with twists, but also logical qubits possible only in fermionic models.
As such, comparing twist-based encodings to hole-based ones~\cite{MjFerm_Surface_Code,Roadmap_to_MSCs}, we can store roughly twice the number of logical qubits in an equivalently sized lattice.

We can fault-tolerantly compute with both of these types of twist qubit -- BT and FT qubits -- simultaneously. 
We have shown that Clifford gates can be applied to BT qubits by braiding twists via code deformation (Section~\ref{subsec:Logical_qubit_braids}) or alternatively by lattice surgery (Section~\ref{subsec:Hybrid_Approaches}).
This contrasts to techniques for performing single-qubit Clifford gates on hole-encoded logical qubits in the MSC, which involve the introduction of an ancilla qubit and multiple Clifford gates and measurements on target and ancilla qubits~\cite{MjFerm_Surface_Code,Roadmap_to_MSCs}.
Instead, our Clifford gate implementation is similar to that in bosonic codes with twists~\cite{Bombin_twists_code_deformation,Holes_Twists_SurfaceCode,Punctures_Raussendorf_2006,Surface_Code_Lattice_Surg}.

The above twist braids would also implement braid operations of the form $\exp(i\frac{\pi}{4}\Upsilon_i\Upsilon_j)$ on logical MZMs stored in the twists.
The effect of braiding twists on FT logical qubits is analogous to that of braiding microscopic MZMs on topological qubits~\cite{MZMs_in_pwave_superconds2001,Majoranas_Tjunctions2011}.
If we wish twist braids to implement a Clifford gate only on the BT logical qubits of fermionic twists (but not on their FT qubits), we can perform the associated braid and classically track the resulting ``error" to the FT logical qubits.

We can enact Clifford gates on FT qubits with measurement-based procedures, for which we require the ability to measure logical fermion parity operators and prepare ancilla states.
We achieve such measurements with novel lattice surgery techniques.
These techniques differ from lattice surgery introduced in previous schemes~\cite{QC_with_MFCs,Maj_Triangle_Code2018}, since the logical measurements performed do not commute with the operator to be measured. 
Instead, lattice surgery measurements can be understood to be ``teleporting" logical MZMs between different odd-length boundaries of the lattice (see Section~\ref{subsec:Bilinear_Mmt} and Table~\ref{table:Four_body_mmt_ops}).
This makes our lattice surgery techniques annalogous to ``anyonic state teleportation" protocols~\cite{Mmt_Only_QC_2008}.

To achieve universal computation, we can inject magic states into the code in two different ways. 
The off-patch method provides a low-overhead approach to preparing magic states in groups of Majorana triangle codes. %
Finally, we can also perform Clifford gates between BT and FT qubits through measurement-based procedures, using lattice surgery.

We also discussed an alternative approach where Clifford gates on BT and FT qubits need not be explicitly implemented.
Rather, in the spirit of Pauli-based computing (PBC)~\cite{Bravyi_PBC2016}, we can classically track their effects on subsequent gates and measurements in the circuit so long as we can measure arbitrary products of logical Pauli operators.
We showed how to measure any operator of the form $\Gamma_F P_B$, where $\Gamma_F$ is a logical MZM product and $P_B$ is a logical Pauli operator for BT qubits, using lattice surgery.
Hence, we can achieve the effects of Clifford gates on BT and FT qubits with zero time overhead (apart from the measurement time and that of efficient classical processing).

Beyond these advantages, we have found that, given a suitably small QPP probability, the fermionic features of the MSC could be exploited to reduce the number of Majoranas required.
Using PBC, we explored a method for expanding the capacity of an $n$-logical-qubit register~\cite{Bravyi_PBC2016}. 
Using Majorana triangle code patches and fermionic ancilla patches, we find potential avenues for a lowered Majorana overhead for logical qubit storage and measurement, depending on the QPP and FPC error rates.
This arises when $d_\Delta$ -- the required minimum weight of logical Majoranas to achieve a desired logical error probability $p_L$ -- is considerably smaller than $d$ -- the bosonic patch distance required to achieve the same $p_L$.
We also find in such cases that the Majorana overhead of measuring multi-logical-qubit operators using a fermionic ancilla patch of length $L$ is reduced from $O(dL)$ to $O(d_\Delta L)$.
For $d_\Delta < d/\sqrt{2}$ we also find an asymptotic reduction in the Majoranas required for logical qubit storage in triangle codes (see Table~\ref{tab:Majorana_overhead}).
Even outside of the PBC context, the same considerations result in smaller spatial overheads of measurement-based Clifford gate implementation.
More in-depth, numerical analysis is required to determine the parameter regimes in which this reduction in resources is possible.

Finally, we explored the consequences of having non-Abelian anyons constitute the building blocks of the MSC.
We found that braiding constituent Majoranas can result in fault-tolerant Clifford gates applied to BT qubits, which can be combined with braids between ($Z_2$) corners to generate the full Clifford group.
This would be a method suitable for either small or ``folded" patches of MSC. 
This ability to achieve Clifford gates with constituent-Majorana braids is likely related to a connection between a folded MSC and the Color Code, which has a large set of transversal Clifford gates~\cite{Kubica_Beverland}.
Future work could investigate this identification of the MSC as an ``unfolded" Color Code~\cite{Maj_Ferm_Codes,Unfolding_color_code}.

One of the advantages of our MSC approach is that it allows different levels of protection for FPC and QPP errors. 
For example, consider a logical qubit encoded in four logical MZMs located at twist defects.
One can protect against FPC errors by increasing the separation between twists, and one can protect against QPP errors by increasing the weights of the logical MZMs.
These quantities may be tuned independently (without sacrificing twist separation), to tailor codes for given FPC and QPP error rates.

Regarding potential realizations, a key requirement for the MSC with twists is the ability to create superpositions of parities of various subsets of microscopic Majoranas. 
This requires a setup where a definite parity is required only for the entire collection of $2n$ Majoranas. 
(This is distinct from MSCs that are effectively bosonic surface codes~\cite{Realistic_MSC,Roadmap_to_MSCs,QC_with_MFCs}
based on qubits sparsely encoded~\cite{MZM_TQC_Review2015} in small groups of Majoranas.)
A proposal for such a MSC setup exists based on the proximity effect in topological insulators~\cite{MjFerm_Surface_Code}, and analogous constructions are expected to be possible for superconducting hybrid structures based on two-dimensional electron gases or spin-orbit nanowires~\cite{Lutchyn_Review_2018,Nanowire_Review_2021}. 
The stabilizer measurement scheme outlined in Ref.~\cite{MjFerm_Surface_Code}, using topological transmon ingredients~\cite{Top_Transmon2011}, can in principle be used to implement the manipulations we used in our scheme to compute with the MSC.

However, one key consideration with these proposed designs is the expected QPP times.
In designs with large-charging-energy, fixed-parity islands the QPP times are expected to be long~\cite{Karzig_Majorana2017,Karzig_QPP_Maj_Qubits_2021}, while the optimal designs without such constraints are less well-specified in the literature.
However, the expected QPP times may still be comparable to the coherence times of, e.g., superconducting qubits~\cite{Karzig_QPP_Maj_Qubits_2021}, hence the Majorana resource efficiency in our designs may present non-trivial trade-offs among designs.
Furthermore, mitigation measures exist that decrease the effect of quasi-particles (QPs) on the coherence times of superconducting qubits, including QP traps~\cite{QP_Traps_1,QP_Traps_2,QP_Traps_3} and pumps~\cite{QP_Pumps}. 
These could be adapted to protect MZMs in bulk superconductors from QPP.
One way this could occur is by utilizing traps to reduce QP density near the center of bulk superconductors, to thereby simulate the lower density of QPs expected from mesoscopic superconductors~\cite{Karzig_QPP_Maj_Qubits_2021}.
Furthermore, QP detection may help greatly reduce the effect of QPP in Majorana systems~\cite{QP_detection}.
While these considerations suggest avenues for realizing the marked benefits of our proposals, further work is needed to combine QPP protection with flexible parity constraints in order to identify the most suitable setups.

Further studies are needed as well to understand the error thresholds for our MSC scheme under various noise models. 
We expect these thresholds to be similarly high to those found in other MSC approaches~\cite{Maj_Triangle_Code2018,Ferm_Error_Corr_2019}.
Future research could also explore other routes for generating encoded gates through braids or other transformations of the constituent Majoranas in MSCs, as we did in Section~\ref{sec:Braids_Small_Codes}.
Studying twist defects in generalizations of the MSC, including parafermion codes~\cite{Parafermion_codes} or models displaying fracton order, such as the Majorana checkerboard model~\cite{Fractons_Majorana_Code}, could also be fruitful.

\acknowledgments
We thank Heidar Moradi, Wilfred Salmon and Dominic Williamson for discussions. This work was supported by an EPSRC Studentship, the EPSRC Grants EP/S019324/1 and EP/V062654/1, and in part by the ERC Starting Grant No. 678795 TopInSy.

\appendix

\section{Details of Anyons and Boundaries in the MSC}\label{app:Anyons}

\subsection{Anyon Statistics}\label{app:Anyon_Statistics}

Given a collection of anyons $\mathcal{A}$, we can assign to them self-exchange and mutual statistics.
The self-exchange statistic $\theta_a$ for $a\in \mathcal{A}$ is the phase acquired when two $a$ particles are clockwise interchanged. 
In general, Abelian anyons can pick up any $U(1)$ phase under self-exchange. 
The mutual statistic, $M_{a,b}$, of two anyons, $a,b\in\mathcal{A}$ with $a\neq b$, is the phase acquired when $a$ is wound completely around $b$ in the clockwise direction.
In the MSC, the particles $\mathsf{R}$, $\mathsf{G}$ and $\mathsf{B}$ all have bosonic ($+1$) self-exchange statistics but mutual semion statistics ($M_{\mathsf{R},\mathsf{G}} = M_{\mathsf{G},\mathsf{B}} = M_{\mathsf{B}, \mathsf{R}} = -1$)~\cite{MjFerm_Surface_Code}. This is due to different-colored string operators anti-commuting with one another, because they overlap at an odd number of vertices (cf. Figure~\ref{fig:Patch}).
$\mathsf{GB}$, $\mathsf{BR}$ and $\mathsf{RG}$ differ from $\mathsf{R}$, $\mathsf{G}$ and $\mathsf{B}$ by the addition of $\mathsf{RGB}$ which has a fermionic ($-1$) self-exchange statistic. 
Hence these three particles also have fermionic self-exchange statistics.
All mutual statistics are preserved under the addition of $\mathsf{RGB}$, as can be seen by splitting it into $\mathsf{R}$, $\mathsf{G}$ and $\mathsf{B}$, and separately moving these particles around anyon $a\in\mathcal{A}$ (for any such $a$, an even number of $M_{\mathsf{R},a}$, $M_{\mathsf{G},a}$ and $M_{\mathsf{B},a}$ are $-1$).

\subsection{Lagrangian Subgroups}\label{app:Lagrangian_subgroups}

In bosonic systems, types of gapped boundaries are in one-to-one correspondence with ``Lagrangian subgroups" of the anyon model, and hence can be fully classified with reference only to the anyons and their properties~\cite{Lagrangian_Subgroups2011,Protected_Edge_Modes}. 
A Lagrangian subgroup is a subset of the collection of anyons $\mathcal{L}\subseteq\mathcal{A}$, such that for all $a,b\in\mathcal{L}$: 
(1) $a\times b\in \mathcal{L}$ and $\bar{a},\bar{b}\in\mathcal{L}$, where $\bar{a}$ refers to the antiparticle of $a$ such that $a\times\bar{a}=\mathbf{1}$; 
(2) $a$ and $b$ are self-bosons ($\theta_a = \theta_b = 1$) and have trivial mutual statistics (i.e., $M_{a,b}=1$); 
(3) every $c\notin\mathcal{L}$ has non-trivial mutual statistics with at least one $a\in\mathcal{L}$. 

To every Lagrangian subgroup we can associate a gapped boundary $G_\mathcal{L}$ (and vice versa). 
The boundary $G_\mathcal{L}$ will be such that all anyons in $\mathcal{L}$ can condense at that boundary. 
This condensation process is achieved by string operators that terminate on the boundary, and hence commute with all plaquette operators around this endpoint.
Such condensation processes are shown in Figure~\ref{fig:MSC_bdrys}.

The properties of $\mathcal{L}$ can be understood as follows.
Property (2) ensures that processes that are confined to the boundary $G_\mathcal{L}$ and that commute with the Hamiltonian act trivially on the ground state. 
In topological codes, these processes correspond to stabilizer elements, which must mutually commute - hence the associated anyons must have trivial statistics. 
Property (1) ensures that $\mathcal{L}$ is a group under the fusion operation, while (3) ensures that this group is maximal: no other anyon can be added to it without violating either (1) or (2).

In the MSC there exist three different Lagrangian subgroups: $\mathcal{L}_\mathsf{R}$, $\mathcal{L}_\mathsf{G}$ and $\mathcal{L}_\mathsf{B}$, where $\mathcal{L}_\mathsf{C} = \lbrace\mathbf{1}, \mathsf{C}\rbrace$ for $\mathsf{C}=\mathsf{R},\mathsf{G},\mathsf{B}$. 
Hence, each boundary can be colored according to the color of the anyon that it condenses (cf. Figure~\ref{fig:Patch}). 
These boundaries are closely related to a subset of gapped boundaries of the Color Code \cite{BBrown_Twists_CC, Bombin_Color_Codes}. 
However, as described in the main text, fermionic systems have an extra $Z_2$ fermion parity grading along their one-dimensional boundaries, allowing for a further classification.

\section{Properties of Twists and Domain Walls}\label{app:Twist_Properties}

\subsection{Gauge Dependence of Domain Walls and Twist Sub-Types}\label{app:Gauge_Dependence}
\begin{figure*}
\centering
\subfloat[\label{subfig:Gauge_dependence_a}]{\includegraphics[width=0.45\linewidth]{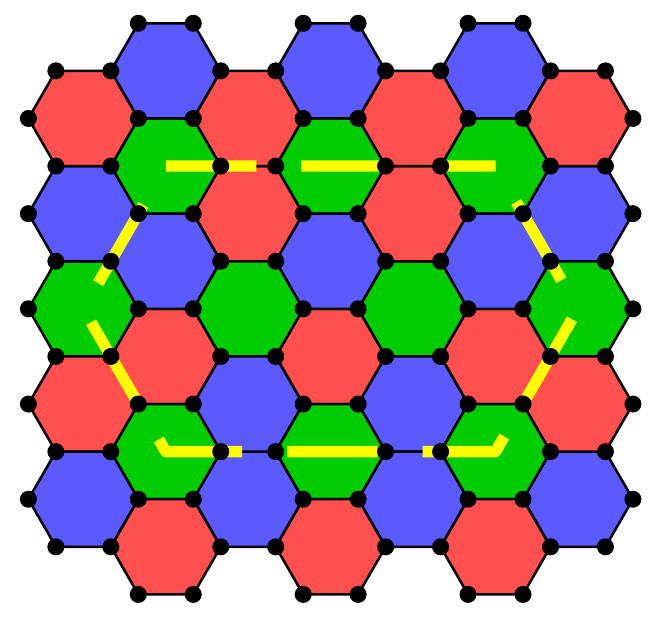}}\hfill
\subfloat[\label{subfig:Gauge_dependence_b}]{\includegraphics[width=0.45\linewidth]{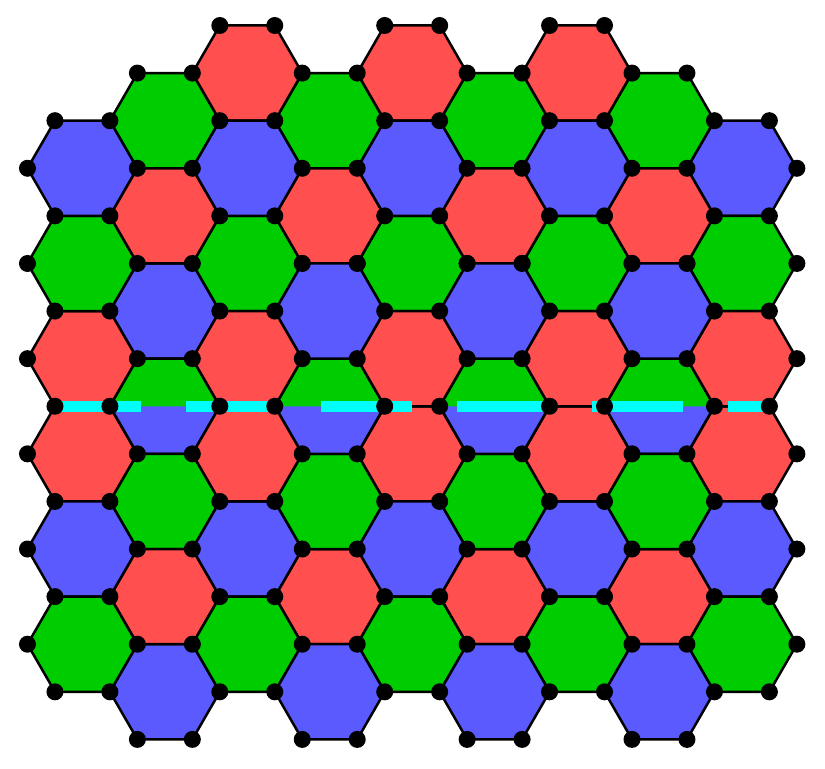}}
\caption{Domain walls can be inserted into the lattice via a gauge transformation (a change of plaquette coloration).  In (a) a closed loop of $Z_2$ domain wall of type $T_\mathsf{G}$ is illustrated by a yellow dashed line. In (b) a line of $Z_3$ domain wall of type $T_\mathsf{RGB}$ (for anyons travelling from top to bottom across the wall) is illustrated by a blue line.}\label{fig:Gauge_dependence}
\end{figure*}

Domain walls can be moved, or a closed loop of domain wall can be artificially introduced, by simply relabeling the colors of plaquettes. 
Thus domain walls are gauge-dependent objects. 
While the locations of twists are gauge-independent, twists do have gauge-dependent features.
Consider, for example, wrapping a domain wall $\psi$ around a pair of twists $\varphi$ and $\varphi^{-1}$, with $\psi$ and $\varphi$ both symmetry transformations of the anyon labels.
Then an anyon $a$ encircling the twist $\varphi$ from outside the domain wall $\psi$ is mapped to $\psi^{-1}\circ\varphi\circ\psi (a) \neq \varphi(a)$. 
Hence, twists are only defined up to conjugation by other permutation group elements.
$S_3$ has two conjugacy classes, one containing all $Z_2$ twist permutations, the other containing those of the $Z_3$ twists.

To illustrate the gauge-dependence of domain walls, Figure~\ref{fig:Gauge_dependence} shows two domain walls that have been inserted into the lattice by simply changing plaquette labels. 
In Figure~\ref{subfig:Gauge_dependence_a}, a closed loop of $Z_2$ domain wall (of type $T_\mathsf{G}$) is shown.
Note that 3-colorability of the lattice is violated along this wall.
In Figure~\ref{subfig:Gauge_dependence_b}, a $Z_3$ domain wall enacting the permutation $T_\mathsf{RGB}$ on anyons travelling across it from top to bottom is shown.

\subsection{Twist-Corner Correspondence}\label{app:Twists_Corners}

As explained in the main text, twists and domain walls are closely associated with boundaries and corners of a model respectively~\cite{Defects_Abelian_states}.
For example, creating a domain wall $\varphi$ along a boundary of the model can change the boundary type. 
If the original boundary was $G_\mathcal{L}$, for Lagrangian subgroup $\mathcal{L}$, it now becomes $G_{\varphi(\mathcal{L})}$, where $\varphi(\mathcal{L})$ is also a Lagrangian subgroup, owing to the fact that $\varphi$ is a symmetry.
Twists in these cases are then associated with corners between boundaries of different types. 

\subsection{Twist Fusion Rules}\label{app:Twists_fusion}

\begin{figure}[t]
    \centering
    \includegraphics[width=0.9\linewidth]{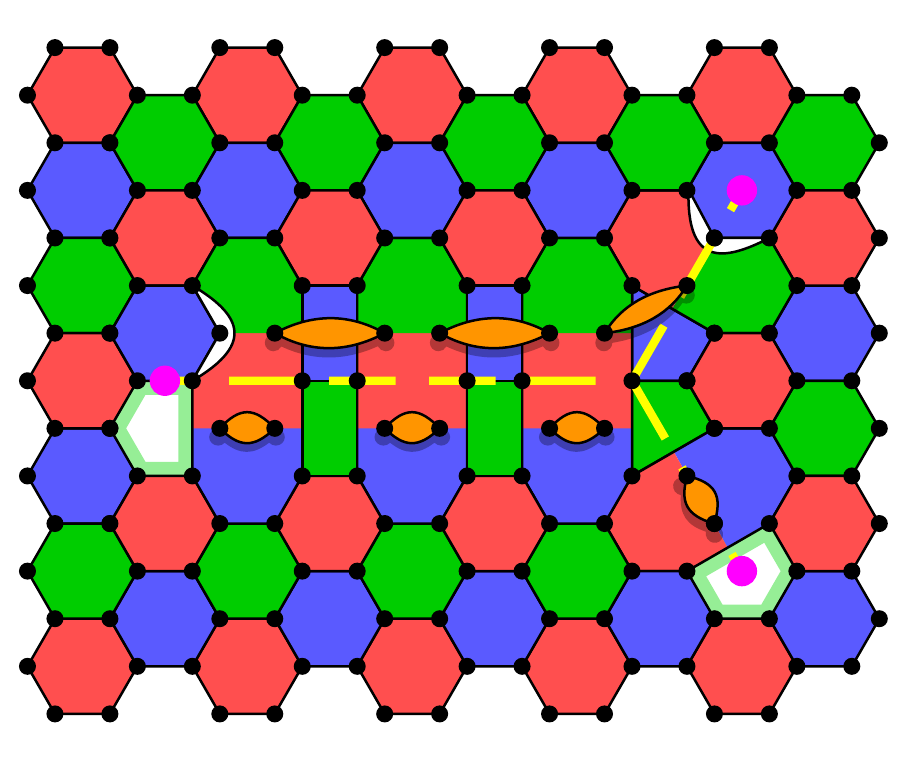}
    \caption{An example of a $Z_3$ twist having split into two $Z_2$ twists. On the left we have one fermionic $Z_3$ twist. On the right, we have a bosonic $T_\mathsf{B}$ twist (top) and a fermionic $T_\mathsf{G}$ twist (bottom).}
    \label{fig:twist_split}
\end{figure}

Similarly to anyons, twists obey fusion rules dictating how they can combine and split.
Consider an anyon encircling both a $T_\mathsf{R}$ twist and a $T_\mathsf{B}$ twist, located in close proximity to one another.
This anyon first undergoes the transformation $T_\mathsf{R}$, and then the transformation $T_\mathsf{B}$, which is equivalent to the transformation $T_\mathsf{RGB}$.
Thus we can consider the combination of $T_\mathsf{R}$ and $T_\mathsf{B}$ (with a suitable ordering of the domain walls) to be $T_\mathsf{RGB}$. 
We can split a $Z_3$ twist into any two $Z_2$ twists of different sub-types.
We demonstrate this at the lattice level in Figure~\ref{fig:twist_split}, where a $T_\mathsf{RGB}$ domain wall terminates at a (fermionic) $T_\mathsf{RGB}$ twist on the left, and splits into two $Z_2$ domain walls on the right. 
The domain walls shown terminate at a (bosonic) $T_\mathsf{B}$ twist and a (fermionic) $T_\mathsf{G}$ twist.
In general, twists obey the same relations as members of $S_3$.
The overall fermion grading of the twists must be preserved when combining or splitting them. 
Thus, for example, bosonic twists can split into two bosonic or two fermionic twists, whereas fermionic twists must split into one fermionic and one bosonic twist.

\subsection{Calculating Quantum Dimensions of Twists}\label{app:Quantum_Dimensions}

We can calculate the quantum dimensions of twists of a given type by determining the number of ground states that a code with those twists possesses. 
We begin with bosonic $T_\mathsf{B}$ twists.
First, consider the code to contain $2n$ Majorana operators, and to be embedded on a torus.
With zero twists present, we have the following non-trivial relationship between plaquette operators and the total fermion parity operator $\Gamma$:
\begin{align}\label{eqn:stab_dependence_app}
    \prod_{p\in\mathcal{R}}\mathcal{O}_p = \prod_{p\in\mathcal{G}}\mathcal{O}_p = \prod_{p\in\mathcal{B}}\mathcal{O}_p = \Gamma
\end{align}
where $\mathcal{R}$, $\mathcal{G}$ and $\mathcal{B}$ are sets of red, green and blue plaquettes respectively.
Thus the degeneracy of the ground state is:
\begin{align}
    D_0 = 2^{n - [3(2n/6) - 2]} = 4
\end{align}
because there are $2n/6$ plaquettes of any given color and so the number of independent stabilizer generators [considering Equation~\eqref{eqn:stab_dependence_app}] is $n-2$.
Hence the code contains two logical qubits, owing to the topology of the lattice.

We now add a pair of bosonic $T_\mathsf{B}$ twists into the lattice. 
In this case, we only have the following non-trivial relations between plaquette operators:
\begin{align}\label{eqn:twists_dependency}
    \prod_{p\in \mathcal{RG}} \mathcal{O}_p = \mathds{1},\quad \prod_{p\in\mathcal{BO}}\mathcal{O}_p = \Gamma
\end{align}
where $\mathcal{RG}$ is the set of all red/green plaquette operators (including those that span the domain wall, colored both red and green in Figure~\ref{subfig:Z2_Z3_Twists_bosonic}), and $\mathcal{BO}$ is the set of all blue and orange operators.
From Figure~\ref{subfig:Z2_Z3_Twists_bosonic}, it can also be seen that introducing a pair of bosonic $Z_2$ twists lowers the number of stabilizer generators along the domain wall by 1 (note that we also count the orange operators as stabilizer generators).
Hence, counting the degeneracy of the ground state when two twists are present, we still obtain:
\begin{align}
    D_2 = 2^{n-[3(2n/6)-1-1]} = 4.
\end{align}
However, as we continue to add pairs of twists, the number of independent stabilizer generators is decreased by 1 each time, while Equations~\eqref{eqn:twists_dependency} remain unchanged.
Thus, $D_4 = 8$, $D_6 = 16$, etc.
So for large twist numbers, introducing an additional pair of twists doubles the number of ground states, indicating that the quantum dimension of these twists is $d^B_2 = \sqrt{2}$. 
Similarly, introducing a pair of $Z_3$ twists lowers the number of stabilizer generators along the domain wall by 2 and hence, for large twist numbers, introducing a pair of them quadruples the number of ground states. 
Thus, their quantum dimension is $d^B_3 = 2$.
In both cases, it can be seen from Figures~\ref{subfig:Z2_Z3_Twists_bosonic} and~\ref{subfig:Z2_Z3_Twists_fermionic} that the fermionic counterparts of these twists have 1 fewer stabilizer generators again, indicating that the quantum dimensions of fermionic twists are $d^F_2 = 2$ and $d^F_3 = 2\sqrt{2}$.

\section{Logical MZMs in Systems with Boundary}\label{app:Logical_Maj_Boundary}

\begin{figure}
\centering
    \subfloat[\label{subfig:Boundary_ops_a}]{%
    \includegraphics[width=0.45\linewidth]{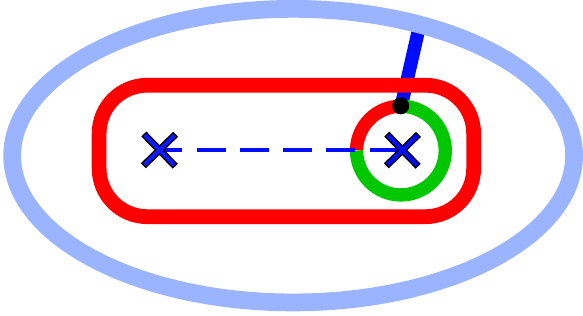}}\hspace{10pt}
    \subfloat[\label{subfig:Boundary_ops_b}]{%
    \includegraphics[width=0.45\linewidth]{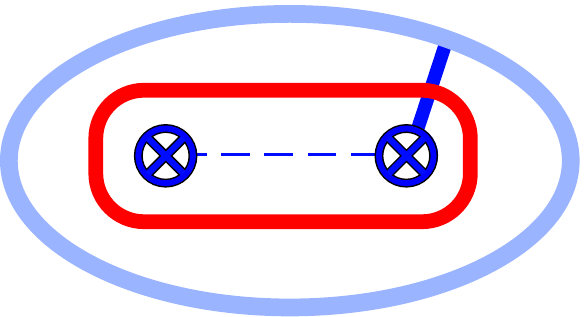}}
    \caption{Logical MZMs in the MSC with twist defects and boundaries. In (a) a logical operator (red string operator, equivalent to $\Gamma$) and a logical MZM (red, green and blue operator) are shown, for a code with two bosonic $T_\mathsf{B}$ twists (blue crosses) and a blue boundary (outer blue ring). In (b), the twists are replaced by fermionic twists, allowing for a qubit to be stored in a single parity sector. The logical operators are the blue and red strings indicated.}
\end{figure}

Logical MZMs can be found in systems with only bosonic twists when boundaries are introduced to the patch of MSC.
Without loss of generality, we can consider the boundary as having a single color and all twists as being in the bulk, since corners are equivalent to twists.

To begin with $Z_2$ twists, the ground state is non-degenerate when the boundary is colored differently to the twist type.
For example, in a single parity sector ($\Gamma = +1$) and in the case of a blue boundary and $T_\mathsf{R}$ twists, we have the non-trivial relation $\prod_{p\in \mathcal{RO}}\mathcal{O}_p = \Gamma$.
Whereas without twists there are two: $\prod_{p\in\mathcal{R}}\mathcal{O}_p = \prod_{p\in\mathcal{G}}\mathcal{O}_p = \Gamma$. 
Thus, introducing two bosonic $T_\mathsf{R}$ twists lowers the number of constraints by 1 and the number of plaquette operators by 1, thus not changing the number of unconstrained degrees of freedom.

However, when two $T_\mathsf{B}$ twists are introduced to a patch with a blue boundary, the total fermion parity is no longer a product of plaquette (and orange) operators,
whereas it still commutes with all $s\in\mathcal{S}$.
So we can expect to find a fermion-parity-odd logical MZM which also commutes with all members of the stabilizer group.
Indeed, if we count both fermion-parity odd and even states, the only non-trivial constraint on the plaquette operators before and after the introduction of twists is $\prod_{p\in\mathcal{RG}} \mathcal{O}_p = \mathds{1}$.
So introducing two $T_\mathsf{B}$ twists increases the number of unconstrained degrees of freedom by 1 (since it decreases the number of independent stabilizer generators along the domain wall), and $\Gamma = \pm 1$ states are now degenerate.
The logical MZM (along with the logical operator $\Gamma$) is shown in Figure~\ref{subfig:Boundary_ops_a}.
The logical MZM is equivalent to a process of producing an $\mathsf{RGB}$ anyon, splitting this into $\mathsf{RG}\times \mathsf{B}$, localizing the $\mathsf{RG}$ anyon at one of the twists and condensing the $\mathsf{B}$ anyon at the boundary.
We can similarly find logical MZMs whenever we have two $Z_3$ twists in a patch with a boundary, since $Z_3$ twists can localize $\mathsf{RG}$, $\mathsf{GB}$ and $\mathsf{BR}$ anyons.
In order to encode a qubit with this logical MZM (which requires working with a single total parity sector~\cite{MZM_TQC_Review2015}), we can either encode it into two copies of this code (in which case $\Gamma$ above becomes the patch fermion parity instead of the total fermion parity, leading to a setup equivalent to Figure~\ref{subfig:Z2_Twist_logical_ops} upon merging boundaries), or we can use fermionic twists instead, which support fermion-parity-preserving logical operators as shown in Figure~\ref{subfig:Boundary_ops_b}.

\section{Generating Encoded Clifford Gates by Braiding Twists}\label{app:Braiding}

\begin{figure}
\centering
\includegraphics[width=\linewidth]{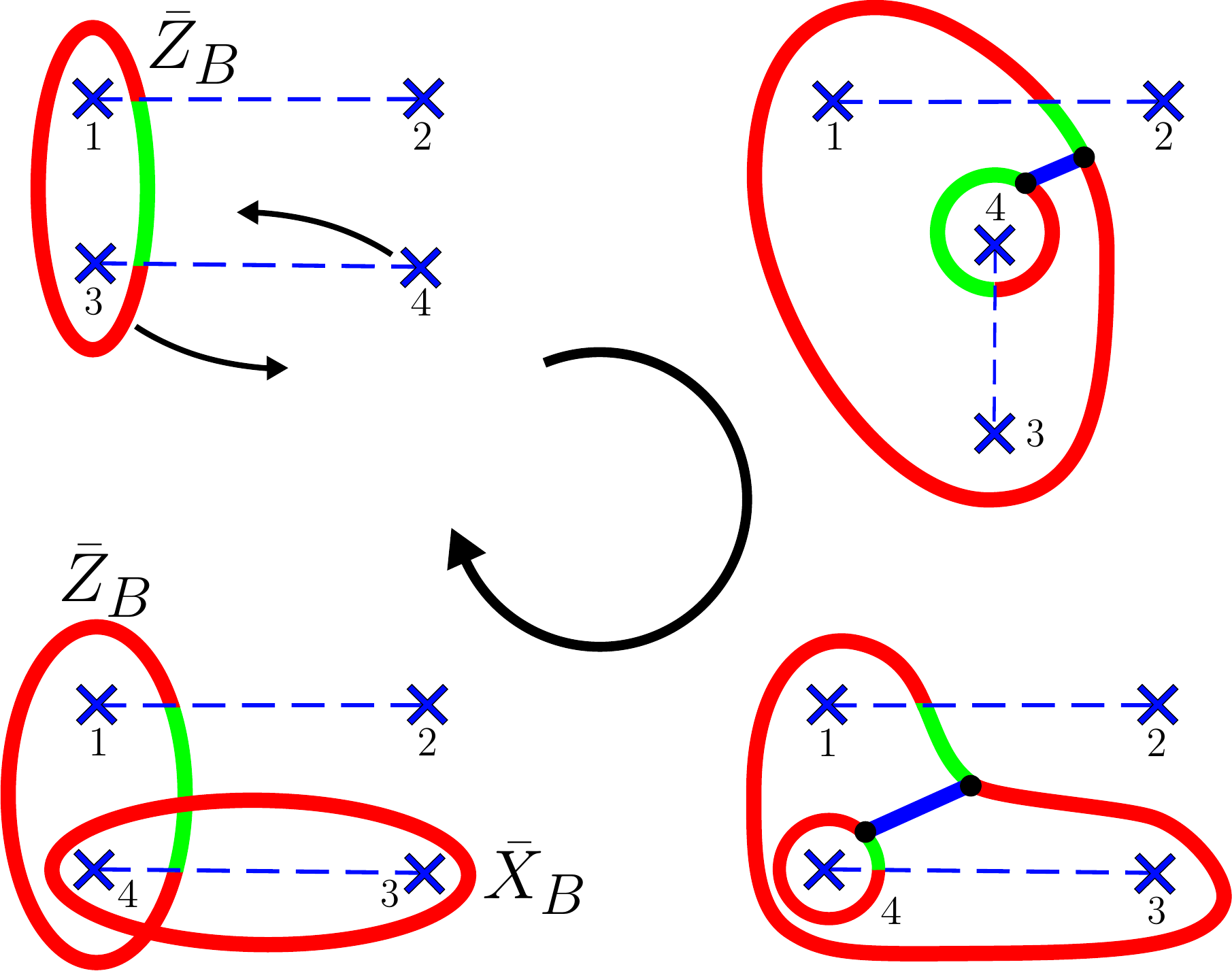}
\caption{Braiding twists $T_3$ and $T_4$ transforms $\bar{Z}_B$ into $\bar{Y}_B$. $\bar{X}_B$, which is a red string operator encircling $T_1$ and $T_2$, is not affected by this braid. The steps of the braid proceed according to the panels, clockwise from the top-left.}\label{fig:Twist_Braid}
\end{figure}

Consider a BT qubit stored between four blue twists, $T_1,T_2,T_3,T_4$, with logical operators $\bar{X}$ and $\bar{Z}$.
The resulting setup is shown in Figure~\ref{fig:Twist_Braid}.
By exchanging twists $T_3$ and $T_4$ (or $T_1$ and $T_2$) we send $\bar{Z}\mapsto \bar{Y}$ and $\bar{Y}\mapsto \bar{Z}$ up to a phase, while preserving $\bar{X}$, as shown in Figure~\ref{fig:Twist_Braid}.

This braiding operation, which we label $\mathcal{B}_1$, can be written in terms of encoded logic gates as $\mathcal{B}_1 = \bar{H}\bar{S}\bar{H}$ (where $\bar{H}$ and $\bar{S}$ are encoded Hadamard and phase gates respectively) up to a logical Pauli operator, since this reproduces the transformations of the Pauli operators found for $\mathcal{B}_1$ (where $\sim$ indicates equality up to a phase):
\begin{align}
&(\bar{H}\bar{S}\bar{H}) \bar{Z}(\bar{H}\bar{S}^\dagger \bar{H}) = -\bar{Y} \sim \mathcal{B}_1 \bar{Z} \mathcal{B}_1^\dagger,\\
&(\bar{H} \bar{S}\bar{H})\bar{Y}(\bar{H}\bar{S}^\dagger\bar{H}) = \bar{Z} \sim \mathcal{B}_1 \bar{Y} \mathcal{B}_1^\dagger ,\\
&(\bar{H}\bar{S}\bar{H})\bar{X}(\bar{H}\bar{S}^\dagger\bar{H}) = \bar{X} \sim \mathcal{B}_1\bar{X} \mathcal{B}_1^\dagger.
\end{align}
The phase can be set by a careful definition of the logical Pauli operators and plaquette operators in the model.

In the same way, exchanging twists $T_2$ and $T_4$ preserves the operator $\bar{Z}$ while sending $\bar{X}\mapsto \bar{Y}$ and $\bar{Y} \mapsto \bar{X}$. 
This has the equivalent action as the logical phase gate $\bar{S}$. 
Thus we have two independent generators of the single-qubit Clifford group and so we can use these to generate all such gates. 

\begin{figure*}
    \centering
    \subfloat[\label{subfig:Z3_Braid_Extras_a}]{\includegraphics[width=0.75\linewidth]{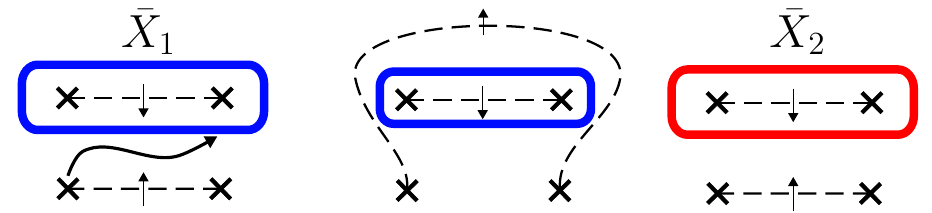}}\\
    \subfloat[\label{subfig:Z3_Braid_Extras_b}]{\includegraphics[width=0.8\linewidth]{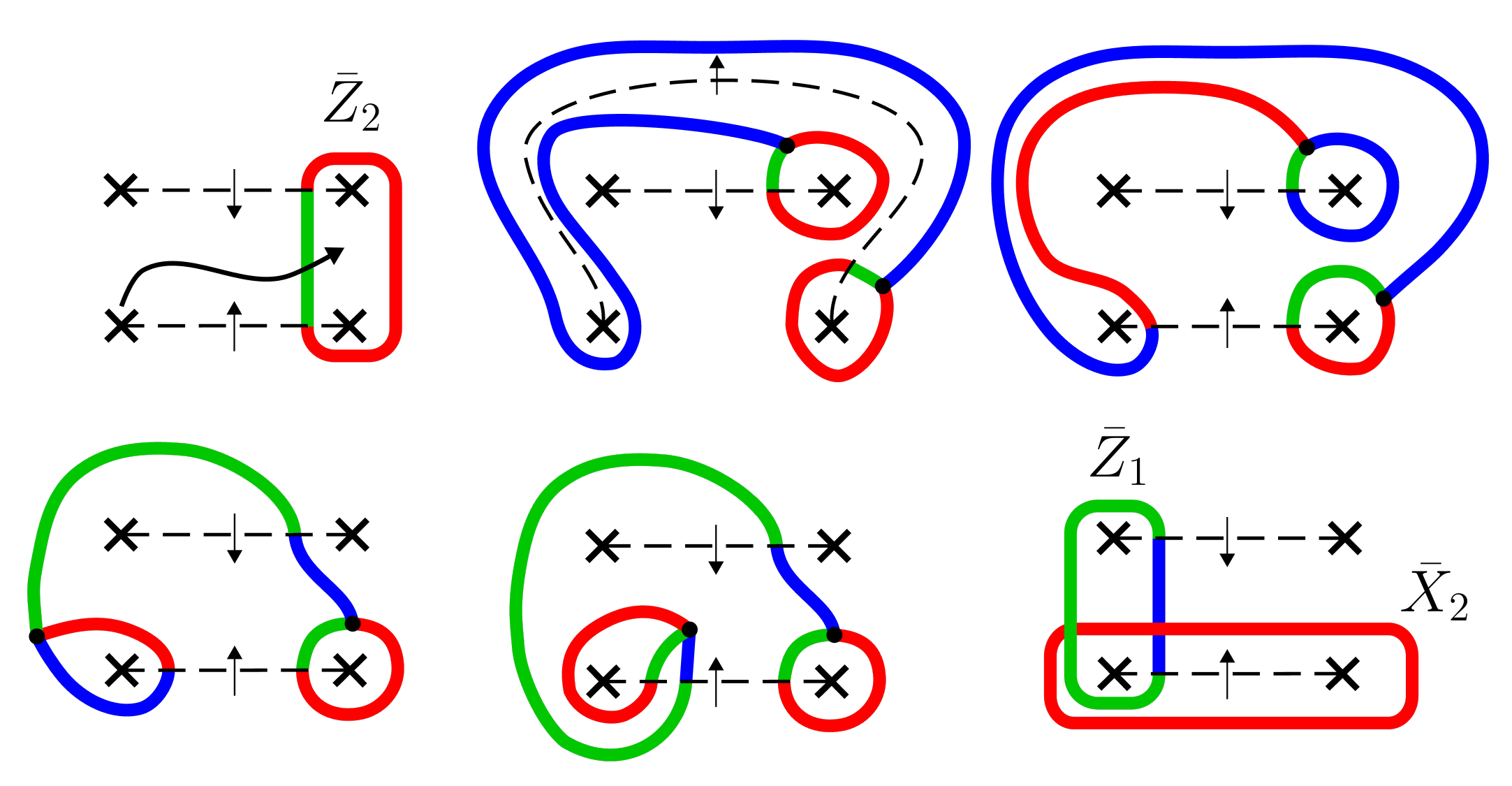}}\\
    \subfloat[\label{subfig:Z3_Braid_Extras_c}]{\includegraphics[width=0.9\linewidth]{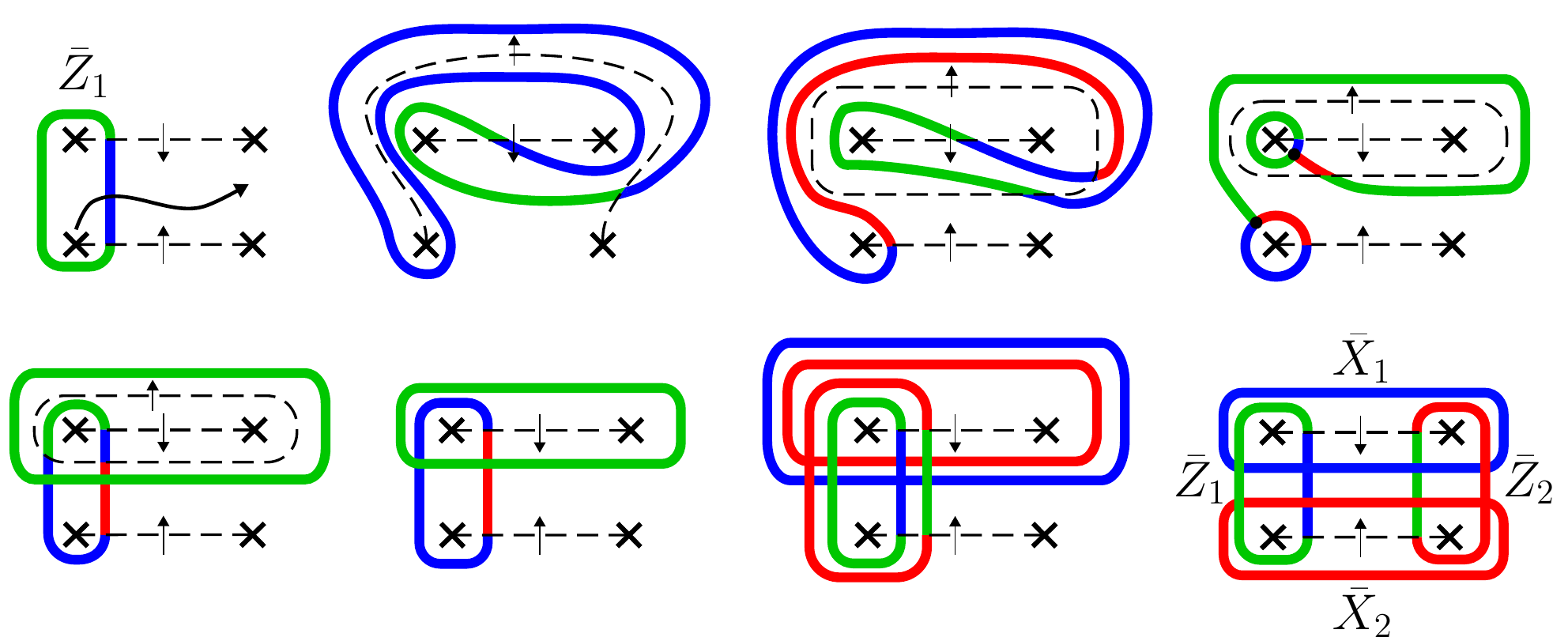}}
    \caption{Braiding $Z_3$ twists results in the following transformations of Pauli operators: (a) $\bar{X}_1\mapsto \bar{X}_2$, (b) $\bar{Z}_2 \mapsto \bar{Z}_1\bar{X}_2$, and (c) $\bar{Z}_1 \mapsto \bar{Y}_1\bar{Y}_2$.}
    \label{fig:Z3_Braid_Extras}
\end{figure*}

In Figure~\ref{fig:Z3_Braid_Extras}, we show how braiding two $Z_3$ twists in a quartet affects the logical operators $\bar{X}_1$, $\bar{Z}_2$ and $\bar{Z}_1$ for the two BT qubits.
The effect on logical operator $\bar{X}_2$ is shown in Figure~\ref{fig:Z3_CNOT_Braid} of the main text.

\section{Lattice Surgery for Large-Weight Logical MZM Bilinear Measurement}\label{app:Lattice_Surg}

\begin{figure*}[t]
\centering
\subfloat[]{\includegraphics[width=0.95\textwidth]{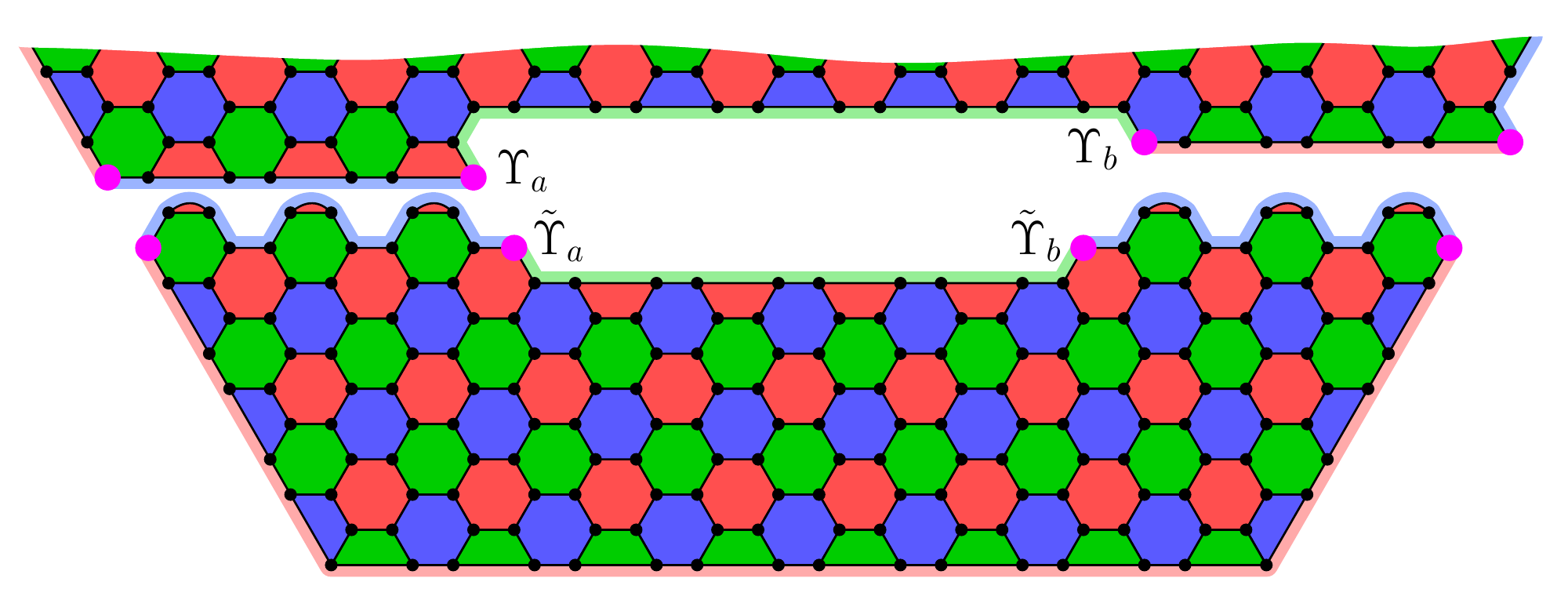}}\\
\subfloat[\label{subfig:Bilinear_lattice_surg_b}]{\includegraphics[width=0.95\textwidth]{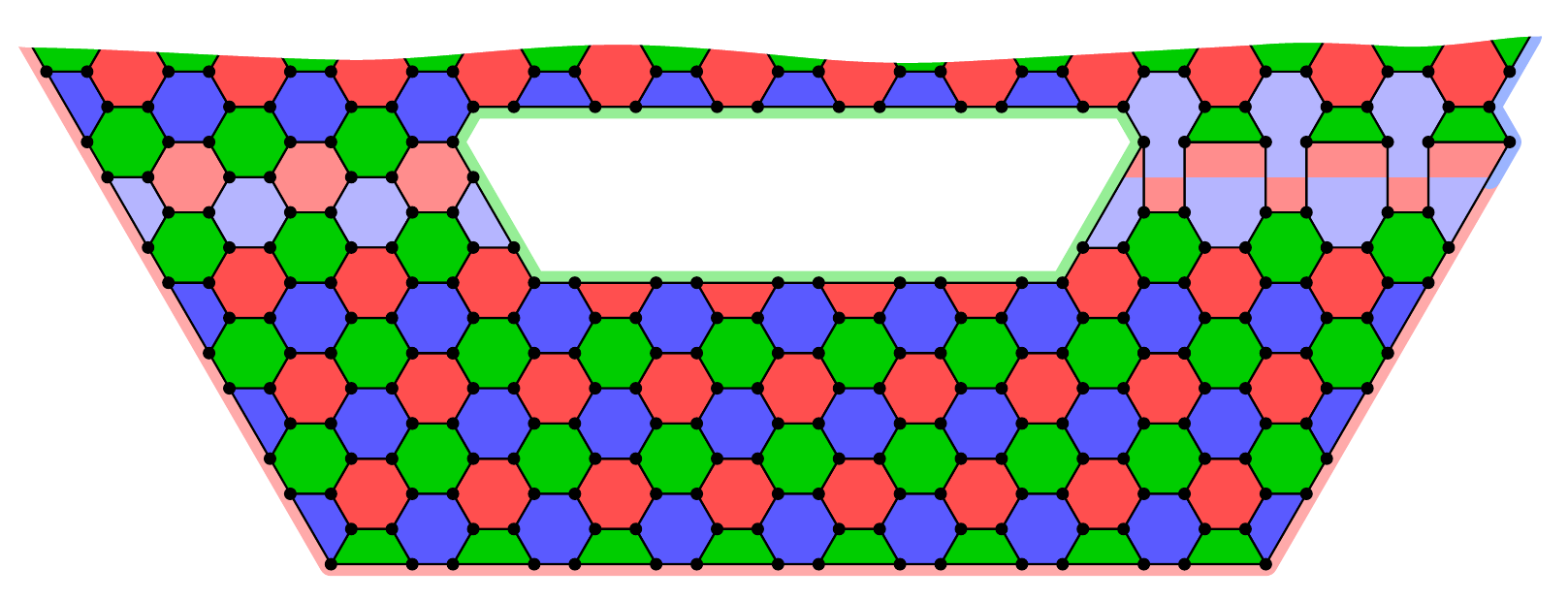}}
\caption{Procedure for measuring a high-weight logical MZM bilinear, $i\Upsilon_a\Upsilon_b$. In (a), a small section of surface code patch is shown in the top-half of the figure, with the color of all boundaries indicated. Logical MZMs $\Upsilon_a$ and $\Upsilon_b$ are supported on odd-length boundaries of color blue and red respectively. Ancilla logical MZMs $\tilde{\Upsilon}_a$ and $\tilde{\Upsilon}_b$ are supported on blue boundaries. The ancilla patch is prepared in the $+1$ eigenstate of $i\tilde{\Upsilon}_a\tilde{\Upsilon}_b$. (b) Plaquette operators connecting the surface code patch with the ancilla patch are then measured. The product of light blue plaquettes (left) and four- and six-body plaquettes (right) are the operators $i\Upsilon_a\tilde{\Upsilon}_a$ and $i\Upsilon_b\tilde{\Upsilon}_b$ respectively. Once these operators are measured non-destructively, destructive measurement of $i\tilde{\Upsilon}_a\tilde{\Upsilon}_b$ is performed. The product of the first two measurements gives the measurement outcome of $i\Upsilon_a\Upsilon_b$ while the final measurement determines the correction operation required.}\label{fig:Bilinear_High_weight}
\end{figure*}

In Figure~\ref{fig:Bilinear_High_weight}, we illustrate the lattice surgery procedure for measuring the bilinear operator $i\Upsilon_a\Upsilon_b$, where these two logical MZMs have arbitrarily high weights.
In the example, $\Upsilon_a$ is supported on a blue boundary and $\Upsilon_b$ on a red boundary.
Similarly to in Section~\ref{subsec:Bilinear_Mmt}, we prepare a patch, supporting ancilla logical MZMs $\tilde{\Upsilon}_a$ and $\tilde{\Upsilon}_b$, in the state $i\tilde{\Upsilon}_a\tilde{\Upsilon}_b = +1$. 
These logical MZMs will have a similar weight to $\Upsilon_a$ and $\Upsilon_b$ respectively, and will be supported on blue boundaries.
We measure the extra plaquette operators shown in Figure~\ref{subfig:Bilinear_lattice_surg_b} connecting the two patches.
Note that the product of the eigenvalues of those plaquette operators connecting $\Upsilon_a$ and $\tilde{\Upsilon}_a$ is equal to the eigenvalue of $i\Upsilon_a\tilde{\Upsilon}_a$, and similarly for $i\Upsilon_b\tilde{\Upsilon}_b$. 
We then measure $i\tilde{\Upsilon}_a\tilde{\Upsilon}_b$ destructively (by measuring all blue bond operators) and perform/track a correction operation dependent on this measurement outcome (cf. Section~\ref{subsec:Bilinear_Mmt}).

%

%

%

\section{Calculating the Correction Operation for Logical Fermion Parity Measurement}\label{app:Correction_Operation}

Here, we go through the details of Section~\ref{subsec:Arb_Ferm_Parity_Mmt}, specifically showing how to determine the correction operation that needs to be tracked when measuring a logical fermion parity operator.

To determine this operation, we need to assess whether each logical bilinear operator has been preserved by the measurement-based procedure outlined.
We deal with the example of $i\Upsilon_1\Upsilon_3$ specifically. 
We can use the final column of Table~\ref{table:Four_body_mmt_ops} to determine that after the lattice surgery measurements, this bilinear is mapped to $\lambda_1\lambda_3\bar{Z}_B$ (it is unchanged by the final measurement of $i\Upsilon_4\tilde{\Upsilon}_4$).
The final step of the procedure from Section~\ref{subsec:Arb_Ferm_Parity_Mmt} is to measure all blue bond operators on the ancilla patch, from which we deduce the measurement outcome of operator $\tilde{\Gamma}_4$ to be $\eta_1$, $i\tilde{\Upsilon}_1\tilde{\Upsilon}_2\, \bar{X}_B$ to be $\eta_2$ and $i\tilde{\Upsilon}_2\tilde{\Upsilon}_4\, \bar{Z}_B$ to be $\eta_3$ (these measurements do not commute with $\bar{Z}_B$).
The post-measurement state of the logical qubits can be obtained by acting three logical operators $R_1$, $R_2$ and $R_3$ on the pre-measurement state. 
This is equivalent to replacing projectors with logical braids in Section~\ref{subsec:Bilinear_Mmt}.

We now determine $R_1$, $R_2$ and $R_3$ for this example.
Note the logical operators stabilizing the pre-measurement state are those that were just measured via lattice surgery: $\lambda_j i\Upsilon_j\tilde{\Upsilon}_j$ for $j=1,2,3,4$.
We consider first the measurement of $\tilde{\Gamma}_4$ (it does not matter which we choose to consider first, since the measurements all commute).
A logical braid that enacts the same transformation as this measurement is (see Section~\ref{subsec:Bilinear_Mmt}):
\begin{align}
    R_1 &= \exp{\left(\eta_1\lambda_1 \frac{\pi}{4}\tilde{\Gamma}_4 i\Upsilon_1\tilde{\Upsilon}_1\right)}\\
    &= \exp{\left(i\eta_1\lambda_1 \frac{\pi}{4} \Upsilon_1\tilde{\Upsilon}_2\tilde{\Upsilon}_3\tilde{\Upsilon}_4\right)}.
\end{align}
The measurement operator anti-commutes with all $\lambda_j i\Upsilon_j\tilde{\Upsilon}_j$ and so only products of pairs of these operators will stabilize the post-measurement state.
The full list of independent operators stabilizing this state is: $\eta_1\tilde{\Gamma}_4$, and $\lambda_1\lambda_j \Upsilon_1\Upsilon_j\tilde{\Upsilon}_1\tilde{\Upsilon}_j$ for $j=2,3,4$.
Note also that $\Gamma_4 = \eta_1\lambda_1\lambda_2\lambda_3\lambda_4$ on the post-measurement state.

We then consider the measurement of $i\tilde{\Upsilon}_1 \tilde{\Upsilon}_2\, \bar{X}_B$. 
This anti-commutes with operators $\lambda_1\lambda_3\Upsilon_1\Upsilon_3\tilde{\Upsilon}_1\tilde{\Upsilon}_3$ and $\lambda_1\lambda_4\Upsilon_1\Upsilon_4\tilde{\Upsilon}_1 \tilde{\Upsilon}_4$, so we can choose:
\begin{align}
    R_2 &=  \exp{\left(\eta_2\lambda_1\lambda_3 \frac{\pi}{4} (i \tilde{\Upsilon}_1 \tilde{\Upsilon}_2 \bar{X}_B) ( \Upsilon_1\Upsilon_3\tilde{\Upsilon}_1\tilde{\Upsilon}_3)\right)}\\
    &= \exp{\left(\eta_2\lambda_1\lambda_3 \frac{\pi}{4} i \tilde{\Upsilon}_2 \Upsilon_3\Upsilon_1\tilde{\Upsilon}_3 \bar{X}_B\right)}.
\end{align}
We then have the stabilizing operators: $\eta_1 \tilde{\Gamma}_4$, $\eta_2 i\tilde{\Upsilon}_1\tilde{\Upsilon}_2\, \bar{X}_B$, $\lambda_1\lambda_2\Upsilon_1\Upsilon_2\tilde{\Upsilon}_1 \tilde{\Upsilon}_2$ and $\eta_1\lambda_1\lambda_2\lambda_3\lambda_4 \Gamma_4$.
Finally we consider the measurement of $i\tilde{\Upsilon}_2\tilde{\Upsilon}_4 \bar{Z}_B$.
We can choose:
\begin{align}
    R_3 &= \exp\left( \eta_3 \lambda_1\lambda_2 \frac{\pi}{4} (i \tilde{\Upsilon}_2\tilde{\Upsilon}_4 \bar{Z}_B) ( \Upsilon_1 \Upsilon_2 \tilde{\Upsilon}_1\tilde{\Upsilon}_2)\right).
\end{align}
We have the final stabilizing operators: $\eta_1\tilde{\Gamma}_4$, $\eta_2 i\tilde{\Upsilon}_1\tilde{\Upsilon}_2\, \bar{X}_B$, $\eta_3 i\tilde{\Upsilon}_2\tilde{\Upsilon}_4\, \bar{Z}_B$ and $\eta_1\lambda_1\lambda_2\lambda_3\lambda_4 \Gamma_4$.

The operator $\lambda_1\lambda_3\bar{Z}_B$ is therefore updated by the measurements to:
\begin{align}
    &R_3 R_2 R_1 (\lambda_1\lambda_3\bar{Z}_B) R_1^\dagger R_2^\dagger R_3^\dagger = i\eta_2 \tilde{\Upsilon}_2 \Upsilon_3 \Upsilon_1 \tilde{\Upsilon}_3 \bar{X}_B \bar{Z}_B.
\end{align}
We can then multiply this operator by the stabilizers $\eta_2 i\tilde{\Upsilon}_1\tilde{\Upsilon}_2\, \bar{X}_B$, $\eta_3 i\tilde{\Upsilon}_2\tilde{\Upsilon}_4\,\bar{Z}_B$ and $\eta_1 \tilde{\Gamma}_4$ to find the operator that acts equivalently on the post-measurement state.
This operator is $-\eta_1\eta_3 i\Upsilon_1\Upsilon_3$.
Hence, we find that the bilinear $i\Upsilon_1\Upsilon_3$ is preserved whenever $-\eta_1\eta_3 = 1$.
Otherwise, we need to apply/track a correction operation which flips (i.e., anti-commutes with) operator $i\Upsilon_1\Upsilon_3$.

We do the same for all other bilinear operators. 
The appropriate correction operation is one that anti-commutes with all bilinears that need to be flipped and commutes with all bilinears that do not need to be flipped.

\section{Arbitrary String Operator Measurement Details}\label{app:Measurement_Details}

\begin{figure*}[t]
    \centering
    \subfloat[\label{subfig:arb_mmt_app_1}]{\includegraphics[width=0.95\textwidth]{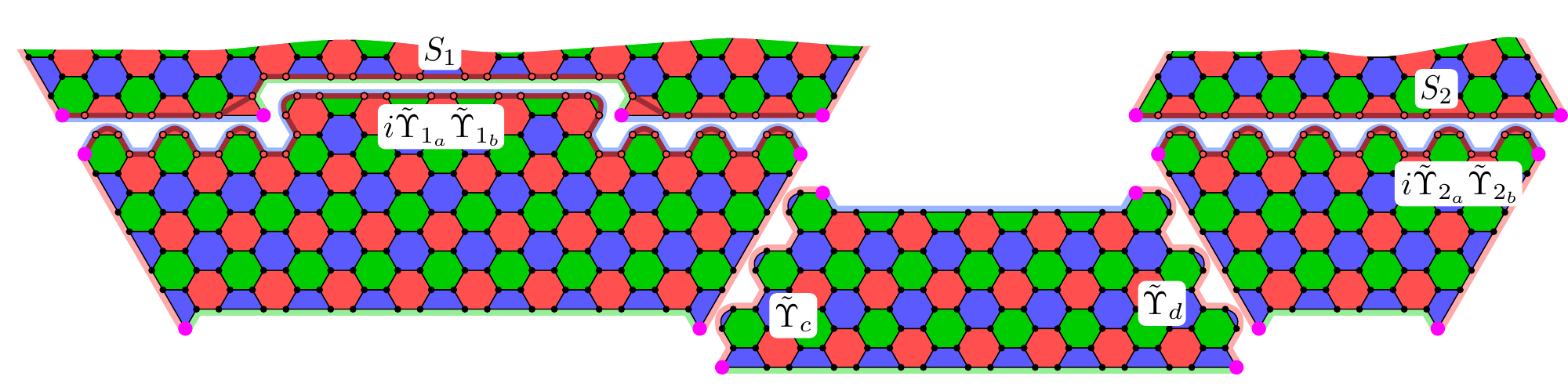}}\\
    \subfloat[\label{subfig:arb_mmt_app_2}]{\includegraphics[width=0.95\textwidth]{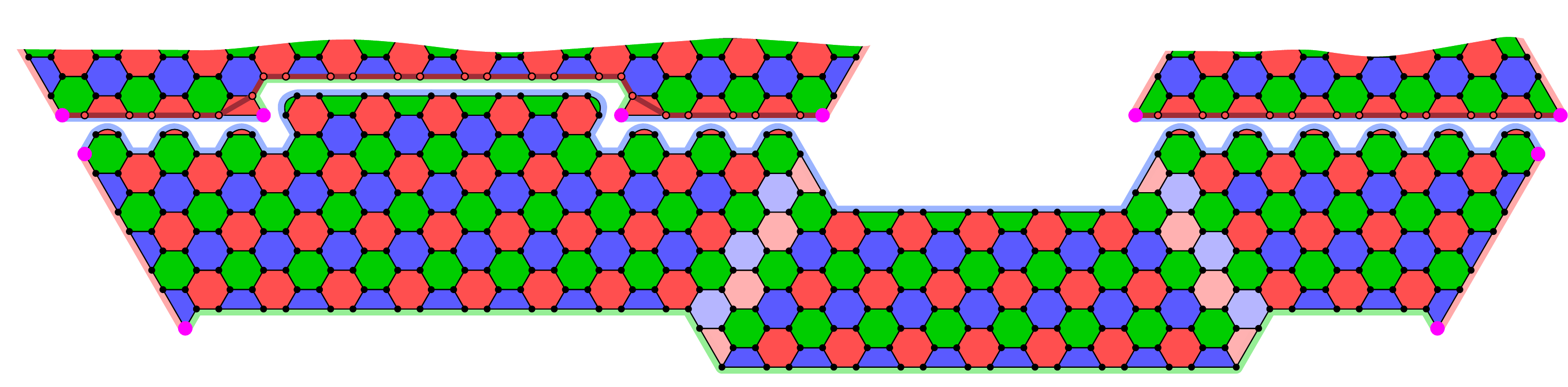}}\\
    \subfloat[\label{subfig:arb_mmt_app_3}]{\includegraphics[width=0.95\textwidth]{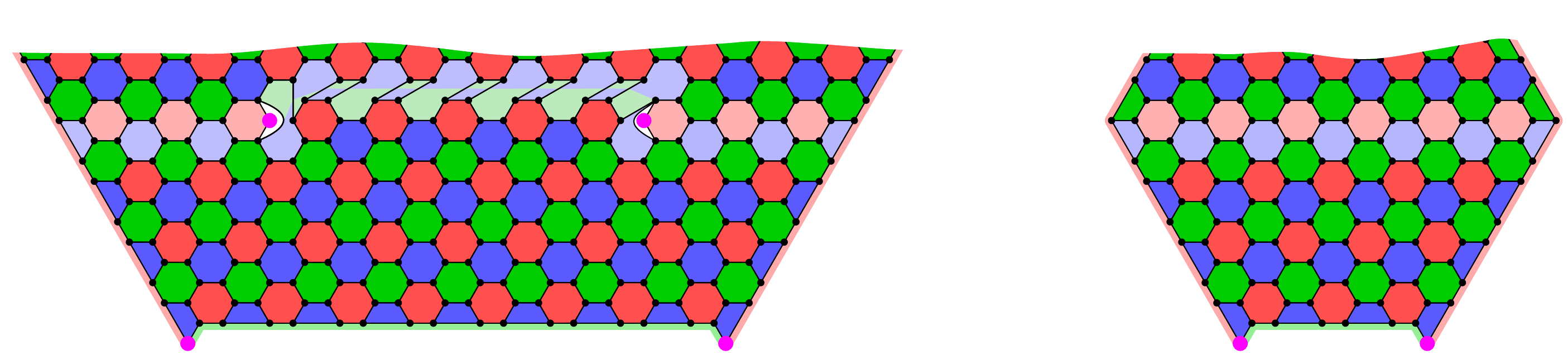}}
    \caption{Example setup for measuring a product of string operators $S_1S_2$, here represented as red string operators. (a) Ancilla patches host logical MZMs $\tilde{\Upsilon}_{1_a}$ and $\tilde{\Upsilon}_{1_b}$ (left), $\tilde{\Upsilon}_{c}$ and $\tilde{\Upsilon}_{d}$ (centre), $\tilde{\Upsilon}_{2_a}$ and $\tilde{\Upsilon}_{2_b}$ (right). (b) Ancilla patches are fused with lattice surgery to measure $i\tilde{\Upsilon}_{1_b}\tilde{\Upsilon}_{2_a}$. (c) Lattice surgery for measuring $S_1 (i\tilde{\Upsilon}_{1_a}\tilde{\Upsilon}_{1_b})$ and $S_2(i\tilde{\Upsilon}_{2_a}\tilde{\Upsilon}_{2_b})$.}
    \label{fig:arb_mmt_app}
\end{figure*}

Here we extend the procedures of Section~\ref{subsec:Hybrid_Approaches} to measuring arbitrary products of fermionic and bosonic string operators. 
We assume we wish to measure a product of $k$ operators, labeled $S_1,\ldots,S_k$ and colored either red or blue.
We start by introducing $k$ ancilla patches, each hosting two logical MZMs, $\tilde{\Upsilon}_{j_a}$ and $\tilde{\Upsilon}_{j_b}$ for $j=1,\ldots,k$, and we prepare each patch in the $i\tilde{\Upsilon}_{j_a}\tilde{\Upsilon}_{j_b} = +1$ state.
This also prepares the $+1$ eigenstate of operator $\tilde{\Gamma} = \prod_{j=1}^k i\tilde{\Upsilon}_{j_a} \tilde{\Upsilon}_{j_b}$.
An example of this setup is shown in Figure~\ref{subfig:arb_mmt_app_1} for $k=2$. 
In this example, $S_1$ is a bosonic string operator encircling a pair of fermionic twists hosting logical MZMs on blue boundaries. Both $S_1$ and $S_2$ are red string operators.
We then measure $i\tilde{\Upsilon}_{j_b}\tilde{\Upsilon}_{(j+1)_a}$ for all $j<k$, to entangle these ancilla patches, while preserving the state of $\tilde{\Gamma}$.
To achieve this, we can bridge the gap between ancilla patches with other ancilla patches.
In Figure~\ref{subfig:arb_mmt_app_1}, we introduce a patch hosting $\tilde{\Upsilon}_c$ and $\tilde{\Upsilon}_d$. 
The measurement of $i\tilde{\Upsilon}_{1_b}\tilde{\Upsilon}_{2_a}$ is performed via lattice surgery, shown in Figure~\ref{subfig:arb_mmt_app_2} (see also Section~\ref{subsec:Bilinear_Mmt} for an explanation of this procedure).
After this measurement, the middle patch hosting $\tilde{\Upsilon}_c$ and $\tilde{\Upsilon}_d$ is discarded.

We then measure $S_j(i\tilde{\Upsilon}_{j_a}\tilde{\Upsilon}_{j_b})$ for all $j$, using lattice surgery.
This does not reveal any information about the individual $S_j$ but we can use the fact that $\tilde{\Gamma} = 1$ to learn the eigenvalue of $\prod_{j=1}^k S_j$. 
This measurement is shown in Figure~\ref{subfig:arb_mmt_app_3}.
Finally we measure $i\tilde{\Upsilon}_{j_2}\tilde{\Upsilon}_{(j+1)_1}$ for all $j$ once again (with $k+1 \equiv 1$), to disentangle the ancilla patches from the code lattice.
We can then discard the ancilla patches.
Our code patch is left in an eigenstate of $\prod_{j=1}^k S_j$.
We may need to apply/track a correction operation that is dependent on the final ancilla measurements.

Notice that the logical MZMs encircled by $S_1$ in the example of Figure~\ref{fig:arb_mmt_app} commute with all lattice surgery measurements. 
Interestingly, there are only bosonic twists present during the lattice surgery phase (see pink dots in Figure~\ref{subfig:arb_mmt_app_3}) and hence the logical MZMs are temporarily of the form discussed in Appendix~\ref{app:Logical_Maj_Boundary}.

\end{document}